\documentclass[preprint,10pt]{elsarticle}
\pdfoutput=1


\makeatletter
\def\ps@pprintTitle{%
   \let\@oddhead\@empty
   \let\@evenhead\@empty
   \def\@oddfoot{\reset@font\hfil\thepage\hfil}
   \let\@evenfoot\@oddfoot
}
\makeatother

\usepackage{titlesec}
\usepackage{xcolor}
\usepackage{enumitem}
\usepackage{hyperref}
\hypersetup{colorlinks=true,urlcolor=black}
\usepackage[T1]{fontenc}\usepackage{lmodern}

\usepackage{mathtools,amsfonts,amssymb}
\usepackage{caption}
\usepackage{subcaption}
\usepackage{booktabs}
\usepackage{pifont}
\usepackage{placeins}

\definecolor{green}{rgb}{0,0.5,0}

\usepackage{tikz}
\usetikzlibrary{calc}
\usepackage{tkz-euclide}
\usepackage{calc}
\usepackage{tikz,pgfplots}
\pgfplotsset{compat=1.12}
\usetikzlibrary{arrows.meta,3d,matrix,positioning,decorations.markings}

\newcommand{\pz}{\phantom{0}}
\newcommand{\vpad}{\vphantom{\bigg(}}
\newcommand{\ntriangles}{n_t}

\newcommand{\nbasis}{n_b}
\newcommand{\srcidx}{j}
\newcommand{\testidx}{i}

\makeatletter
\tikzoption{canvas is xy plane at z}[]{%
  \def\tikz@plane@origin{\pgfpointxyz{0}{0}{#1}}%
  \def\tikz@plane@x{\pgfpointxyz{1}{0}{#1}}%
  \def\tikz@plane@y{\pgfpointxyz{0}{1}{#1}}%
  \tikz@canvas@is@plane
}
\makeatother

\makeatletter
\tikzoption{canvas is plane}[]{\@setOxy#1}
\def\@setOxy O(#1,#2,#3)x(#4,#5,#6)y(#7,#8,#9)%
  {\def\tikz@plane@origin{\pgfpointxyz{#1}{#2}{#3}}%
   \def\tikz@plane@x{\pgfpointxyz{#4}{#5}{#6}}%
   \def\tikz@plane@y{\pgfpointxyz{#7}{#8}{#9}}%
   \tikz@canvas@is@plane
  }
\makeatother 

\definecolor{darkgray}{gray}{0.3}
\definecolor{rev_red}{RGB}{227,26,28}
\definecolor{rev_green}{RGB}{51,160,44}
\definecolor{rev_blue}{RGB}{31,120,180}
\definecolor{rev_orange}{RGB}{255,127,0}
\definecolor{rev_purple}{RGB}{106,61,154}

\definecolor{orange}{rgb}{1,0.5,0}
\definecolor{green}{rgb}{0,0.5,0}
\definecolor{purple}{rgb}{0.5,0,0.5}
\newcommand{\reviewerOne}[1]{#1}

\newcommand{\reviewerTwo}[1]{#1}
\usepackage[margin=1in]{geometry}
\biboptions{sort&compress}
\begin{document}

\begin{frontmatter}
\title{Code-Verification Techniques for the Method-of-Moments Implementation of the Magnetic-Field Integral Equation}
\author[freno]{Brian A.\ Freno}
\ead{bafreno@sandia.gov}
\author[freno]{Neil R.\ Matula}
%
\address[freno]{Sandia National Laboratories, Albuquerque, NM 87185}
\begin{abstract}
\reviewerTwo{For computational physics simulations, code verification plays a major role in establishing the credibility of the results by assessing the correctness of the implementation of the underlying numerical methods.  In computational electromagnetics, surface integral equations, such as the method-of-moments implementation of the magnetic-field integral equation, are frequently used to solve Maxwell's equations on the surfaces of electromagnetic scatterers.  These} electromagnetic surface integral equations yield many code-verification challenges due to the various sources of numerical error and their possible interactions.  In this paper, we provide approaches to separately measure the numerical errors arising from these different error sources.  We demonstrate the effectiveness of these approaches for cases with and without coding errors.
\end{abstract}

\end{frontmatter}

\section{Introduction}

In computational electromagnetics, surface integral equations are frequently used to solve Maxwell's equations on the surfaces of electromagnetic scatterers.  The electric-field integral equation (EFIE) relates the surface current to the scattered electric field that arises from an incident electric field, whereas the magnetic-field integral equation (MFIE) relates the surface current to the scattered magnetic field that arises from an incident magnetic field.  

Through the method-of-moments (MoM) implementation of these surface integral equations, the surface of the electromagnetic scatterer is discretized using planar or curvilinear mesh elements, and four-dimensional integrals are evaluated over two-dimensional source and test elements.  However, the presence of a Green's function in these equations yields singularities when the test and source elements share one or more edges or vertices, and near-singularities when they are otherwise close.  Many approaches have been developed to address the singularity and near-singularity for the inner, source-element integral~\cite{graglia_1993,wilton_1984,rao_1982,khayat_2005,fink_2008,khayat_2008,vipiana_2011,vipiana_2012,botha_2013,rivero_2019}, as well as for the outer, test-element integral~\cite{vipiana_2013,polimeridis_2013,wilton_2017,rivero_2019b,freno_em}.

Code verification is an important step towards establishing the credibility of the results of a computational physics simulation~\reviewerTwo{\cite{roache_1998,knupp_2022,oberkampf_2010}}.  Through code verification, the correctness of the implementation of the numerical methods is assessed.  The discretization of differential, integral, or integro-differential equations incurs some truncation error, and thus the approximate solutions produced from the discretized equations will incur an associated discretization error.  If the discretization error tends to zero as the discretization is refined, the consistency of the code is verified~\cite{roache_1998}.  This may be taken a step further by examining not only consistency, but the rate at which the error decreases as the discretization is refined, thereby verifying the order of accuracy of the discretization scheme.  The correctness of the numerical-method implementation may then be verified by comparing the expected and observed orders of accuracy obtained from numerous test cases with known solutions. 

Because exact solutions are generally limited and may not sufficiently exercise the capabilities of the code, manufactured solutions~\cite{roache_2001} are a popular alternative, permitting the construction of problems of arbitrary complexity with known solutions.  Through the method of manufactured solutions (MMS), a solution is manufactured and substituted directly into the governing equations to yield a residual term, which is added as a source term to coerce the solution to the manufactured solution. 

However, integral equations yield an additional challenge.  While analytical differentiation is straightforward, analytical integration is not always possible.  Therefore, the residual source term arising from the manufactured solution may not be representable in closed form, and its implementation may be accompanied by numerical techniques that carry their own \reviewerTwo{numerical errors}.  Furthermore, in many applications, such as the MoM implementation of the MFIE, singular integrals appear, which can further complicate the numerical evaluation of the source term.  Therefore, many of the benefits associated with MMS are lost when applied to integral equations in a straightforward manner.

Code verification has been performed on computational physics codes associated with several physics disciplines, including aerodynamics~\cite{nishikawa_2022}, fluid dynamics~\cite{roy_2004,bond_2007,veluri_2010,oliver_2012,eca_2016,hennink_2021,freno_2021}, solid mechanics~\cite{chamberland_2010}, fluid--structure interaction~\cite{etienne_2012}, heat transfer in fluid--solid interaction~\cite{veeraragavan_2016}, multiphase flows~\cite{brady_2012,lovato_2021}, radiation hydrodynamics~\cite{mcclarren_2008}, electrostatics~\cite{tranquilli_2022}, electrodynamics~\cite{ellis_2009}, and ablation~\cite{amar_2008,amar_2009,amar_2011,freno_ablation,freno_ablation_2022}.  
For surface integral equations in computational electromagnetics, code-verification activities that employ manufactured solutions have been limited to the EFIE~\cite{marchand_2013,marchand_2014,freno_em_mms_2020,freno_em_mms_quad_2021}.

\reviewerTwo{As with the EFIE, the numerical solution to the MoM implementation of the MFIE incurs numerical error from three sources:}
%
\begin{enumerate}[leftmargin=*,labelindent=\parindent]
\item \label{err:dom_disc} \textbf{Domain discretization.} While planar surfaces can be represented exactly by planar elements, the approximation of sufficiently smooth curved surfaces with planar elements introduces a second-order numerical error~\cite[Chap.~3]{warnick_2008}.  This error can be reduced by employing curved elements~\cite{graglia_1997}.

\item \label{err:sol_disc} \textbf{Solution discretization.} Common in the solution to differential, integral, and integro-differential equations, the approximation of the solution in terms of a finite number of basis functions, or alternatively the approximation of the underlying equation operators in terms of a finite amount of solution queries, is the most common contributor to the numerical error.  For sufficiently smooth solutions, this error can be reduced by employing higher-order basis functions~\cite{graglia_1997} or stencils.

\item \label{err:num_int}\textbf{Numerical integration.}
The analytical evaluation of the integrals in integral equations is usually not possible.  For well-behaved integrals, quadrature rules or other integration methods can be used, with the expectation that the associated numerical error is at least of the same order as that arising from the solution-discretization error.  A less rigorous expectation is that the error from numerical integration decreases as the fidelity of the numerical integration algorithm is increased (e.g., increasing the number of quadrature points).  However, for singular or nearly singular integrals, such convergence is not assured~\cite{freno_quad}.

\end{enumerate}

For the EFIE, Marchand et al.~\cite{marchand_2013,marchand_2014} compute the MMS source term using additional quadrature points.  Freno et al.\ manufacture the Green's function, permitting the numerical-integration error to be eliminated and the solution-discretization error to be isolated~\cite{freno_em_mms_2020}.  They also provide approaches to isolate the numerical-integration error~\cite{freno_em_mms_quad_2021}.
Unlike the EFIE, however, the MFIE is used to model only closed surfaces, introducing additional constraints on the ability to manufacture solutions.

In this paper, we present code-verification techniques for the MoM implementation of the MFIE that measure these three error sources separately, to the extent possible.  As in~\cite{freno_em_mms_2020}, to eliminate the numerical-integration error, we manufacture the Green's function.  When the manufactured Green's function makes the equations practically singular, we select a unique solution by minimizing the error in two different norms, each with their own trade-offs.  To isolate the numerical-integration error, we present two approaches and show how the numerical-integration error is related to the discretization error.  Finally, to address domains with curvature, we present \reviewerTwo{code-verification} approaches that account for and neglect the curvature.

\reviewerOne{It is important to note that, by manufacturing the Green's function, we avoid the challenges associated with evaluating the aforementioned (nearly) singular integrals.  Given the computational expense of computing accurate reference solutions, assessing these integral evaluations is best accomplished through extensive unit testing as complementary code verification.  Examples for the EFIE are included in~\cite{freno_quad,freno_em}.}

This paper is organized as follows.  In Section~\ref{sec:mfie}, we describe the MoM implementation of the MFIE.  In Section~\ref{sec:mms}, we describe the challenges of using MMS with the MoM implementation of the MFIE, and we describe our approach to mitigating them.  In Section~\ref{sec:results}, we demonstrate the effectiveness of our approaches for cases with and without coding errors and with and without curvature.  In Section~\ref{sec:conclusions}, we summarize this work.

\section{The Method-of-Moments Implementation of the MFIE}
\label{sec:mfie}

In time-harmonic form, the scattered magnetic field $\mathbf{H}^\mathcal{S}$ due to induced surface currents on a scatterer can be computed by~\reviewerTwo{\cite{harrington_2001}}
\begin{align}
\mathbf{H}^\mathcal{S}(\mathbf{x}) = \frac{1}{\mu}\nabla\times\mathbf{A}(\mathbf{x}), 
\label{eq:Hs}
\end{align}
where the magnetic vector potential $\mathbf{A}$ is defined by
\begin{align}
\mathbf{A}(\mathbf{x})= \mu \int_{S'} \mathbf{J} (\mathbf{x}')G(\mathbf{x},\mathbf{x}')dS'.
\label{eq:A}
\end{align}
In~\eqref{eq:Hs} and~\eqref{eq:A}, the integration domain is the closed surface $S$ of a perfectly conducting scatterer.  Additionally, $\mathbf{J}$ is the electric surface current density, and $G$ is the Green's function 
\begin{align}
G(\mathbf{x},\mathbf{x}') = \frac{e^{-jkR}}{4\pi R},
\label{eq:G}
\end{align}
where $R=|\mathbf{x}-\mathbf{x}'|$, $k=\omega\sqrt{\mu\epsilon}$ is the wavenumber, and $\mu$ and $\epsilon$ are the permeability and permittivity of the surrounding medium.  

The total magnetic field $\mathbf{H}$ is the sum of the incident magnetic field $\mathbf{H}^\mathcal{I}$ and $\mathbf{H}^\mathcal{S}$.  On $S$,
\begin{align}
\mathbf{n}\times \mathbf{H} = \mathbf{J},
\label{eq:mfie_bc}
\end{align}
where $\mathbf{n}$ is the unit vector normal to $S$.  
From~\eqref{eq:Hs} and~\eqref{eq:A}, and noting that $\nabla\times\big[\mathbf{J}(\mathbf{x}')G(\mathbf{x},\mathbf{x}')\big]=\nabla G(\mathbf{x},\mathbf{x}')\times\mathbf{J}(\mathbf{x}')$ and $\nabla G(\mathbf{x},\mathbf{x}') = -\nabla' G(\mathbf{x},\mathbf{x}')$,
\begin{align*}
\mathbf{H}^\mathcal{S}(\mathbf{x}) = \int_{S'} \big[\mathbf{J}(\mathbf{x}')\times\nabla' G(\mathbf{x},\mathbf{x}')\big]dS'
\end{align*}
when $\mathbf{x}$ is just outside of $S$.  Therefore, at $S$,
\begin{align}
\mathbf{n}\times\mathbf{H}^\mathcal{S} = \lim_{\mathbf{x}\to S}\mathbf{n}\times \int_{S'} \big[\mathbf{J}(\mathbf{x}')\times\nabla' G(\mathbf{x},\mathbf{x}')\big]dS' = \frac{1}{2}\mathbf{J} + \mathbf{n}\times \int_{S'} \big[\mathbf{J}(\mathbf{x}')\times\nabla' G(\mathbf{x},\mathbf{x}')\big]dS',
\label{eq:limit}
\end{align}
where the final term is evaluated through principal value integration.
From~\eqref{eq:mfie_bc} and~\eqref{eq:limit} the MFIE at a point on the surface of the scatterer is~\reviewerTwo{\cite{chew_1995, balanis_2012}}
\begin{align}
\frac{1}{2}\mathbf{J} - \mathbf{n}\times \int_{S'} \big[\mathbf{J}(\mathbf{x}')\times\nabla' G(\mathbf{x},\mathbf{x}')\big]dS' = \mathbf{n}\times \mathbf{H}^\mathcal{I}.
\label{eq:mfie}
\end{align}

Projecting~\eqref{eq:mfie} onto an appropriate space $\mathbb{V}$ containing vector fields that are tangent to $S$ yields the variational form: find $\mathbf{J}\in\mathbb{V}$, such that
\begin{align}
\frac{1}{2}\int_S \bar{\mathbf{v}}\cdot\mathbf{J} dS - \int_S \bar{\mathbf{v}}(\mathbf{x})\cdot \bigg(\mathbf{n}(\mathbf{x})\times \int_{S'}\big[ \mathbf{J}(\mathbf{x}')\times\nabla' G(\mathbf{x},\mathbf{x}')\big]dS'\bigg) dS = \int_S \bar{\mathbf{v}}\cdot \big(\mathbf{n}\times \mathbf{H}^\mathcal{I}\big) dS
\label{eq:variational}
\end{align} 
for all $\mathbf{v}\in\mathbb{V}$, where the bar notation denotes complex conjugation.
We can write~\eqref{eq:variational} more succinctly as
\begin{align}
a(\mathbf{J},\mathbf{v}) = b\big(\mathbf{H}^\mathcal{I}, \mathbf{v}\big),
\label{eq:var_sesquilinear}
\end{align}
where the sesquilinear forms are defined by
\begin{align}
a(\mathbf{u}, \mathbf{v}) &{}= \frac{1}{2}\int_S \bar{\mathbf{v}}(\mathbf{x})\cdot\mathbf{u}(\mathbf{x}) dS - \int_S \bar{\mathbf{v}}(\mathbf{x})\cdot \bigg(\mathbf{n}(\mathbf{x})\times \int_{S'}\big[ \mathbf{u}(\mathbf{x}')\times\nabla' G(\mathbf{x},\mathbf{x}')\big]dS'\bigg) dS, \label{eq:a}
\\
b(\mathbf{u}, \mathbf{v}) &{}= \int_S \bar{\mathbf{v}}(\mathbf{x})\cdot [\mathbf{n}(\mathbf{x})\times \mathbf{u}(\mathbf{x})] dS. \label{eq:b}
\end{align}

To solve the variational problem~\eqref{eq:var_sesquilinear}, we discretize $S$ with a mesh composed of triangular elements and approximate $\mathbf{J}$ with $\mathbf{J}_h$ in terms of the Rao--Wilton--Glisson (RWG) basis functions $\boldsymbol{\Lambda}_{\srcidx}(\mathbf{x})$~\cite{rao_1982}:
\begin{align}
\mathbf{J}_h(\mathbf{x}) = \sum_{\srcidx=1}^{\nbasis} J_{\srcidx} \boldsymbol{\Lambda}_{\srcidx}(\mathbf{x}),
\label{eq:J_h}
\end{align}%
where $\nbasis$ is the total number of basis functions. 
\reviewerOne{There are other suitable basis function choices for the MFIE such as curl-conforming basis functions~\cite{peterson_2002,ergul_2006}; however, given that the MFIE is often combined with the EFIE to obtain the combined-field integral equation, we restrict the scope of this work to the RWG basis functions.}
The RWG basis functions are second-order accurate~\cite[pp.\ 155--156]{warnick_2008}, and are defined for a triangle pair by
\begin{align*}
\boldsymbol{\Lambda}_{\srcidx}(\mathbf{x}) = \left\{
\begin{matrix}
\displaystyle\frac{\ell_{\srcidx}}{2A_{\srcidx}^+}\boldsymbol{\rho}_{\srcidx}^+, & \text{for }\mathbf{x}\in T_{\srcidx}^+ \\[1em]
\displaystyle\frac{\ell_{\srcidx}}{2A_{\srcidx}^-}\boldsymbol{\rho}_{\srcidx}^-, & \text{for }\mathbf{x}\in T_{\srcidx}^- \\[1em]
\mathbf{0}, & \text{otherwise}
\end{matrix}
\right.,
\end{align*}
where $\ell_{\srcidx}$ is the length of the edge shared by the triangle pair, and $A_{\srcidx}^+$ and $A_{\srcidx}^-$ are the areas of the triangles $T_{\srcidx}^+$ and $T_{\srcidx}^-$ associated with basis function $\srcidx$.  $\boldsymbol{\rho}_{\srcidx}^+$ denotes the vector from the vertex of $T_{\srcidx}^+$ opposite the shared edge to $\mathbf{x}$, and $\boldsymbol{\rho}_{\srcidx}^-$ denotes the vector to the vertex of $T_{\srcidx}^-$ opposite the shared edge from $\mathbf{x}$.

These basis functions ensure that $\mathbf{J}_h$ is tangent to the mesh when using planar triangular elements.
Additionally, along the shared edge of the triangle pair, the component of $\boldsymbol{\Lambda}_{\srcidx}(\mathbf{x})$ normal to that edge is unity.  Therefore, for a triangle edge shared by only two triangles, the component of $\mathbf{J}_h$ normal to that edge is $J_\srcidx$.  The solution is considered most accurate at the midpoint of the edge~\cite[pp.\ 155--156]{warnick_2008}; therefore, we measure the solution at the midpoints.

Defining $\mathbb{V}_h$ to be the span of RWG basis functions associated with the mesh on $S$, the Galerkin approximation of~\eqref{eq:var_sesquilinear} is now: find $\mathbf{J}_h\in\mathbb{V}_h$, such that
\begin{align}
a(\mathbf{J}_h,\boldsymbol{\Lambda}_{\testidx}) = b\big(\mathbf{H}^\mathcal{I}, \boldsymbol{\Lambda}_{\testidx}\big)
\label{eq:proj_disc}
\end{align}
for $i=1,\hdots,\nbasis$.  
Letting $\mathbf{J}^h$ denote the vector of coefficients used to construct $\mathbf{J}_h$~\eqref{eq:J_h}, \eqref{eq:proj_disc} can be written in matrix form as $\mathbf{Z}\mathbf{J}^h = \mathbf{V}$,
%
%
where $J_{\srcidx}^h = J_{\srcidx}$, 
%
$V_{\testidx} =b\big(\mathbf{H}^\mathcal{I}, \boldsymbol{\Lambda}_{\testidx}\big)$,
%
and
%
$Z_{\testidx,\srcidx} = a(\boldsymbol{\Lambda}_{\srcidx},\boldsymbol{\Lambda}_{\testidx})$. 
\section{Manufactured Solutions} 
\label{sec:mms}

We define the residual functional for each test basis function as
\begin{align}
r_{\testidx}(\mathbf{u}) = a(\mathbf{u},\boldsymbol{\Lambda}_{\testidx}) -b\big(\mathbf{H}^\mathcal{I}, \boldsymbol{\Lambda}_{\testidx}\big).
\label{eq:res_func}
\end{align}
%
We can write the variational form~\eqref{eq:var_sesquilinear} in terms of~\eqref{eq:res_func} as
\begin{align}
r_{\testidx}(\mathbf{J}) = a(\mathbf{J},\boldsymbol{\Lambda}_{\testidx}) -b\big(\mathbf{H}^\mathcal{I}, \boldsymbol{\Lambda}_{\testidx}\big) = 0.
\label{eq:res}
\end{align}
Similarly, we can write the discretized problem~\eqref{eq:proj_disc} in terms of~\eqref{eq:res_func} as
\begin{align}
r_{\testidx}(\mathbf{J}_h) = a(\mathbf{J}_h,\boldsymbol{\Lambda}_{\testidx}) -b\big(\mathbf{H}^\mathcal{I}, \boldsymbol{\Lambda}_{\testidx}\big) = 0.
\label{eq:res_disc}
\end{align}
%

The method of manufactured solutions modifies~\eqref{eq:res_disc} to be
\begin{align}
r_{\testidx}(\mathbf{J}_h) = r_{\testidx}(\mathbf{J}_\text{MS}),
\label{eq:mms}
\end{align}
where $\mathbf{J}_\text{MS}$ is the manufactured solution, and $r_{\testidx}(\mathbf{J}_\text{MS})$ is computed exactly.
%

%

%
Inserting~\eqref{eq:res} and~\eqref{eq:res_disc} into~\eqref{eq:mms} yields
\begin{align}
a(\mathbf{J}_h,\boldsymbol{\Lambda}_{\testidx}) = a(\mathbf{J}_\text{MS},\boldsymbol{\Lambda}_{\testidx}).
\label{eq:proj_disc_mms}
\end{align}
%
%
However, instead of solving~\eqref{eq:proj_disc_mms}, we can equivalently solve~\eqref{eq:proj_disc} by setting 
\begin{align}
b\big(\mathbf{H}^\mathcal{I}, \boldsymbol{\Lambda}_{\testidx}\big) = a(\mathbf{J}_\text{MS},\boldsymbol{\Lambda}_{\testidx}).
\label{eq:H_mms_1}
\end{align}
Equation~\eqref{eq:H_mms_1} is satisfied by
\begin{align}
\mathbf{n}\times \mathbf{H}^\mathcal{I} 
= 
\frac{1}{2}\mathbf{J}_\text{MS} - \mathbf{n}\times \int_{S'}\big[ \mathbf{J}_\text{MS}(\mathbf{x}')\times\nabla' G(\mathbf{x},\mathbf{x}')\big]dS'.
\label{eq:H_mms_2}
\end{align}
Noting that $\mathbf{J}_\text{MS} - (\mathbf{J}_\text{MS}\cdot\mathbf{n})\mathbf{n} = \mathbf{n}\times(\mathbf{J}_\text{MS}\times \mathbf{n})$ and that $\mathbf{J}_\text{MS}$ must be tangential to the surface ($\mathbf{J}_\text{MS}\cdot\mathbf{n}=0$), $\mathbf{J}_\text{MS} = \mathbf{n}\times(\mathbf{J}_\text{MS}\times \mathbf{n})$ can be inserted into~\eqref{eq:H_mms_2}, and factoring out the cross product with $\mathbf{n}$ yields
\begin{align}
\mathbf{H}^\mathcal{I} = \frac{1}{2}\mathbf{J}_\text{MS}\times\mathbf{n} - \int_{S'}\big[\mathbf{J}_\text{MS}(\mathbf{x}')\times\nabla'G(\mathbf{x},\mathbf{x}')\big] dS',
\label{eq:H_i}
\end{align}
which we can use to solve~\eqref{eq:proj_disc}.

However, as described in the introduction, integrals containing the Green's function~\eqref{eq:G} or its derivatives, such as those appearing in~\eqref{eq:H_i}, \reviewerTwo{cannot be computed analytically. Additionally,} the singularity when $R\to 0$ complicates their accurate approximation, potentially contaminating convergence studies.  Therefore, as is done in~\cite{freno_em_mms_2020}, we manufacture the Green's function, using the form
\begin{align}
G_\text{MS}(\mathbf{x},\mathbf{x}') = \bigg(1 - \frac{R^2}{R_m^2}\bigg)^d,
\label{eq:G_mms}
\end{align}
where $R_m=\max_{\mathbf{x},\mathbf{x}'\in S} R$ is the maximum possible distance between two points on $S$, and $d\in\mathbb{N}$.  The form of~\eqref{eq:G_mms} is chosen for two reasons: 1) the even powers of $R$ permit the integrals in~\eqref{eq:proj_disc} and~\eqref{eq:H_i} to be computed analytically for many choices of $\mathbf{J}_\text{MS}$, avoiding contamination from additional error, and 2) $G_\text{MS}$ increases when $R$ decreases, as with the actual Green's function~\eqref{eq:G}.

We update~\eqref{eq:proj_disc} to account for the modifications arising from $\mathbf{J}_\text{MS}$ and $G_\text{MS}$: 
\begin{align}
a(\mathbf{J}_h,\boldsymbol{\Lambda}_{\testidx}) = b\big(\mathbf{H}^\mathcal{I}_\text{MS}, \boldsymbol{\Lambda}_{\testidx}\big),
\label{eq:mms_problem}
\end{align}
%
where
\begin{align*}
\mathbf{H}^\mathcal{I}_\text{MS} = \frac{1}{2}\mathbf{J}_\text{MS}\times\mathbf{n} - \int_{S'}\big[\mathbf{J}_\text{MS}(\mathbf{x}')\times\nabla'G_\text{MS}(\mathbf{x},\mathbf{x}')\big] dS',
\end{align*}
and, for notational simplicity, $a(\cdot,\cdot)$ uses $G_\text{MS}$~\eqref{eq:G_mms} in~\eqref{eq:a}, instead of $G$~\eqref{eq:G}.

\subsection{Solution-Discretization Error} 
\label{sec:sde}

In~\eqref{eq:mms_problem}, if the integrals in $a(\cdot,\cdot)$ are evaluated exactly, the only contribution to the discretization error is the solution-discretization error.  Solving for $\mathbf{J}^h$ enables us to compute the discretization error 
\begin{align}
\mathbf{e}_\mathbf{J} = \mathbf{J}^h - \mathbf{J}_n,
\label{eq:solution_error}
\end{align}
where $J_{n_\srcidx}$ denotes the component of $\mathbf{J}_\text{MS}$ flowing from $T_\srcidx^+$ to $T_\srcidx^-$.  The norm of~\eqref{eq:solution_error} has the property $\|\mathbf{e}_\mathbf{J}\|\le C_\mathbf{J} h^{p_\mathbf{J}}$, where
$C_\mathbf{J}$ is a function of the solution derivatives, $h$ is representative of the mesh size, and $p_\mathbf{J}$ is the order of accuracy.  By performing a mesh-convergence study of the norm of the discretization error, we can ensure the expected order of accuracy is obtained.  For the RWG basis functions, the expectation is second-order accuracy $(p_\mathbf{J}=2)$.

\subsubsection{Solution Uniqueness} 
\label{sec:su}

For the EFIE, the manufactured Green's function $G_\text{MS}$~\eqref{eq:G_mms} yields a matrix that is practically singular, admitting infinite solutions $\mathbf{J}^h$.  For the MFIE, the matrix arising from the first term of~\eqref{eq:a}, which does not contain a Green's function, is nonsingular.  Additionally, the matrix arising from the first and second terms is nonsingular when $G_\text{MS}$ is used.  However, the matrix arising from only the second term is practically singular.  

In~\cite{freno_em_mms_2020,freno_short_note_2022}, a mitigation approach is presented to select $\mathbf{J}^h$, by solving the optimization problem
\begin{align}
\begin{array}{r l}
\text{minimize} \quad & {\|\mathbf{e}_\mathbf{J}\|}_2
\\[.5em]
\text{subject to} \quad & a(\mathbf{J}_h,\boldsymbol{\Lambda}_{\testidx}) = b\big(\mathbf{H}^\mathcal{I}_\text{MS}, \boldsymbol{\Lambda}_{\testidx}\big).
\end{array}
\label{eq:opt_L2}
\end{align}
The solution to~\eqref{eq:opt_L2} is 
\begin{align*}
\mathbf{J}^h=\mathbf{J}_n + \mathbf{Q}_1\big(\mathbf{u}-\mathbf{Q}^H_1\mathbf{J}_n\big),
\end{align*}
where $\mathbf{R}^H_1\mathbf{u}=\mathbf{P}^T\mathbf{V}$ and $\mathbf{Q}_1$, $\mathbf{R}_1$, and $\mathbf{P}$ arise from the pivoted QR factorization of $\mathbf{Z}^H$.

\reviewerTwo{In formulating the optimization problem~\eqref{eq:opt_L2}, we selected ${\|\mathbf{e}_\mathbf{J}\|}_2$ as the cost function because it has the benefit of a closed form solution.  However, in order to assess the rate of convergence, we must select a second norm that is used to report the error.  ${\|\mathbf{e}_\mathbf{J}\|}_\infty$ is often preferred for error reporting in code verification because it is more sensitive and therefore more rigorous.  However, the mismatch between the minimized norm ($L^2$) and the reported norm ($L^\infty$) may require finer meshes to enter the asymptotic region.}  
%
Therefore, as a trade-off, in this paper, we additionally solve the optimization problem
\begin{align}
\begin{array}{r l}
\text{minimize} \quad & {\|\mathbf{e}_\mathbf{J}\|}_\infty
\\[.5em]
\text{subject to} \quad & a(\mathbf{J}_h,\boldsymbol{\Lambda}_{\testidx}) = b\big(\mathbf{H}^\mathcal{I}_\text{MS}, \boldsymbol{\Lambda}_{\testidx}\big).
\end{array}
\label{eq:opt_Linf}
\end{align}
\reviewerTwo{The $L^\infty$-norm of the error arising from the solution to~\eqref{eq:opt_Linf}} reaches the asymptotic region faster than \reviewerTwo{that arising from}~\eqref{eq:opt_L2} but requires the solution to a linear programming problem, which is more expensive.

\subsection{Numerical-Integration Error} 
\label{sec:nie}

In practice, the integrals in $a(\cdot,\cdot)$~\eqref{eq:a} and $b(\cdot,\cdot)$~\eqref{eq:b} are evaluated numerically, yielding the approximations $a^q(\cdot,\cdot)$ and $b^q(\cdot,\cdot)$.  $a^q(\cdot,\cdot)$ and $b^q(\cdot,\cdot)$ are obtained by integrating over each triangular element using quadrature, and generally incur a numerical-integration error.  Therefore, it is important to measure the numerical-integration error without contamination from the solution-discretization error.
To do this, we build upon the solution-discretization error cancellation and elimination approaches of~\cite{freno_em_mms_quad_2021} by defining the error functionals
\begin{align}
e_a(\mathbf{u}) &{} = a^q(\mathbf{u},\mathbf{u}) - a(\mathbf{u},\mathbf{u}),  \label{eq:a_error}\\
e_b(\mathbf{u}) &{} = b^q\big(\mathbf{H}^\mathcal{I}_\text{MS},\mathbf{u}\big) - b\big(\mathbf{H}^\mathcal{I}_\text{MS},\mathbf{u}\big), \label{eq:b_error}
\end{align}
which have the properties $|e_a| \le C_a h^{p_a}$ and $|e_b| \le C_b h^{p_b}$, where $C_a$ and $C_b$ are functions of the integrand derivatives, and $p_a$ and $p_b$ depend on the quadrature accuracy.

In Section~\ref{sec:sdec}, we describe our approach to canceling the solution-discretization error, and in Section~\ref{sec:sdee}, we describe our approach to eliminating the solution-discretization error.

\subsubsection{Solution-Discretization Error Cancellation}
\label{sec:sdec}
In this paper, we cancel the solution-discretization error and measure the numerical-integration error from
\begin{alignat}{19}
&e_a&(\mathbf{J}_{h_\text{MS}}) = {} && a^q&    (&&\mathbf{J}_{h_\text{MS}}        &,\mathbf{J}_{h_\text{MS}}&    )& {}-{} &a&&    (&&\mathbf{J}_{h_\text{MS}}        &,\mathbf{J}_{h_\text{MS}}&    )&,  \label{eq:a_error_cancel}\\
&e_b&(\mathbf{J}_{h_\text{MS}}) = {} && b^q&\big(&&\mathbf{H}^\mathcal{I}_\text{MS}&,\mathbf{J}_{h_\text{MS}}&\big)& {}-{} &b&&\big(&&\mathbf{H}^\mathcal{I}_\text{MS}&,\mathbf{J}_{h_\text{MS}}&\big)&, \label{eq:b_error_cancel}
\end{alignat}
where $\mathbf{J}_{h_\text{MS}}$ is the basis-function representation of $\mathbf{J}_\text{MS}$, obtained from~\eqref{eq:J_h} by setting the coefficients $J_\srcidx$ equal to the normal component of $\mathbf{J}_\text{MS}$ at the midpoint of each edge associated with $\boldsymbol{\Lambda}_\srcidx(\mathbf{x})$.  Due to the presence of the basis functions in the minuend and subtrahend of~\eqref{eq:a_error_cancel} and~\eqref{eq:b_error_cancel}, we have canceled the solution-discretization error.
This approach is more advantageous than the solution-discretization error cancellation approach of~\cite{freno_em_mms_quad_2021}, as it does not require solving a potentially inconsistent system of equations.

To relate $e_a(\mathbf{J}_{h_\text{MS}})$ to the discretization error $\mathbf{e}_\mathbf{J}$~\eqref{eq:solution_error}, we can solve the optimization problem
\begin{align}
\begin{array}{r l}
\text{minimize} \quad & {\|\mathbf{e}_\mathbf{J}\|}_2
\\[.5em]
\text{subject to} \quad & a^q(\mathbf{J}_h,\mathbf{J}_{h_\text{MS}}) = a(\mathbf{J}_{h_\text{MS}},\mathbf{J}_{h_\text{MS}}).
\end{array}
\label{eq:opt_a}
\end{align}
Similarly, to relate $e_b(\mathbf{J}_{h_\text{MS}})$ to $\mathbf{e}_\mathbf{J}$, we can solve the optimization problem
\begin{align}
\begin{array}{r l}
\text{minimize} \quad & {\|\mathbf{e}_\mathbf{J}\|}_2
\\[.5em]
\text{subject to} \quad & b^q\big(\mathbf{H}^\mathcal{I}_\text{MS},\mathbf{J}_h\big) = b\big(\mathbf{H}^\mathcal{I}_\text{MS},\mathbf{J}_{h_\text{MS}}\big).
\end{array}
\label{eq:opt_b}
\end{align}
In~\eqref{eq:opt_a} and~\eqref{eq:opt_b}, ${\|\mathbf{e}_\mathbf{J}\|}_\infty$ could be minimized instead; however, in this paper, minimizing ${\|\mathbf{e}_\mathbf{J}\|}_2$ is sufficient to reach the asymptotic region for the meshes considered.

\subsubsection{Solution-Discretization Error Elimination}
\label{sec:sdee}
Canceling the solution-discretization error enables us to assess how the numerical integration performs for the integrands arising from the approximated solution.  
\reviewerTwo{The polynomial degrees of these integrands are finite, depending on those of the basis functions and the manufactured Green's function.  Therefore, the ability to perform convergence studies on quadrature rules capable of integrating higher polynomial degrees is limited.}
%
Alternatively, we can assess the performance of the numerical integration without \reviewerTwo{polynomial} degree limits by removing the basis-function restriction.

In this paper, we eliminate the solution-discretization error and measure the numerical-integration error from
\begin{alignat}{19}
&e_a&(\mathbf{J}_\text{MS}) = {}&& a^q&     (&&\mathbf{J}_\text{MS}            &,\mathbf{J}_\text{MS}&    )& {}-{} &a&&    (&&\mathbf{J}_\text{MS}            &,\mathbf{J}_\text{MS}&)&,  \label{eq:a_error_eliminate}\\
&e_b&(\mathbf{J}_\text{MS}) = {}&& b^q& \big(&&\mathbf{H}^\mathcal{I}_\text{MS}&,\mathbf{J}_\text{MS}&\big)& {}-{} &b&&\big(&&\mathbf{H}^\mathcal{I}_\text{MS}&,\mathbf{J}_\text{MS}&\big)&. \label{eq:b_error_eliminate}
\end{alignat}
By eliminating the presence of the basis functions in~\eqref{eq:a_error_eliminate} and~\eqref{eq:b_error_eliminate}, we have eliminated the solution-discretization error.

\subsection{Domain-Discretization Error} 
\label{sec:dde}
For domains with curvature, the discretization of $S$ with planar triangular elements yields the faceted approximation $S_h$.  To address domain-discretization error, we consider two \reviewerTwo{code-verification} approaches: accounting for the curvature by using curved triangular elements and neglecting the curvature by using planar triangular elements.

\subsubsection{Accounting for Curvature}
\label{sec:ac}
We restrict the scope of the basis functions in this paper to planar RWG basis functions.  Therefore, because they rely on basis functions, isolating the solution-discretization error in Section~\ref{sec:sde} and isolating the numerical-integration error by canceling the solution-discretization error in Section~\ref{sec:sdec} will not account for curvature \reviewerTwo{since the faceted approximation to the geometry $S_h$ is used}.

For the solution-discretization error elimination approach of Section~\ref{sec:sdee}, instead of integrating over a planar triangle, we can integrate over a curved triangle that conforms to $S$.  Through this approach, \reviewerTwo{we modify the planar triangular integrals to include the determinant of the transformation between the planar and curved triangles.  By doing so,}
%
we can assess the curvature implementation by measuring the numerical-integration error.

\subsubsection{Neglecting Curvature}
\label{sec:nc}
To neglect curvature, the solution-discretization error cancellation approach of Section~\ref{sec:sdec} enables us to isolate and measure the numerical-integration error by computing the integrals on $S_h$ instead of $S$.

\section{Numerical Examples} 
\label{sec:results}

In this section, we demonstrate the effectiveness of the approaches described in Section~\ref{sec:mms}.  
In Section~\ref{sec:without}, we consider domains without curvature, thereby avoiding domain-discretization error, and we isolate and measure the solution-discretization error (Section~\ref{sec:sde}) and the numerical-integration error (Section~\ref{sec:nie}).
In Section~\ref{sec:with}, we consider a domain with curvature.  To address the domain-discretization error, we (1) account for the curvature, which modifies the integrands, such that we measure the numerical-integration error (Section~\ref{sec:ac}), and (2) neglect the curvature, such that we isolate and measure the numerical-integration error (Section~\ref{sec:nc}).

When measuring a norm of the discretization error $\|\mathbf{e}_\mathbf{J}\|$~\eqref{eq:solution_error}, we nondimensionalize by the constant $\varepsilon_0=1$ A/m.  When measuring the numerical-integration error $e_a(\cdot)$~\eqref{eq:a_error} or $e_b(\cdot)$~\eqref{eq:b_error}, we nondimensionalize by the constant $\varepsilon_0=1$ A$^2$.

\subsection{Without Curvature} 
\label{sec:without}

\begin{figure}
\centering
\includegraphics[scale=.28,clip=true,trim=0in 0in 0in 0in]{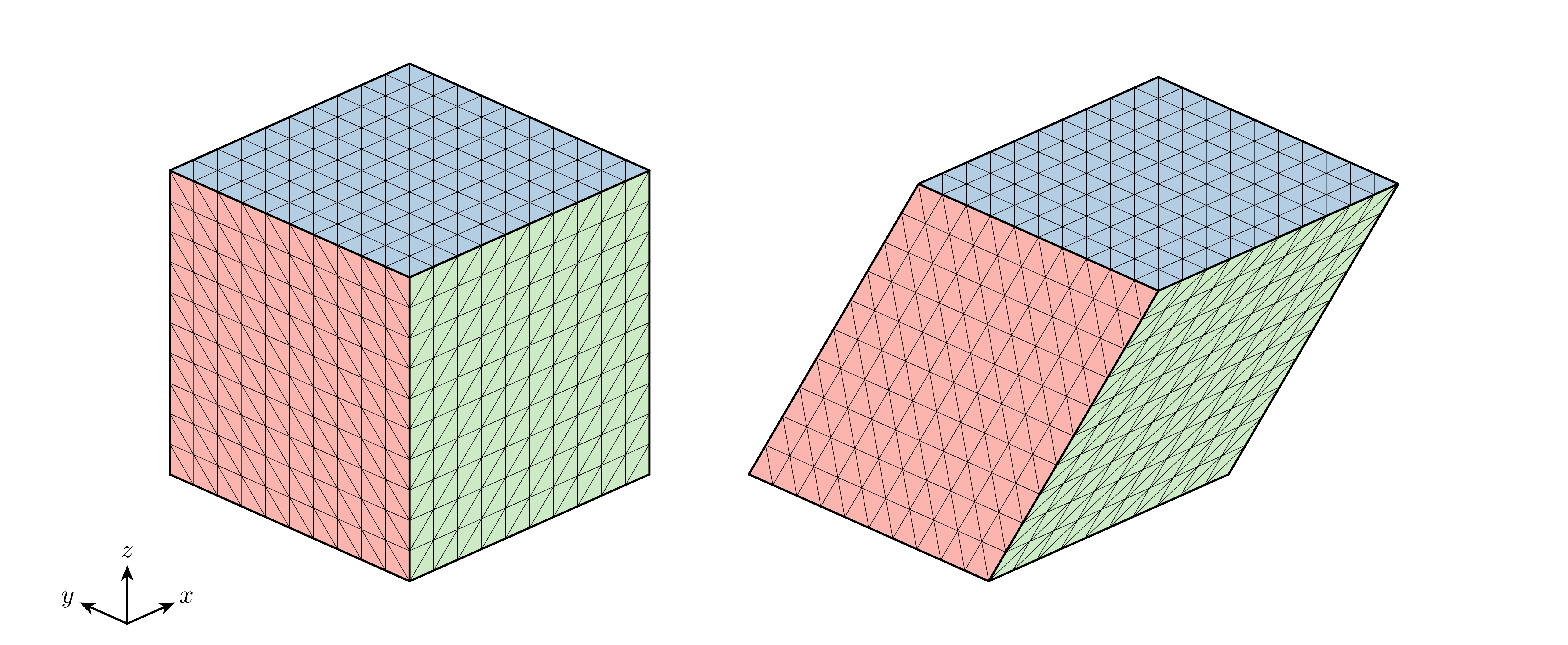}
\caption{Meshes for the cube (left) and the rhombic prism (right), with $\ntriangles=1200$.}
\vskip-\dp\strutbox
\label{fig:3d_domains}
\end{figure}

\begin{figure}
\centering
\includegraphics[scale=.28,clip=true,trim=0in 0in 0in 0in]{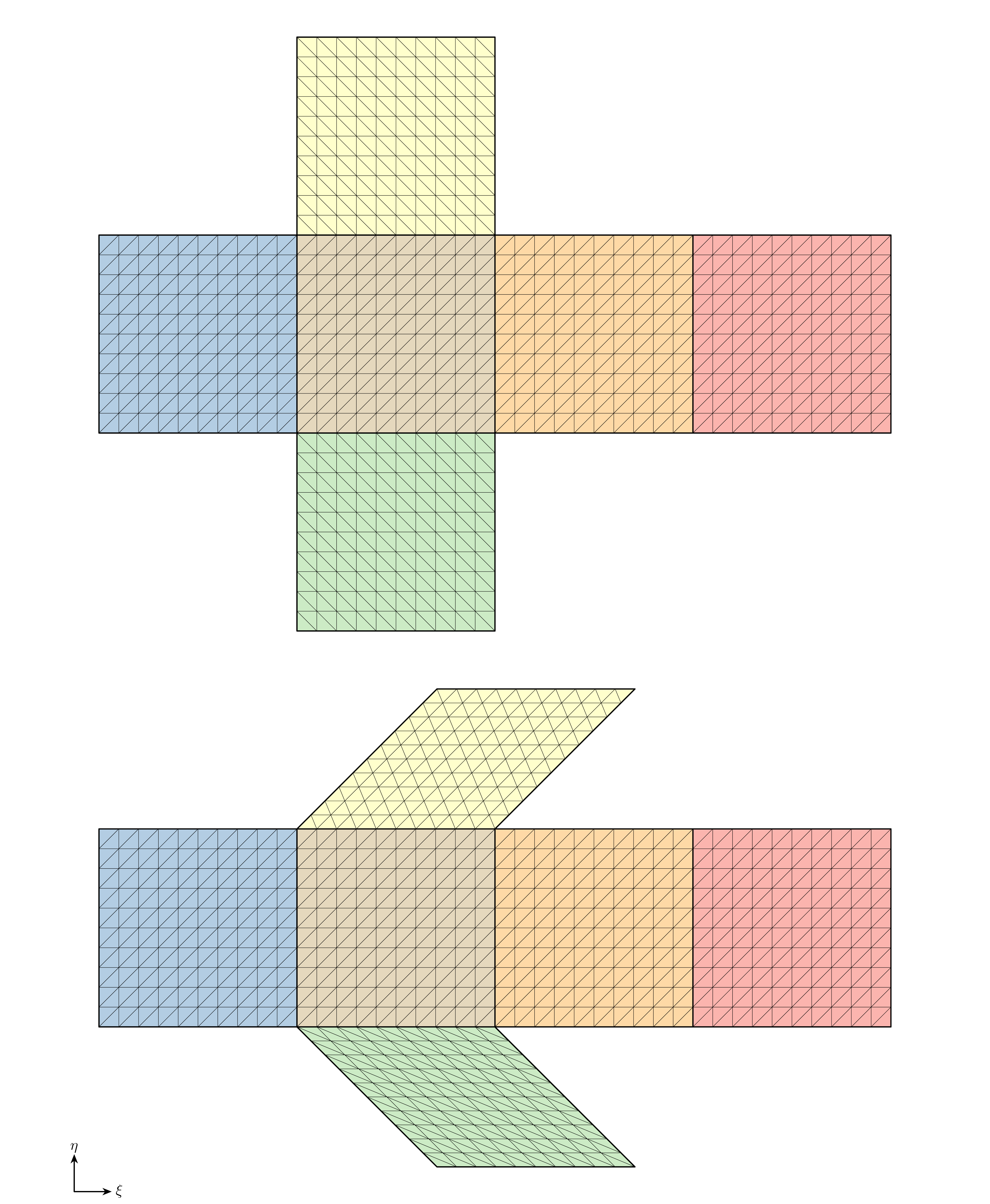}
\caption{Meshes for the cube (top) and the rhombic prism (bottom), with $\ntriangles=1200$.}
\vskip-\dp\strutbox
\label{fig:2d_domains}
\end{figure}

\begin{figure}
\centering
\includegraphics[scale=.28,clip=true,trim=0in 0in 0in 0in]{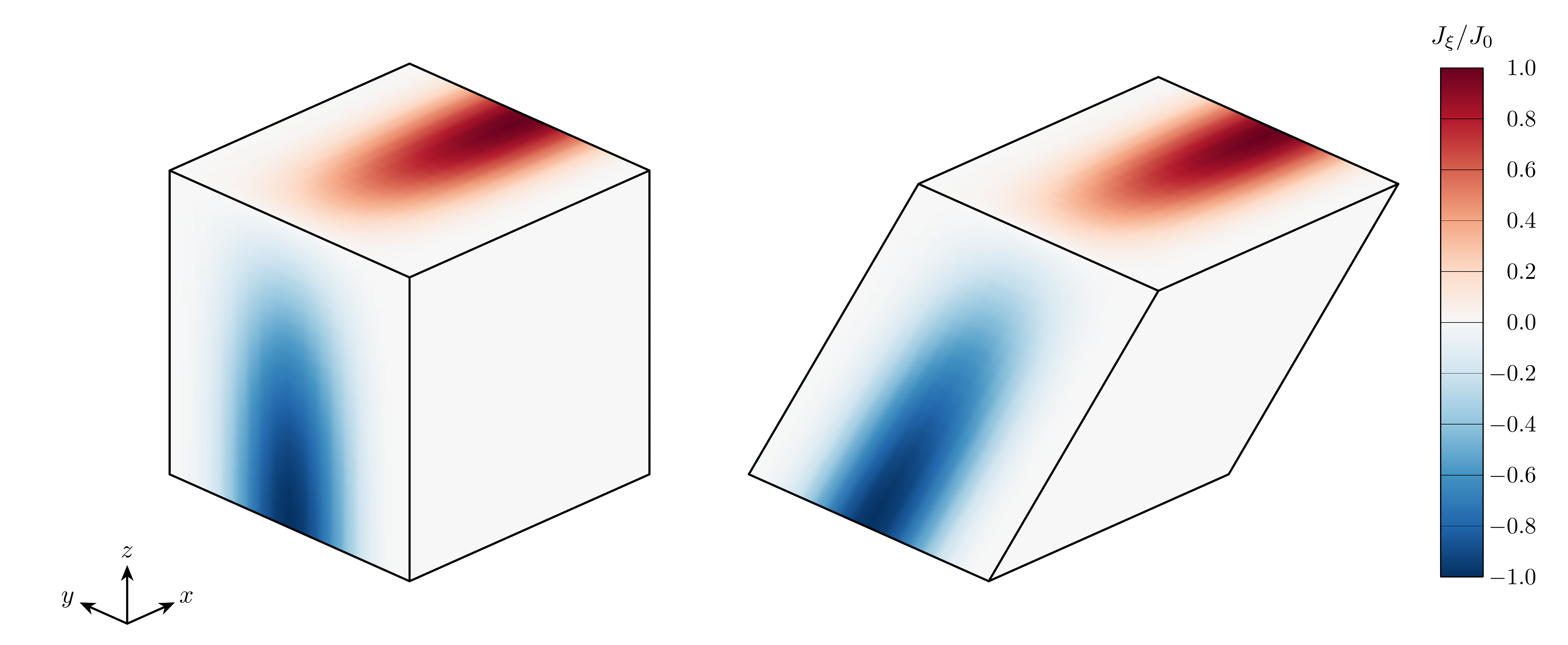}
\caption{\strut Manufactured surface current density $\mathbf{J}_\text{MS}$ for the cube (left) and the rhombic prism (right).}
\vskip-\dp\strutbox
\label{fig:3d_solutions}
\end{figure}

\begin{figure}
\centering
\includegraphics[scale=.28,clip=true,trim=0in 0in 0in 0in]{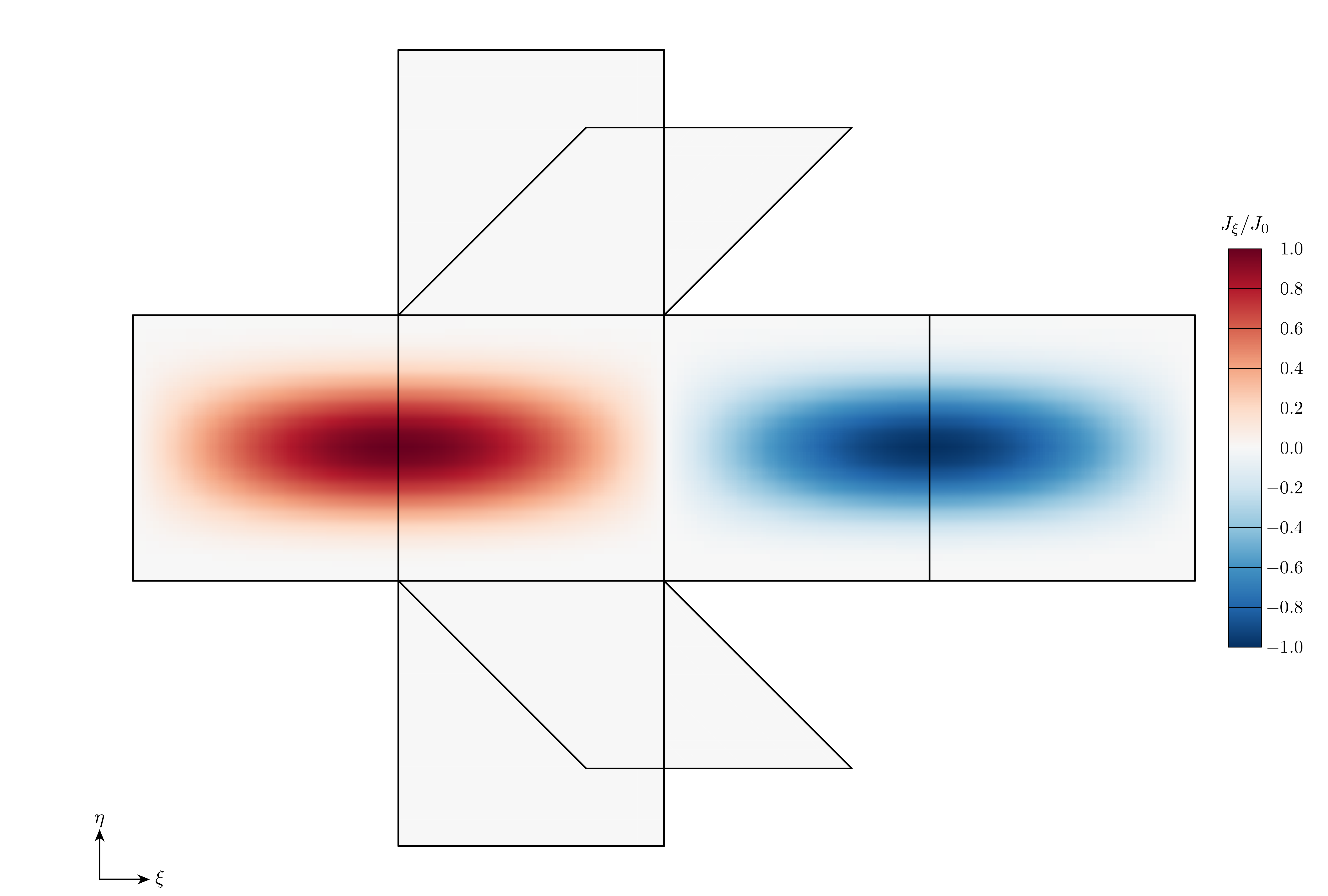}
\caption{\strut Manufactured surface current density $\mathbf{J}_\text{MS}$ for cube and rhombic prism.}
\vskip-\dp\strutbox
\label{fig:2d_solutions}
\end{figure}

\begin{figure}
\centering
\begin{subfigure}[b]{.49\textwidth}
\includegraphics[scale=.64,clip=true,trim=2.3in 0in 2.8in 0in]{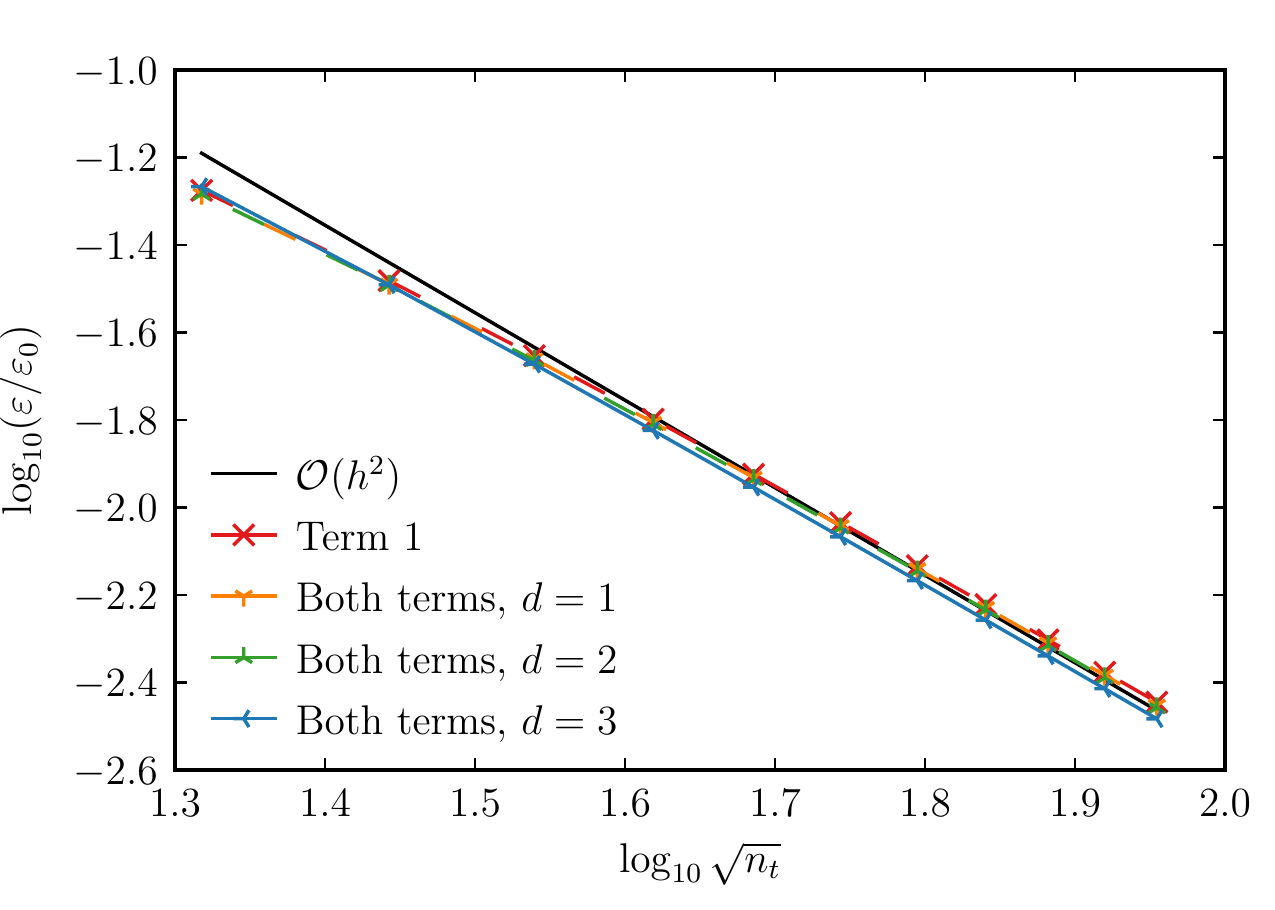}
\caption{Cube\vpad}
\end{subfigure}
\hspace{0.25em}
\begin{subfigure}[b]{.49\textwidth}
\includegraphics[scale=.64,clip=true,trim=2.3in 0in 2.8in 0in]{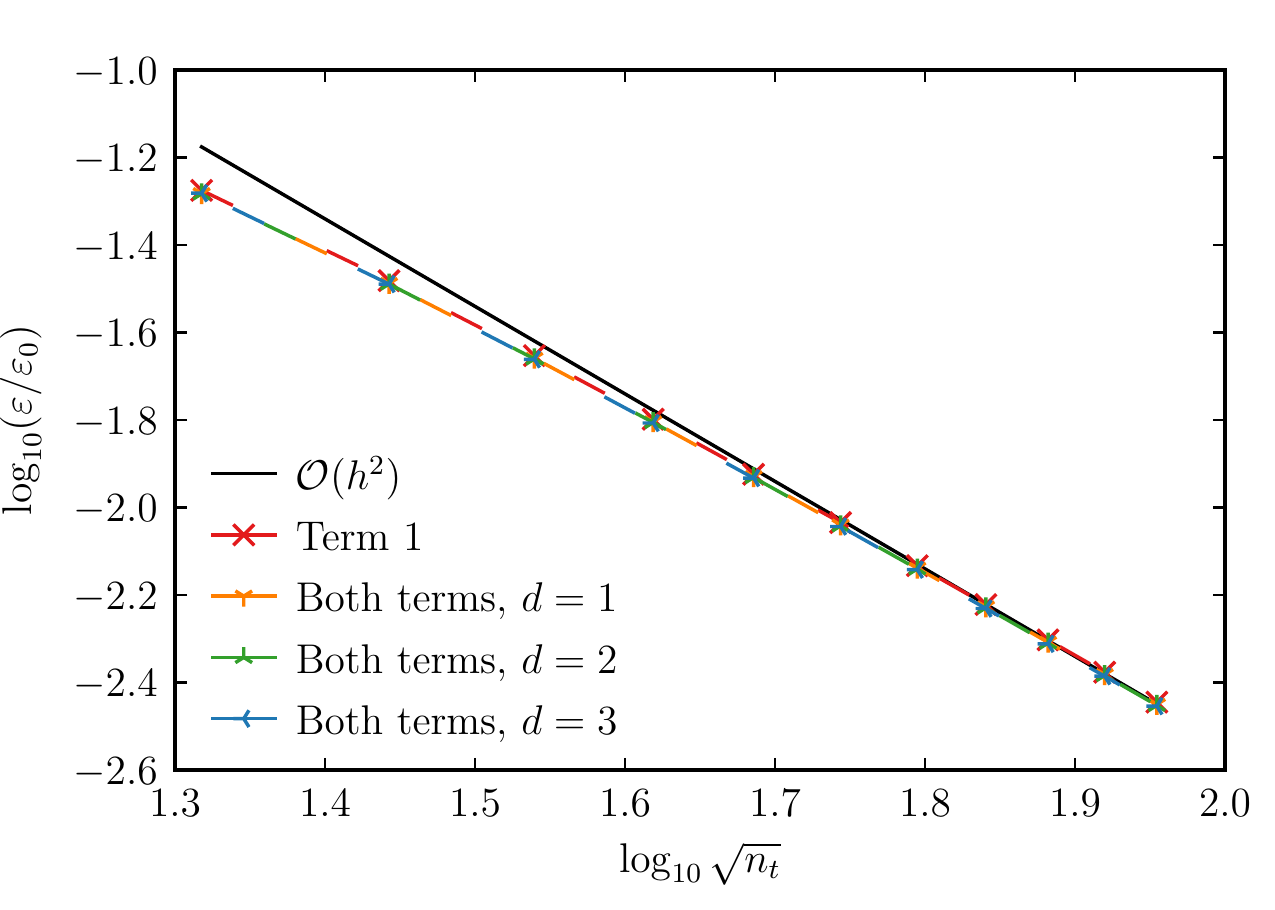}
\caption{Rhombic Prism\vpad}
\end{subfigure}
\caption{Solution-discretization error: $\varepsilon={\|\mathbf{e}_\mathbf{J}\|}_\infty$.}
\vskip-\dp\strutbox
\label{fig:part1a}
\end{figure}

\begin{figure}
\centering
\begin{subfigure}[b]{.49\textwidth}
\includegraphics[scale=.64,clip=true,trim=2.3in 0in 2.8in 0in]{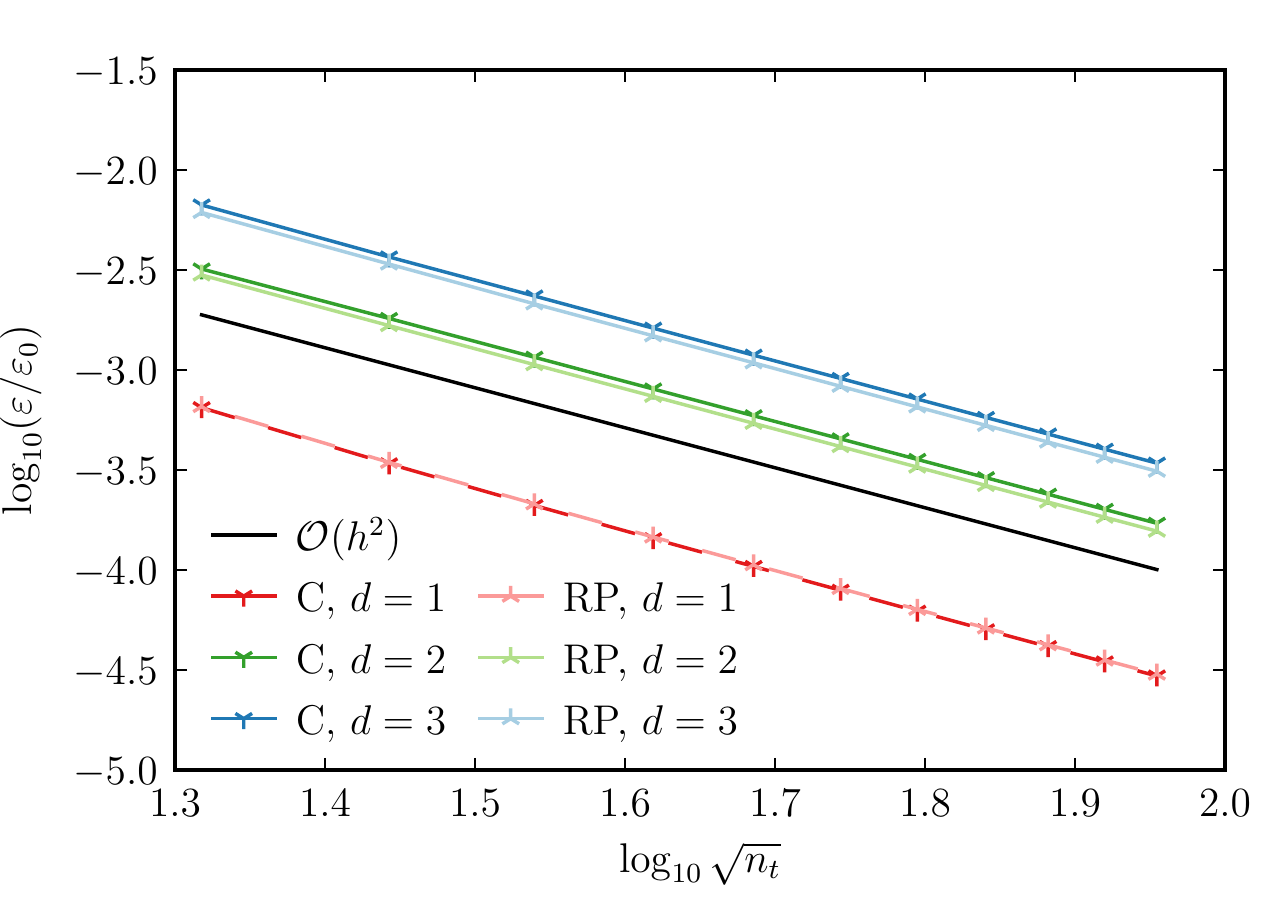}
\caption{$\varepsilon={\|\mathbf{e}_\mathbf{J}\|}_\infty$, $\min_{\mathbf{J}^h} {\|\mathbf{e}_\mathbf{J}\|}_\infty$\vpad}
\label{fig:p1_LinfErr_LinfOpt}
\end{subfigure}
\hspace{0.25em}
\begin{subfigure}[b]{.49\textwidth}
\includegraphics[scale=.64,clip=true,trim=2.3in 0in 2.8in 0in]{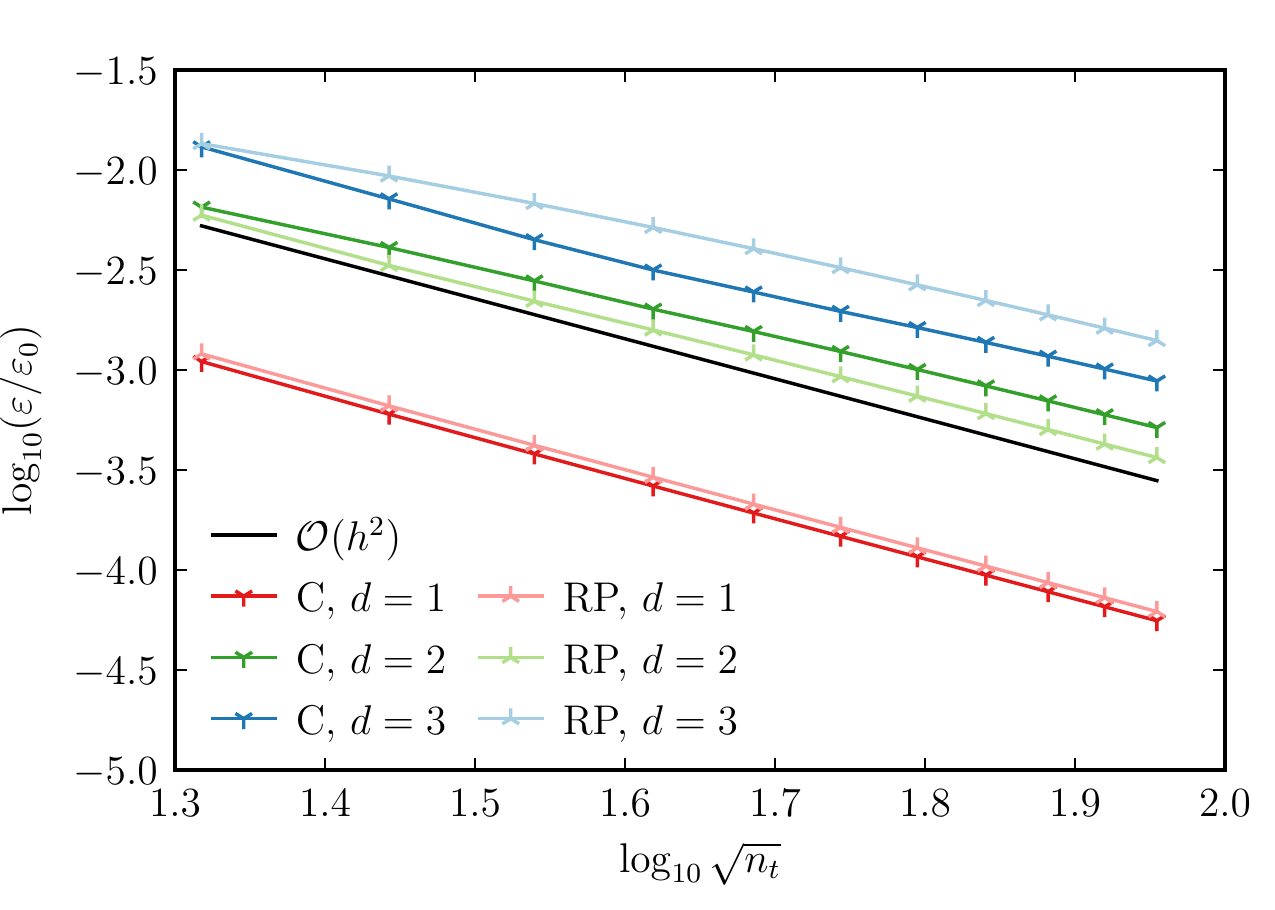}
\caption{$\varepsilon={\|\mathbf{e}_\mathbf{J}\|}_\infty$, $\min_{\mathbf{J}^h} {\|\mathbf{e}_\mathbf{J}\|}_2$\vpad}
\label{fig:p1_LinfErr_L2Opt}
\end{subfigure}
\\
\begin{subfigure}[b]{.49\textwidth}
\includegraphics[scale=.64,clip=true,trim=2.3in 0in 2.8in 0in]{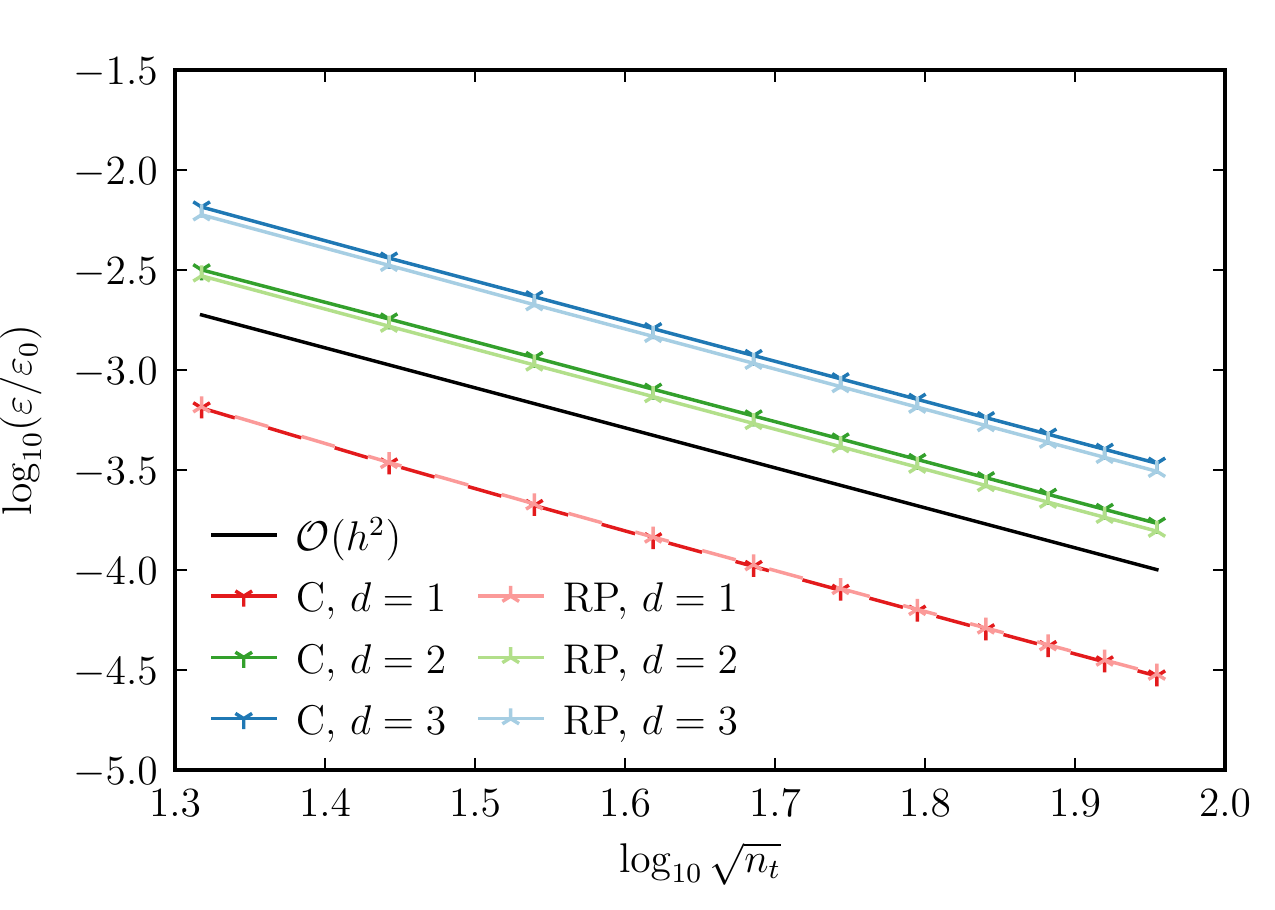}
\caption{$\varepsilon={\|\mathbf{e}_\mathbf{J}\|}_2$, $\min_{\mathbf{J}^h} {\|\mathbf{e}_\mathbf{J}\|}_\infty$\vpad}
\label{fig:p1_L2Err_LinfOpt}
\end{subfigure}
\hspace{0.25em}
\begin{subfigure}[b]{.49\textwidth}
\includegraphics[scale=.64,clip=true,trim=2.3in 0in 2.8in 0in]{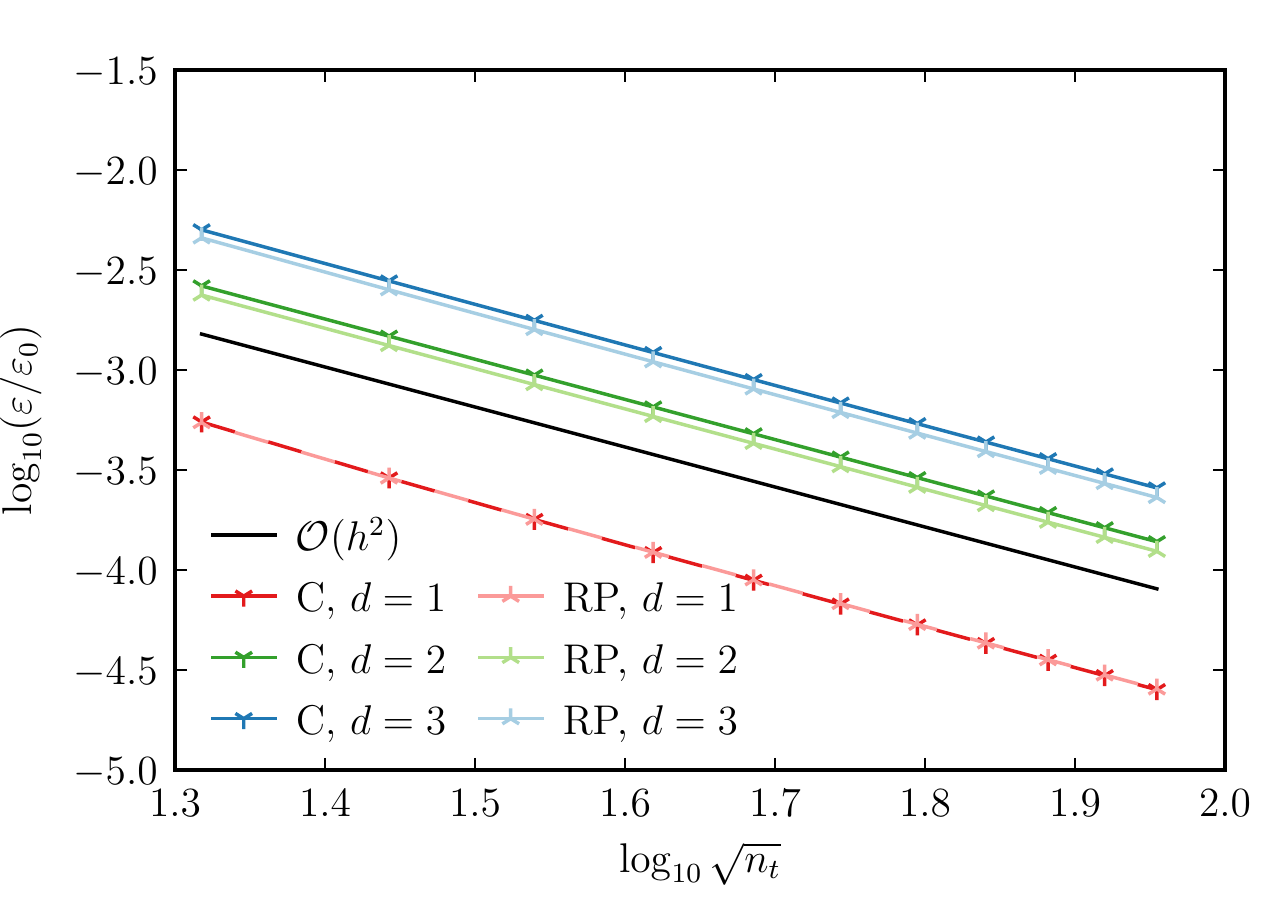}
\caption{$\varepsilon={\|\mathbf{e}_\mathbf{J}\|}_2$, $\min_{\mathbf{J}^h} {\|\mathbf{e}_\mathbf{J}\|}_2$\vpad}
\label{fig:p1_L2Err_L2Opt}
\end{subfigure}
\caption{Solution-discretization error: $\varepsilon=\|\mathbf{e}_\mathbf{J}\|$ for different metric and minimization norms.\\\strut}
\vskip-\dp\strutbox
\label{fig:part1b}
\end{figure}

We consider two domains without curvature: a cube and a rhombic prism, each with all edges of length 1~m, as shown in Figures~\ref{fig:3d_domains} and~\ref{fig:2d_domains} with the total number of triangles $\ntriangles=1200$.  The acute angle of the rhombic prism is $45^\circ$.  We manufacture the surface current density $\mathbf{J}_\text{MS}(\mathbf{x}) = J_\xi(\xi,\eta)\mathbf{e}_\xi$, where 
\begin{align}
J_\xi(\xi,\eta)=J_0\left\{\begin{matrix}
\sin\bigg(\displaystyle\frac{\pi\xi}{2 L_0}\bigg)\sin^3\bigg(\displaystyle\frac{\pi\eta}{L_0}\bigg), & \text{for }\mathbf{n}\cdot\mathbf{e}_y=0 \\[1em]
0, & \text{for }\mathbf{n}\cdot\mathbf{e}_y\ne 0
\end{matrix}\right.,
\label{eq:jxi}
\end{align}
$J_0=1$~A/m, and $L_0=1$~m. $\xi\in[0,\,4]$~m is perpendicular to $\eta=y\in[0,\,1]$~m, wrapping around the surfaces for which $\mathbf{n}\cdot\mathbf{e}_y=0$, beginning at $x=0$~m and $z=1$~m for the cube and $x=z=\sqrt{2}/2$~m for the rhombic prism, as depicted in Figure~\ref{fig:2d_domains}, which shows the nets of these domains.  Equation~\eqref{eq:jxi} is of class $C^2$.  Figures~\ref{fig:3d_solutions} and~\ref{fig:2d_solutions} show plots of~\eqref{eq:jxi}.

\subsubsection{Solution-Discretization Error} 

To isolate and measure the solution-discretization error, we proceed with the assessment described in Section~\ref{sec:sde}.  
\reviewerTwo{%
With $G_\text{MS}$~\eqref{eq:G_mms}, we are able to compute the integral in~\eqref{eq:H_i} analytically, which yields a finite-degree polynomial integrand for $b\big(\mathbf{H}^\mathcal{I}, \boldsymbol{\Lambda}_{\testidx}\big)$~\eqref{eq:proj_disc}.  Because the integrands of $a(\mathbf{J}_h,\boldsymbol{\Lambda}_{\testidx})$ and $b\big(\mathbf{H}^\mathcal{I}, \boldsymbol{\Lambda}_{\testidx}\big)$ in~\eqref{eq:proj_disc} are finite-degree polynomials, they can be integrated exactly with the appropriate amount of polynomial quadrature points.} For $G_\text{MS}$, we consider $d=\{1,\,2,\,3\}$.  
We account for potential disparities in the magnitudes of the contributions to~\eqref{eq:a} from the first (Term 1) and second (Term 2) terms by considering them together and separately.

Figure~\ref{fig:part1a} shows the $L^\infty$ norm of the discretization error ${\|\mathbf{e}_\mathbf{J}\|}_\infty$~\eqref{eq:solution_error} arising from only the solution-discretization error for both geometries and for Term 1 with and without Term 2.  
%
%
The convergence rates are all $\mathcal{O}(h^2)$ as expected.

Figure~\ref{fig:part1b} shows norms of the discretization error with only Term 2.  As stated in Section~\ref{sec:su}, the arising matrix is singular.  Therefore, to compute a unique solution we minimize ${\|\mathbf{e}_\mathbf{J}\|}_\infty$~\eqref{eq:opt_Linf} (\ref{fig:p1_LinfErr_LinfOpt} and \ref{fig:p1_L2Err_LinfOpt}) and ${\|\mathbf{e}_\mathbf{J}\|}_2$~\eqref{eq:opt_L2} (\ref{fig:p1_LinfErr_L2Opt} and \ref{fig:p1_L2Err_L2Opt}), and we measure ${\|\mathbf{e}_\mathbf{J}\|}_\infty$ (\ref{fig:p1_LinfErr_LinfOpt} and \ref{fig:p1_LinfErr_L2Opt}) and ${\|\mathbf{e}_\mathbf{J}\|}_2$ (\ref{fig:p1_L2Err_LinfOpt} and \ref{fig:p1_L2Err_L2Opt}). 
To minimize ${\|\mathbf{e}_\mathbf{J}\|}_\infty$, we use HiGHS~\cite{huangfu_2018}.
Table~\ref{tab:part1b} shows the convergence rates computed between adjacent mesh pairs for these four combinations for the cube (C) and rhombic prism (RP).  Unlike the other three combinations, measuring ${\|\mathbf{e}_\mathbf{J}\|}_\infty$ and minimizing ${\|\mathbf{e}_\mathbf{J}\|}_2$ yields a convergence rate that is not as clearly $\mathcal{O}(h^2)$ for the meshes considered.  On the other hand, by minimizing ${\|\mathbf{e}_\mathbf{J}\|}_\infty$, the convergence rate is clearly $\mathcal{O}(h^2)$ without requiring finer meshes.

\begin{table}
\centering
\begin{tabular}{c c c c c c c c c}
\toprule
& \multicolumn{4}{c}{$\varepsilon={\|\mathbf{e}_\mathbf{J}\|}_\infty$} &
  \multicolumn{4}{c}{$\varepsilon={\|\mathbf{e}_\mathbf{J}\|}_2     $} \\
  \cmidrule(r){2-5} \cmidrule(l){6-9}
& \multicolumn{2}{c}{$\min_{\mathbf{J}^h} {\|\mathbf{e}_\mathbf{J}\|}_\infty$} &
  \multicolumn{2}{c}{$\min_{\mathbf{J}^h} {\|\mathbf{e}_\mathbf{J}\|}_2     $} &
  \multicolumn{2}{c}{$\min_{\mathbf{J}^h} {\|\mathbf{e}_\mathbf{J}\|}_\infty$} &
  \multicolumn{2}{c}{$\min_{\mathbf{J}^h} {\|\mathbf{e}_\mathbf{J}\|}_2     $} \\
\cmidrule(r){2-3} \cmidrule(lr){4-5} \cmidrule(lr){6-7} \cmidrule(l){8-9}
Mesh & C & RP & C & RP & C & RP & C & RP \\ \midrule
   1--2  & 2.0800 & 2.0653 & 2.0811 & 1.2935 & 2.0447 & 2.0229 & 2.0454 & 2.0763 \\ 
   2--3  & 2.0141 & 2.0529 & 2.1055 & 1.4193 & 1.9948 & 2.0323 & 2.0359 & 2.0499 \\ 
   3--4  & 2.0303 & 2.0193 & 1.9159 & 1.5150 & 2.0141 & 1.9999 & 2.0283 & 2.0372 \\ 
   4--5  & 2.0196 & 2.0163 & 1.6421 & 1.5847 & 2.0064 & 2.0093 & 2.0229 & 2.0297 \\ 
   5--6  & 2.0061 & 2.0242 & 1.6677 & 1.6372 & 2.0060 & 2.0102 & 2.0190 & 2.0246 \\ 
   6--7  & 2.0133 & 2.0158 & 1.5800 & 1.6779 & 2.0057 & 2.0097 & 2.0162 & 2.0211 \\ 
   7--8  & 2.0113 & 2.0167 & 1.6282 & 1.7104 & 2.0057 & 2.0122 & 2.0140 & 2.0184 \\ 
   8--9  & 2.0037 & 2.0122 & 1.6664 & 1.7369 & 1.9965 & 2.0076 & 2.0123 & 2.0163 \\ 
\pz9--10 & 2.0086 & 2.0117 & 1.6974 & 1.7589 & 2.0039 & 2.0067 & 2.0110 & 2.0146 \\
  10--11 & 2.0053 & 2.0118 & 1.7231 & 1.7776 & 2.0039 & 2.0094 & 2.0099 & 2.0133 \\
\bottomrule
\end{tabular}
\caption{Solution-discretization error: convergence rates computed between adjacent mesh pairs for $d=3$ in Figure~\ref{fig:part1b}.}
\label{tab:part1b}
\end{table}

\begin{figure}
\centering
\begin{subfigure}[b]{.49\textwidth}
\includegraphics[scale=.64,clip=true,trim=2.3in 0in 2.8in 0in]{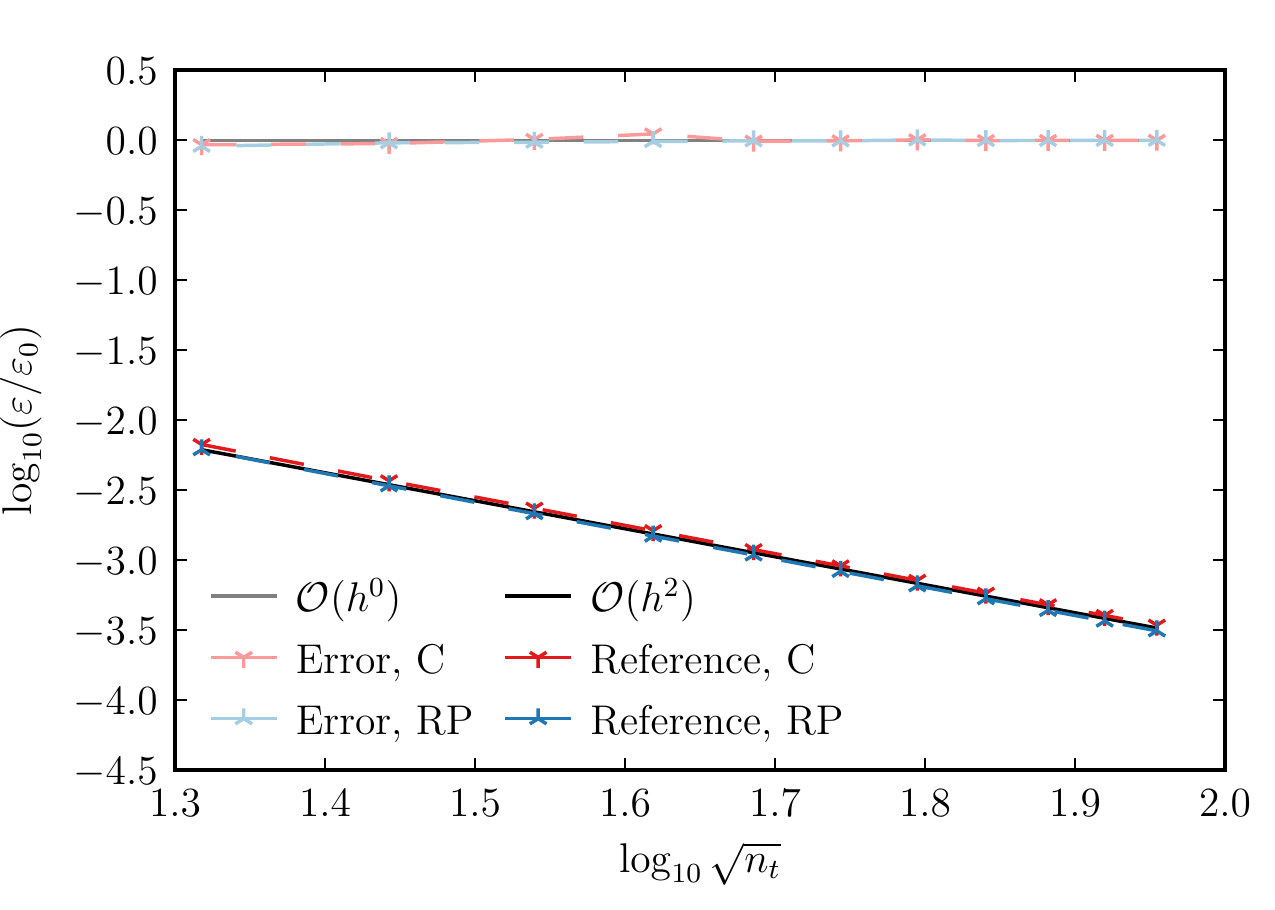}
\caption{$\varepsilon={\|\mathbf{e}_\mathbf{J}\|}_\infty$, $\min_{\mathbf{J}^h} {\|\mathbf{e}_\mathbf{J}\|}_\infty$\vpad}
\end{subfigure}
\hspace{0.25em}
\begin{subfigure}[b]{.49\textwidth}
\includegraphics[scale=.64,clip=true,trim=2.3in 0in 2.8in 0in]{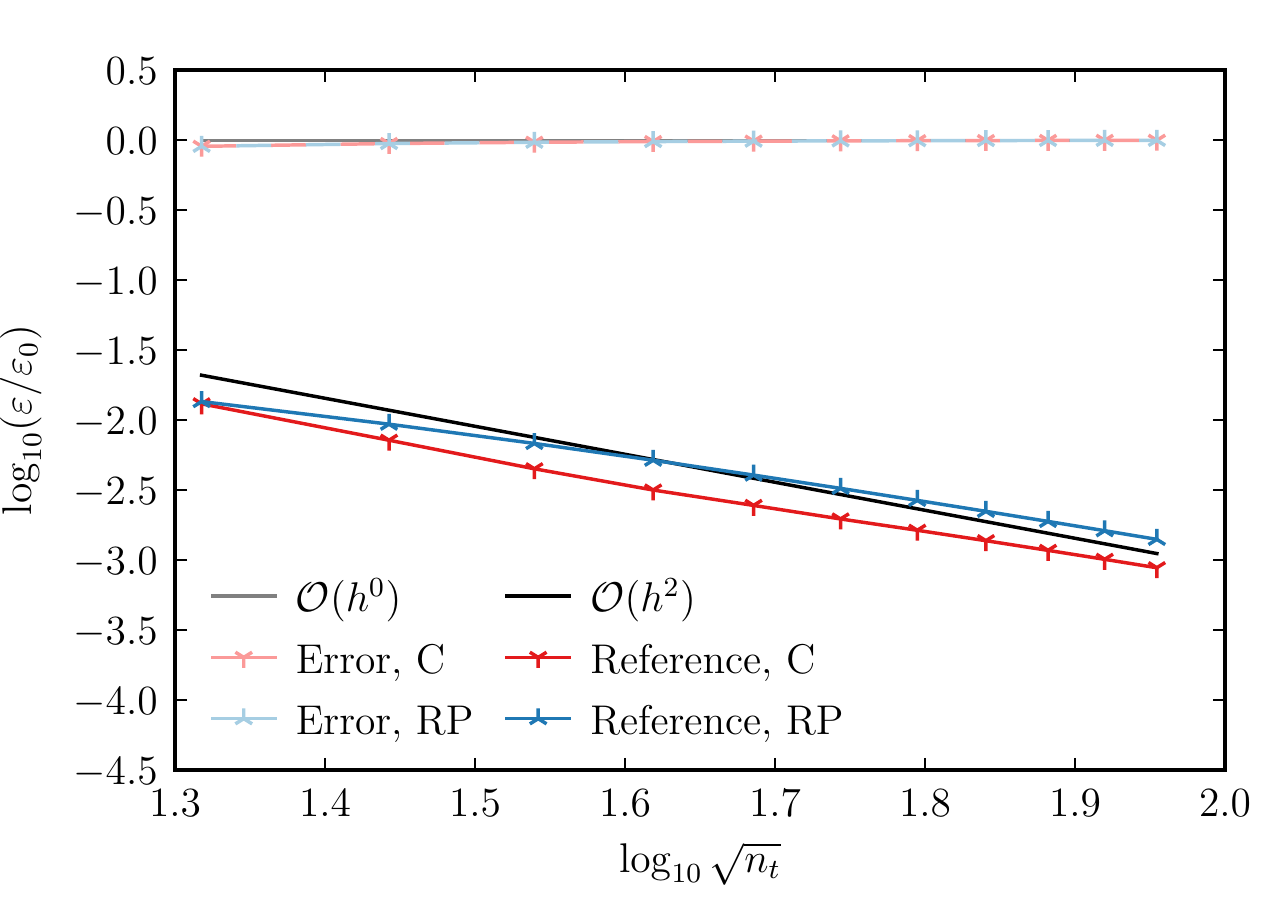}
\caption{$\varepsilon={\|\mathbf{e}_\mathbf{J}\|}_\infty$, $\min_{\mathbf{J}^h} {\|\mathbf{e}_\mathbf{J}\|}_2$\vpad}
\end{subfigure}
\\
\begin{subfigure}[b]{.49\textwidth}
\includegraphics[scale=.64,clip=true,trim=2.3in 0in 2.8in 0in]{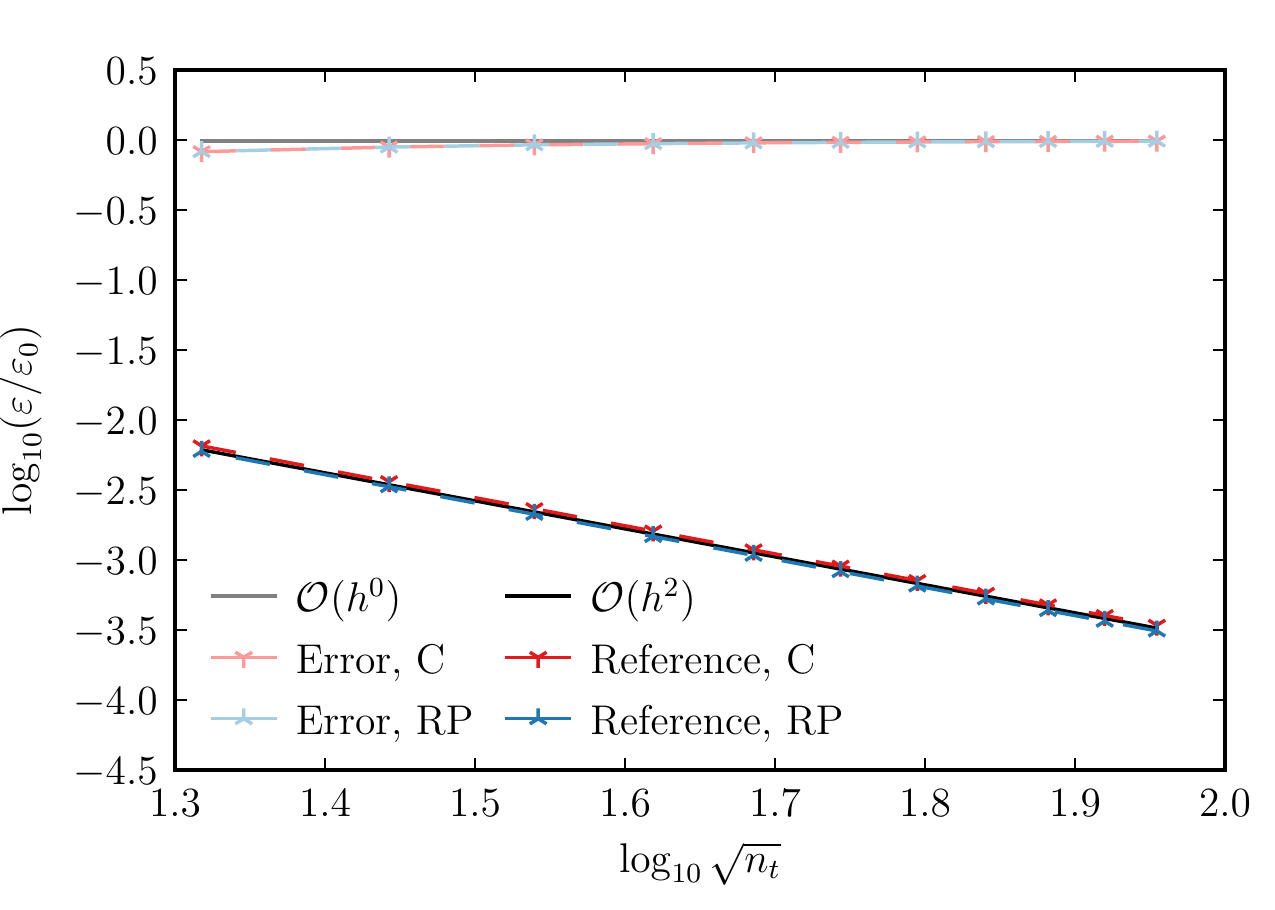}
\caption{$\varepsilon={\|\mathbf{e}_\mathbf{J}\|}_2$, $\min_{\mathbf{J}^h} {\|\mathbf{e}_\mathbf{J}\|}_\infty$\vpad}
\end{subfigure}
\hspace{0.25em}
\begin{subfigure}[b]{.49\textwidth}
\includegraphics[scale=.64,clip=true,trim=2.3in 0in 2.8in 0in]{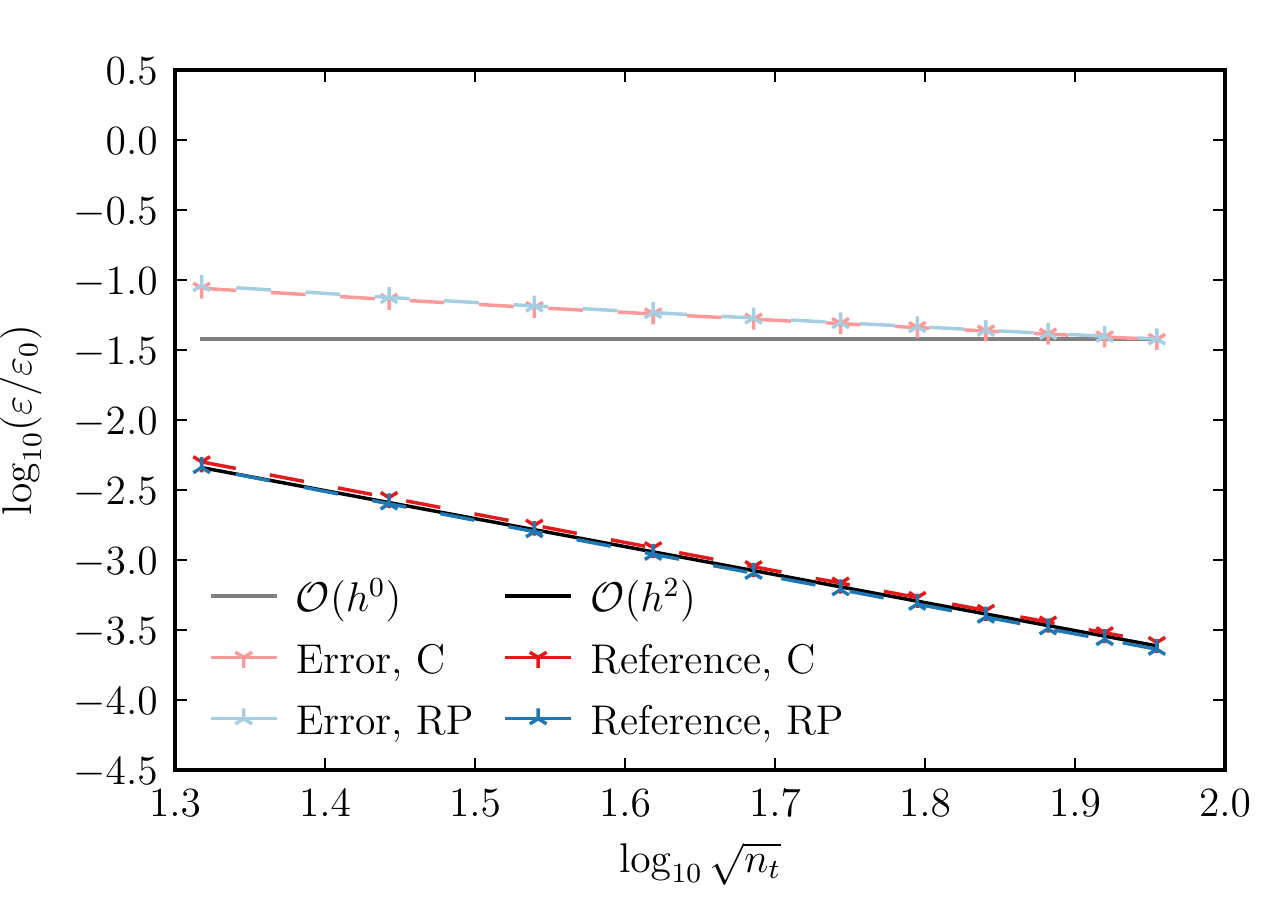}
\caption{$\varepsilon={\|\mathbf{e}_\mathbf{J}\|}_2$, $\min_{\mathbf{J}^h} {\|\mathbf{e}_\mathbf{J}\|}_2$\vpad}
\end{subfigure}
\caption{Solution-discretization error: $\varepsilon=\|\mathbf{e}_\mathbf{J}\|$ for different metric and minimization norms in the presence of coding error for $d=3$.}
\vskip-\dp\strutbox
\label{fig:part1b_bug}
\end{figure}

Figure~\ref{fig:part1b_bug} shows the four metric and minimization combinations shown in Figure~\ref{fig:part1b} for Term 2 and $d=3$ in the presence of a coding error in which the magnitudes of the diagonal elements of the matrix are increased by 1\%.  None of the cases with the coding errors have convergence rates that are $\mathcal{O}(h^2)$; therefore, each of these combinations detects the coding error.



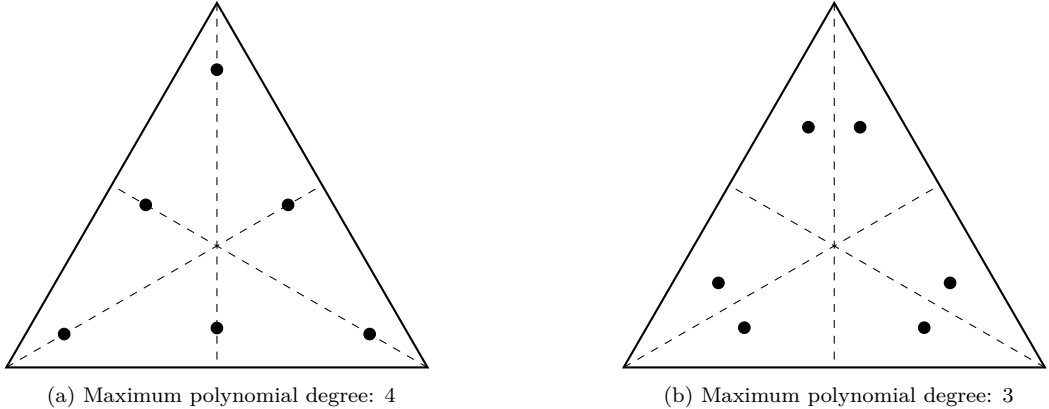
\begin{figure}
\centering
\begin{subfigure}[b]{.49\textwidth}
\centering
\begin{tikzpicture}[scale=0.7, every node/.style={scale=0.7}]

\def\ts{8}; 

\coordinate (A) at (-0.5,{-sqrt(3)/6});
\coordinate (B) at ( 0.5,{-sqrt(3)/6});
\coordinate (C) at ( 0  , {sqrt(1/3)});
\coordinate (D) at ($0.5*(A)+0.5*(B)$);
\coordinate (E) at ($0.5*(B)+0.5*(C)$);
\coordinate (F) at ($0.5*(C)+0.5*(A)$);
\coordinate (O) at (0,0);

\def\tx{0}
\def\ty{{\ts*sqrt(3)/6}}

\coordinate (T1) at ($\ts*(A)+(\tx,\ty)$);
\coordinate (T2) at ($\ts*(B)+(\tx,\ty)$);
\coordinate (T3) at ($\ts*(C)+(\tx,\ty)$);
\coordinate (T4) at ($\ts*(D)+(\tx,\ty)$);
\draw[thick] (T1) -- (T2) -- (T3) -- cycle;

\draw[dashed] ($\ts*(C)+(\tx,\ty)$) -- ($\ts*(D)+(\tx,\ty)$);
\draw[dashed] ($\ts*(A)+(\tx,\ty)$) -- ($\ts*(E)+(\tx,\ty)$);
\draw[dashed] ($\ts*(B)+(\tx,\ty)$) -- ($\ts*(F)+(\tx,\ty)$);

\draw[fill=black] ($\ts*(-0.168922736373948, 0.097527587317747)+(\tx,\ty)$) circle (.11);
\draw[fill=black] ($\ts*( 0.168922736373948, 0.097527587317747)+(\tx,\ty)$) circle (.11);
\draw[fill=black] ($\ts*( 0.000000000000000,-0.195055174635494)+(\tx,\ty)$) circle (.11);
\draw[fill=black] ($\ts*( 0.362635679735344,-0.209367807312963)+(\tx,\ty)$) circle (.11);
\draw[fill=black] ($\ts*(-0.362635679735344,-0.209367807312963)+(\tx,\ty)$) circle (.11);
\draw[fill=black] ($\ts*( 0.000000000000000, 0.418735614625928)+(\tx,\ty)$) circle (.11);

\end{tikzpicture}
\caption{Maximum polynomial degree: 4}
\label{fig:quad_optimal_6}
\end{subfigure}
\begin{subfigure}[b]{.49\textwidth}
\centering
\begin{tikzpicture}[scale=0.7, every node/.style={scale=0.7}]

\def\ts{8}; 

\coordinate (A) at (-0.5,{-sqrt(3)/6});
\coordinate (B) at ( 0.5,{-sqrt(3)/6});
\coordinate (C) at ( 0  , {sqrt(1/3)});
\coordinate (D) at ($0.5*(A)+0.5*(B)$);
\coordinate (E) at ($0.5*(B)+0.5*(C)$);
\coordinate (F) at ($0.5*(C)+0.5*(A)$);
\coordinate (O) at (0,0);

\def\tx{0}
\def\ty{{\ts*sqrt(3)/6}}

\coordinate (T1) at ($\ts*(A)+(\tx,\ty)$);
\coordinate (T2) at ($\ts*(B)+(\tx,\ty)$);
\coordinate (T3) at ($\ts*(C)+(\tx,\ty)$);
\coordinate (T4) at ($\ts*(D)+(\tx,\ty)$);
\draw[thick] (T1) -- (T2) -- (T3) -- cycle;

\draw[dashed] ($\ts*(C)+(\tx,\ty)$) -- ($\ts*(D)+(\tx,\ty)$);
\draw[dashed] ($\ts*(A)+(\tx,\ty)$) -- ($\ts*(E)+(\tx,\ty)$);
\draw[dashed] ($\ts*(B)+(\tx,\ty)$) -- ($\ts*(F)+(\tx,\ty)$);

\draw[fill=black] ($\ts*( 6.1447179740076699e-2, 2.8205952817680885e-1)+(\tx,\ty)$) circle (.11);
\draw[fill=black] ($\ts*( 2.1354712691053079e-1,-1.9424458273421930e-1)+(\tx,\ty)$) circle (.11);
\draw[fill=black] ($\ts*(-2.7499430665060748e-1,-8.7814945442589498e-2)+(\tx,\ty)$) circle (.11);
\draw[fill=black] ($\ts*(-6.1447179740076699e-2, 2.8205952817680885e-1)+(\tx,\ty)$) circle (.11);
\draw[fill=black] ($\ts*(-2.1354712691053079e-1,-1.9424458273421930e-1)+(\tx,\ty)$) circle (.11);
\draw[fill=black] ($\ts*( 2.7499430665060748e-1,-8.7814945442589498e-2)+(\tx,\ty)$) circle (.11);

\end{tikzpicture}
\caption{Maximum polynomial degree: 3}
\label{fig:quad_suboptimal_6}
\end{subfigure}
\caption{6-point quadrature rules.}
\vskip-\dp\strutbox
\label{fig:quadrature}
\end{figure}

\begin{table}
\centering
\begin{tabular}{l c c c c c c c c}
\toprule
Number of points         & 1     & 3     & 4     &     6 &     7 &    12 &    13 \\ \midrule
Maximum integrand degree & 1     & 2     & 3     &     4 &     5 &  \pz6 &  \pz7 \\
Convergence rate         & $\mathcal{O}(h^2)$ & $\mathcal{O}(h^4)$ & $\mathcal{O}(h^4)$ & $\mathcal{O}(h^6)$ & $\mathcal{O}(h^6)$ & $\mathcal{O}(h^8)$ & $\mathcal{O}(h^8)$ \\
\bottomrule
\end{tabular}
\caption{Polynomial triangle quadrature properties.}
\vskip-\dp\strutbox
\label{tab:dunavant_properties}
\end{table}

\subsubsection{Numerical-Integration Error} 

To isolate and measure the numerical-integration error, we continue by performing the assessments described in Section~\ref{sec:nie}.
\reviewerTwo{The numerical integration is performed using polynomial quadrature rules for triangles.  For multiple} quadrature point amounts, Table~\ref{tab:dunavant_properties} lists the maximum \reviewerTwo{polynomial degree of the integrand} the points can integrate exactly~\cite{lyness_1975,dunavant_1985}, as well as the convergence rates of the errors for inexact integrations of nonsingular integrands.  \reviewerTwo{These properties correspond to the optimal point locations and weights.}  For nonsingular integrands, the slowest expected quadrature convergence rate is $\mathcal{O}(h^2)$.  Figure~\ref{fig:quad_optimal_6} shows the \reviewerTwo{optimal 6-point quadrature rule, which can exactly integrate polynomials up to degree 4, whereas Figure~\ref{fig:quad_suboptimal_6} shows a suboptimal 6-point quadrature rule~\cite{papanicolopulos_2015}, which can exactly integrate polynomials up to degree 3.}


\begin{figure}
\centering
\begin{subfigure}[b]{.49\textwidth}
\includegraphics[scale=.64,clip=true,trim=2.3in 0in 2.8in 0in]{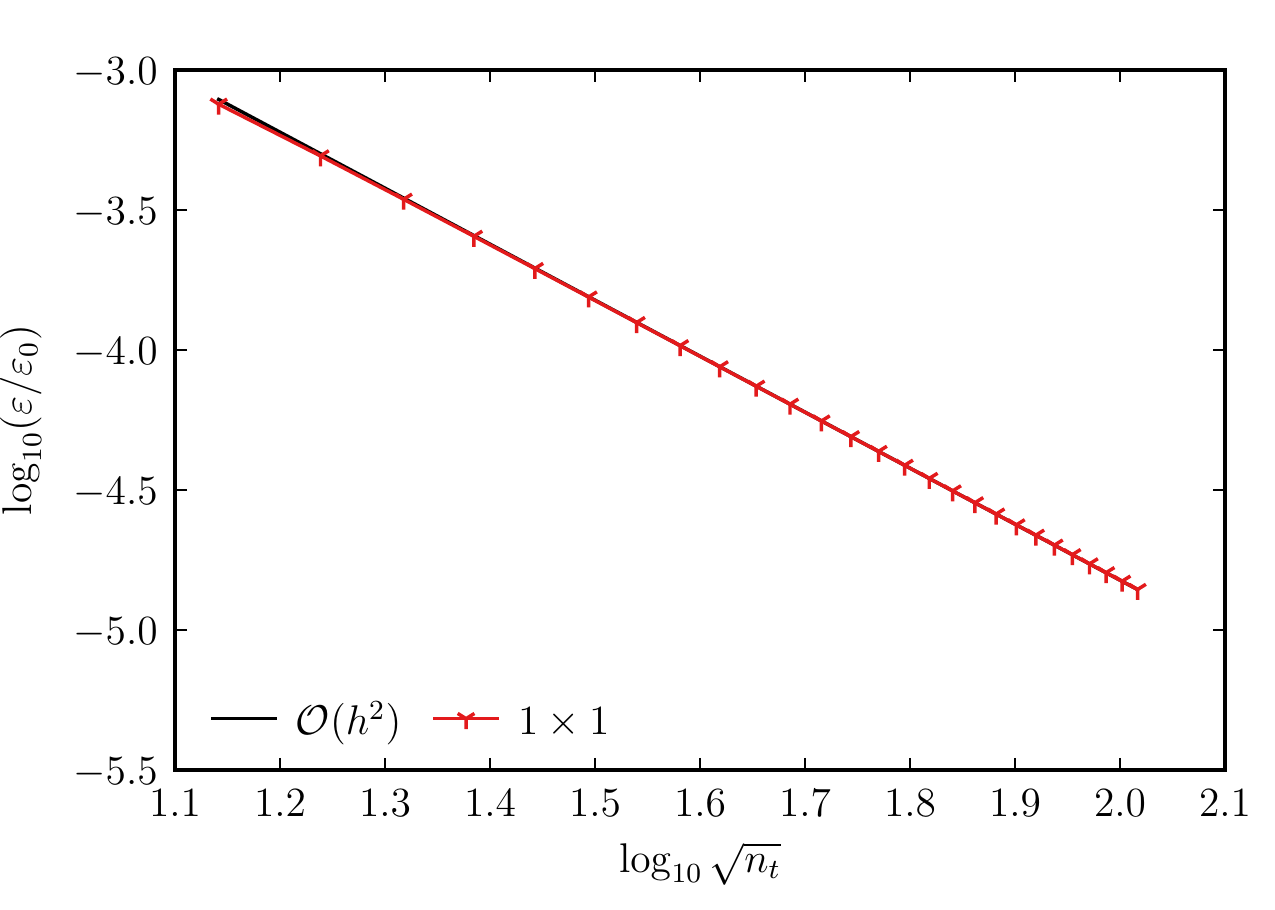}
\caption{Cube, $d=1$\vpad}
\end{subfigure}
\hspace{0.25em}
\begin{subfigure}[b]{.49\textwidth}
\includegraphics[scale=.64,clip=true,trim=2.3in 0in 2.8in 0in]{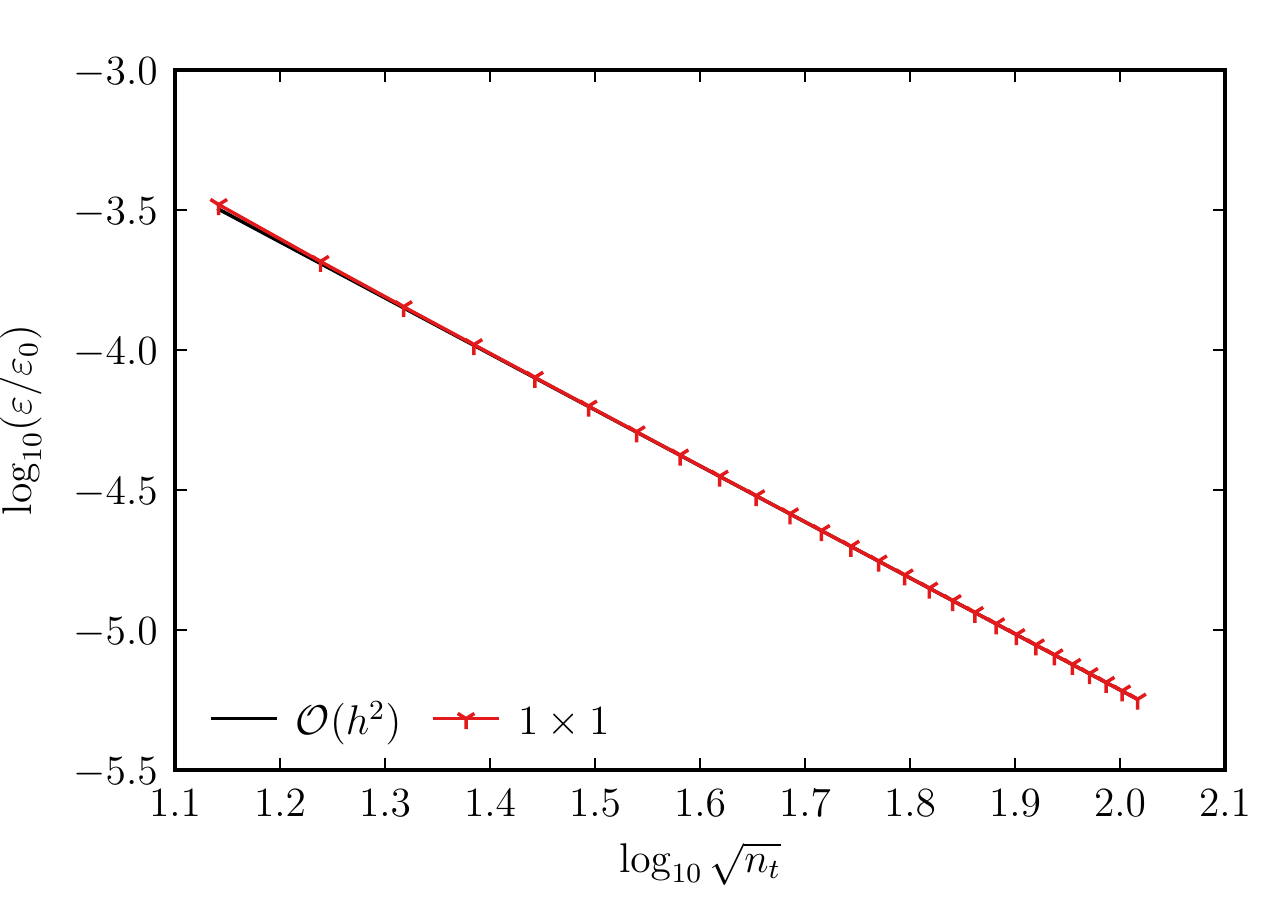}
\caption{Rhombic Prism, $d=1$\vpad}
\end{subfigure}
\\
\begin{subfigure}[b]{.49\textwidth}
\includegraphics[scale=.64,clip=true,trim=2.3in 0in 2.8in 0in]{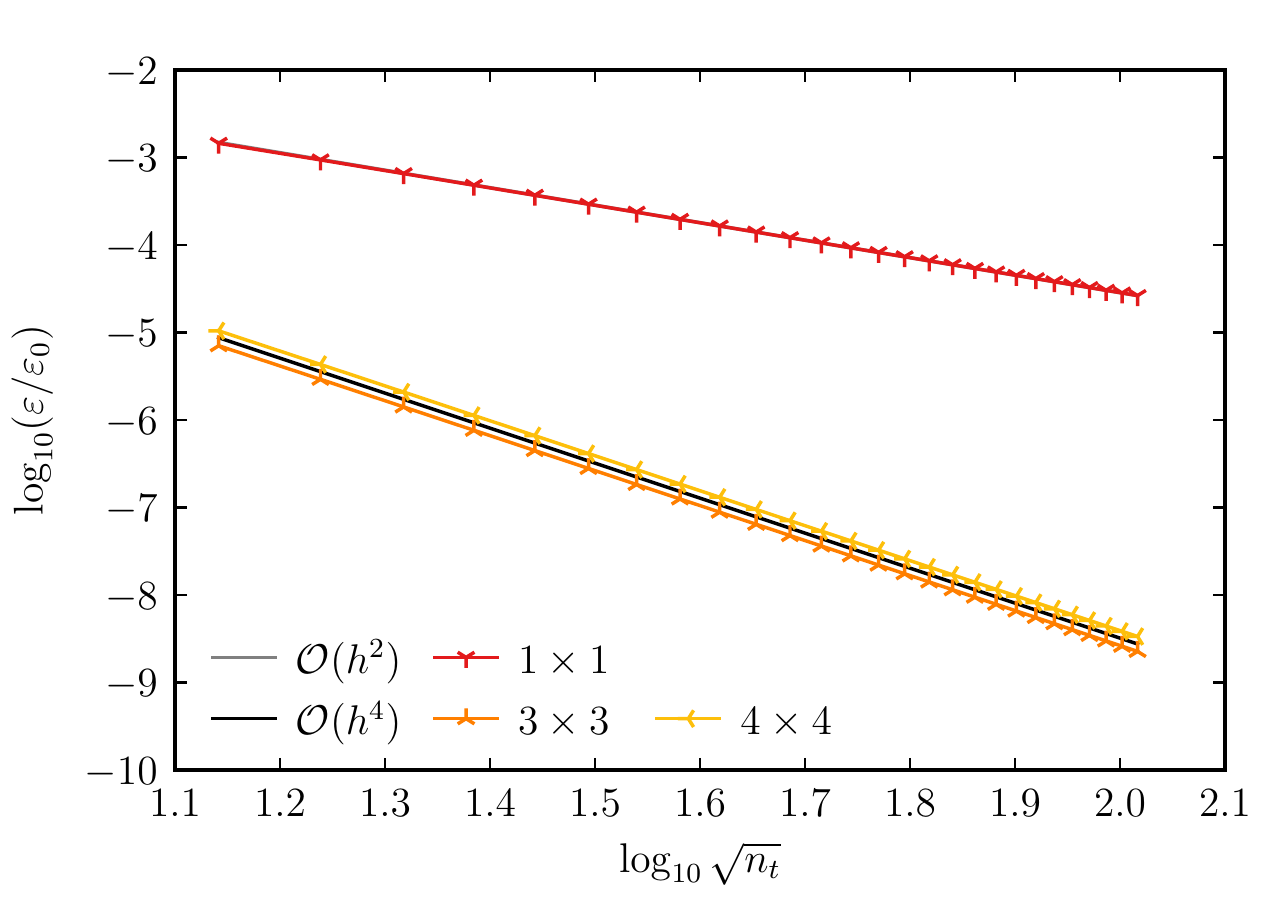}
\caption{Cube, $d=2$\vpad}
\end{subfigure}
\hspace{0.25em}
\begin{subfigure}[b]{.49\textwidth}
\includegraphics[scale=.64,clip=true,trim=2.3in 0in 2.8in 0in]{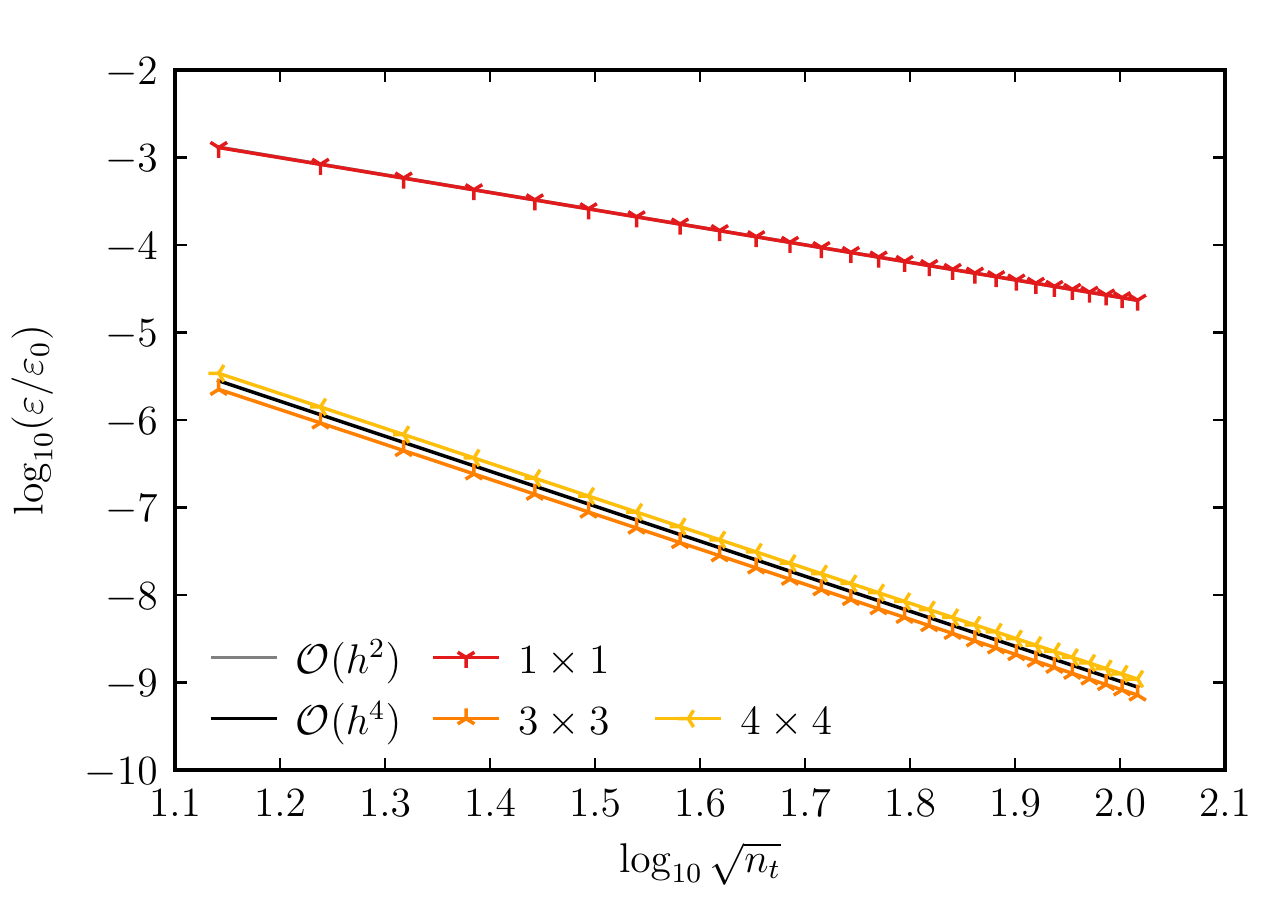}
\caption{Rhombic Prism, $d=2$\vpad}
\end{subfigure}
\\
\begin{subfigure}[b]{.49\textwidth}
\includegraphics[scale=.64,clip=true,trim=2.3in 0in 2.8in 0in]{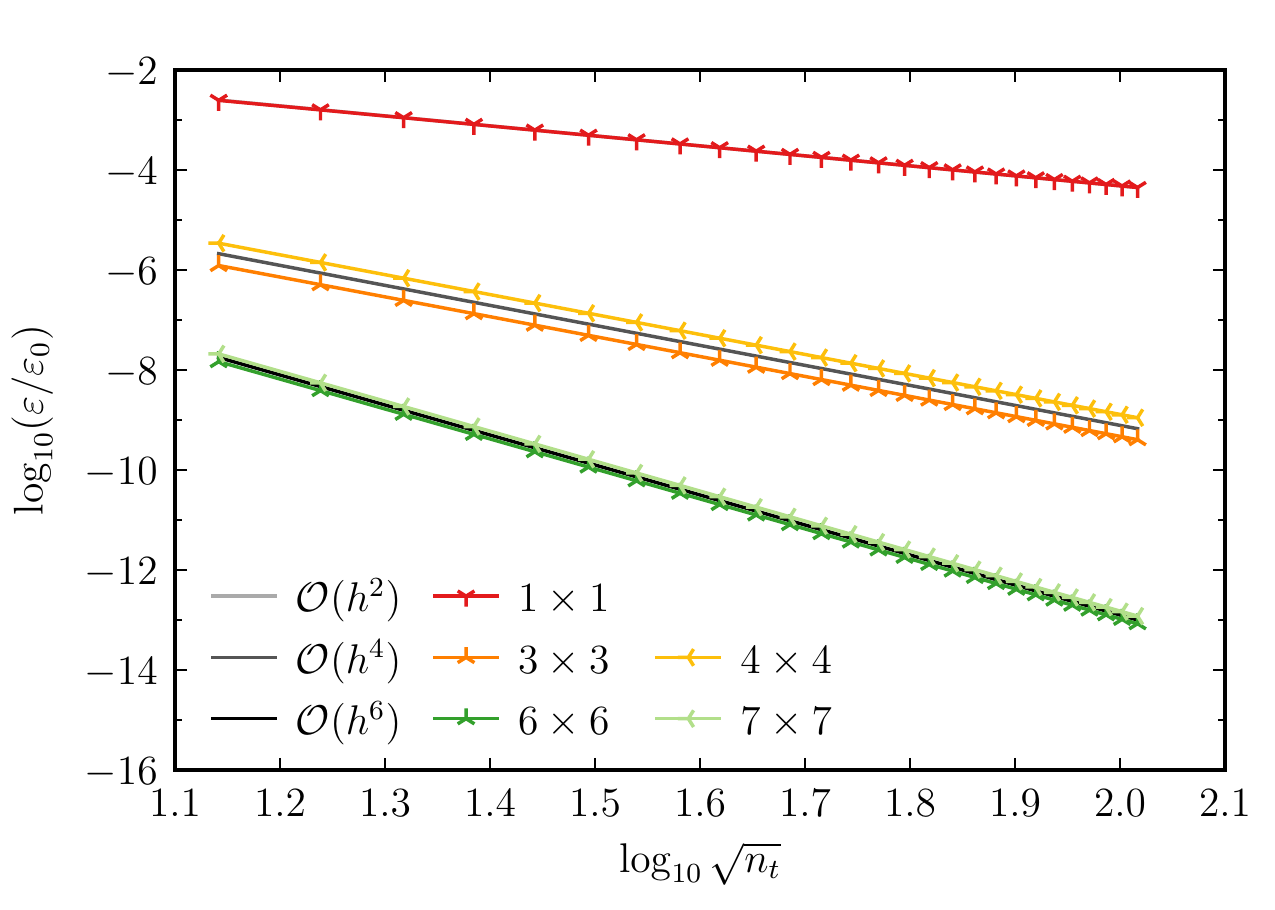}
\caption{Cube, $d=3$\vpad}
\end{subfigure}
\hspace{0.25em}
\begin{subfigure}[b]{.49\textwidth}
\includegraphics[scale=.64,clip=true,trim=2.3in 0in 2.8in 0in]{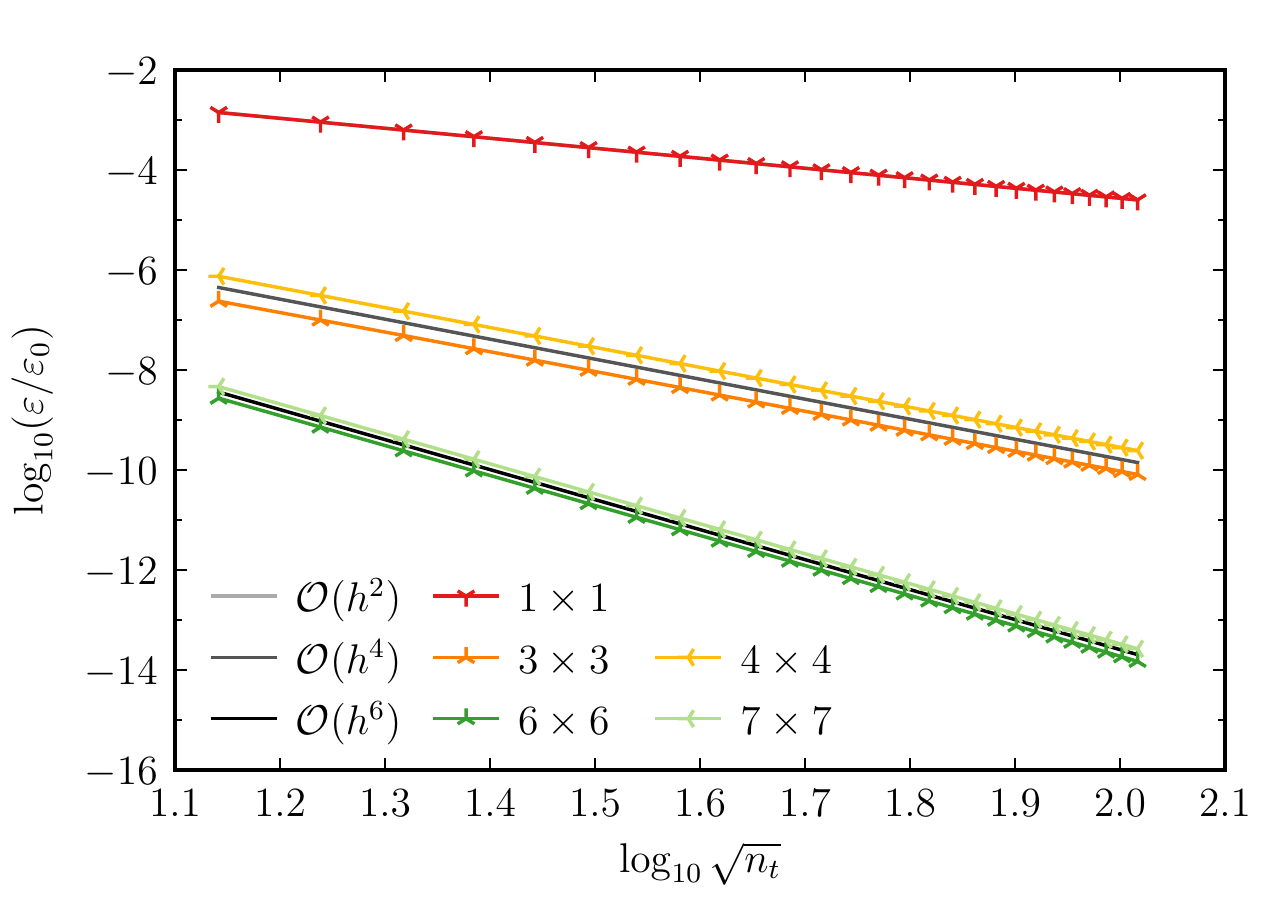}
\caption{Rhombic Prism, $d=3$\vpad}
\end{subfigure}
\caption{Numerical-integration error: $\varepsilon=|e_a(\mathbf{J}_{h_\text{MS}})|$~\eqref{eq:a_error_cancel} for different amounts of quadrature points.}
\vskip-\dp\strutbox
\label{fig:part5}
\end{figure}

\begin{figure}
\centering
\begin{subfigure}[b]{.49\textwidth}
\includegraphics[scale=.64,clip=true,trim=2.3in 0in 2.8in 0in]{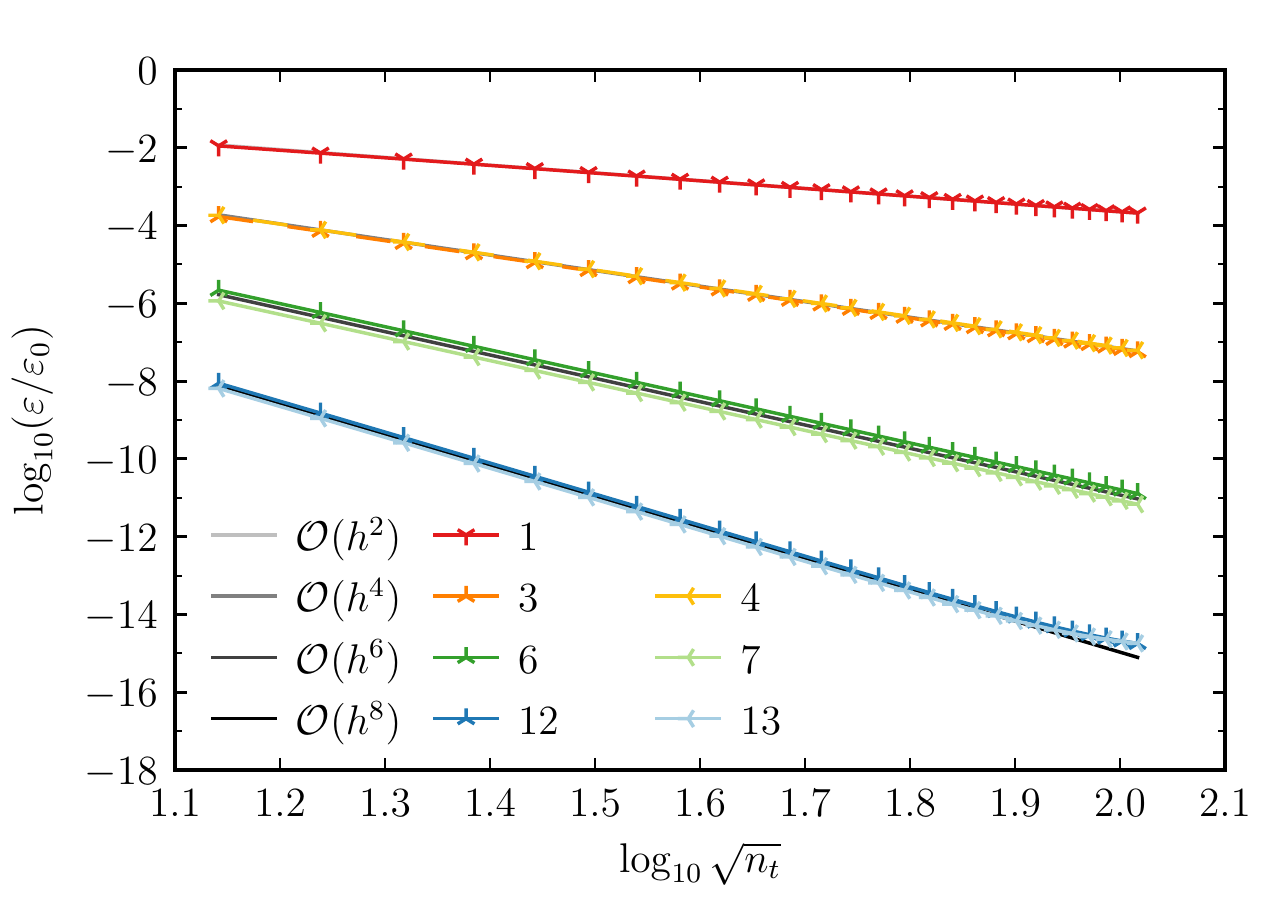}
\caption{Cube, $d=1$\vpad}
\end{subfigure}
\hspace{0.25em}
\begin{subfigure}[b]{.49\textwidth}
\includegraphics[scale=.64,clip=true,trim=2.3in 0in 2.8in 0in]{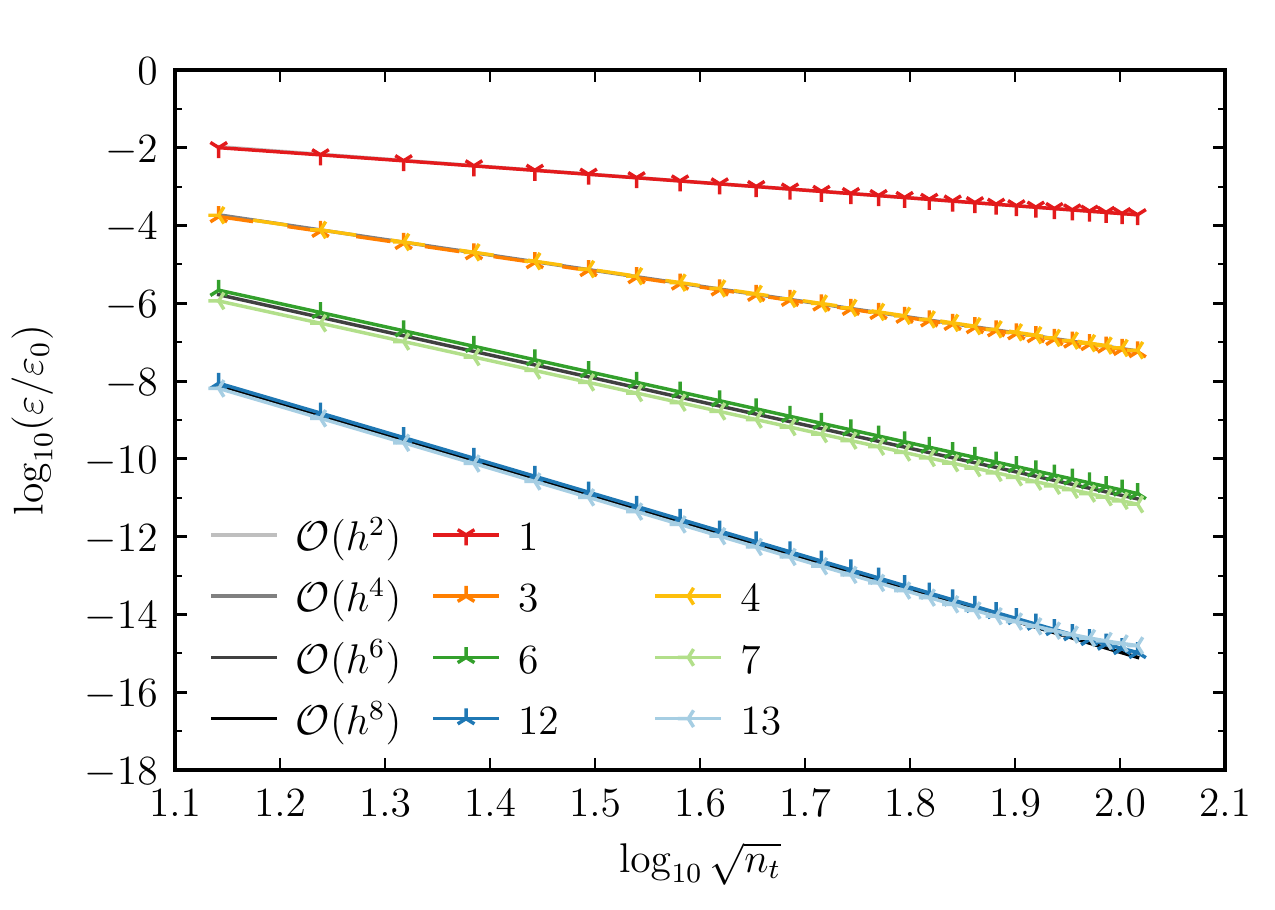}
\caption{Rhombic Prism, $d=1$\vpad}
\end{subfigure}
\\
\begin{subfigure}[b]{.49\textwidth}
\includegraphics[scale=.64,clip=true,trim=2.3in 0in 2.8in 0in]{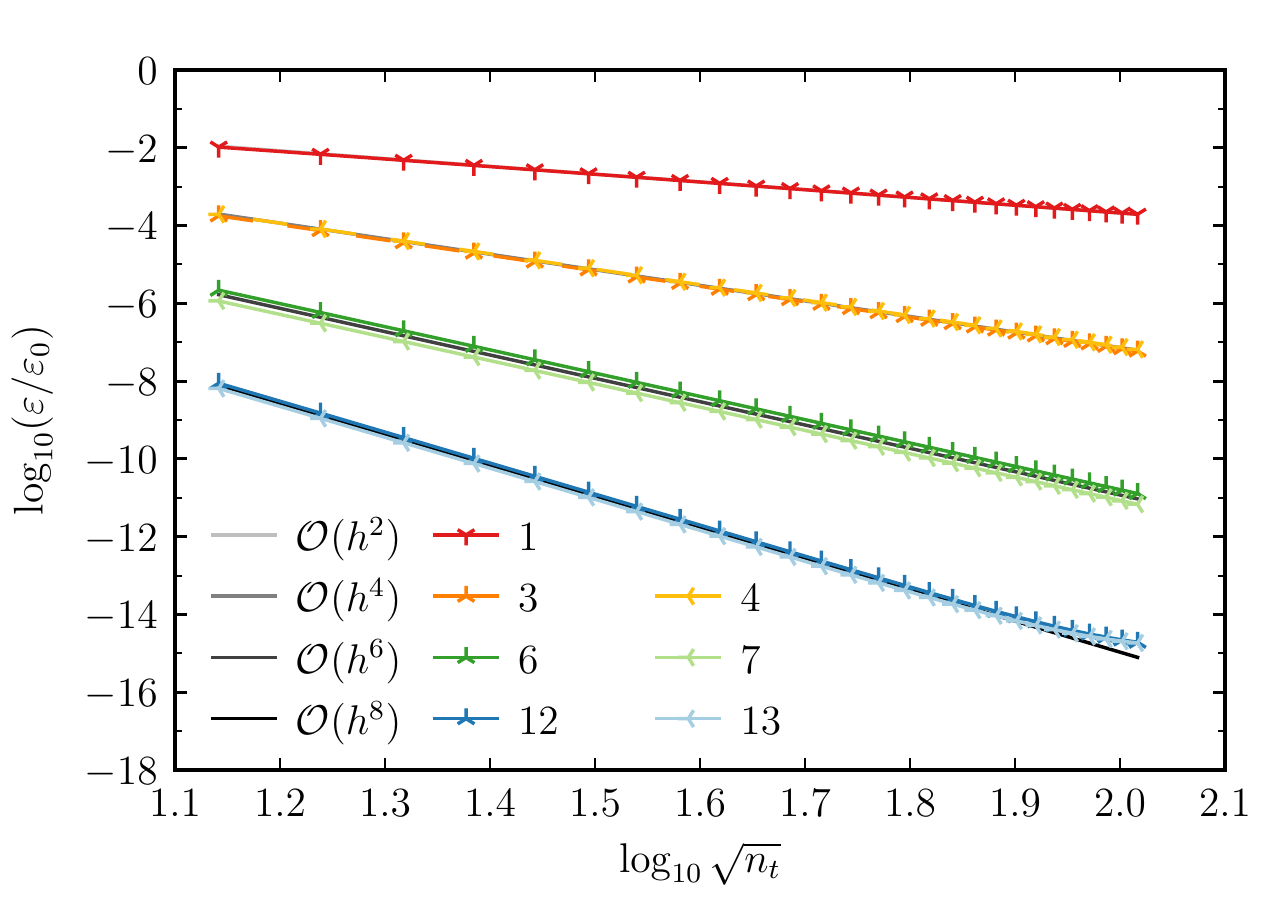}
\caption{Cube, $d=2$\vpad}
\end{subfigure}
\hspace{0.25em}
\begin{subfigure}[b]{.49\textwidth}
\includegraphics[scale=.64,clip=true,trim=2.3in 0in 2.8in 0in]{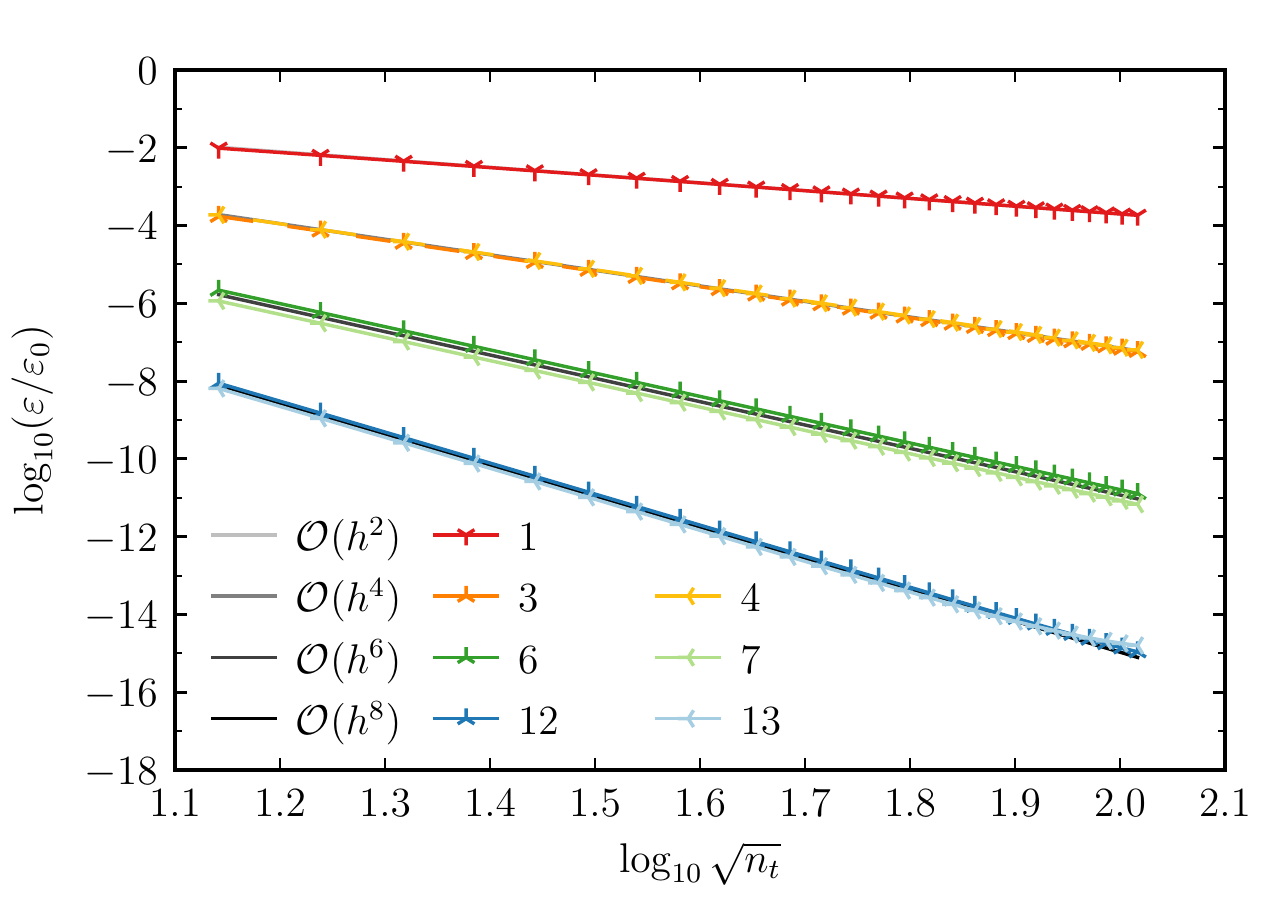}
\caption{Rhombic Prism, $d=2$\vpad}
\end{subfigure}
\\
\begin{subfigure}[b]{.49\textwidth}
\includegraphics[scale=.64,clip=true,trim=2.3in 0in 2.8in 0in]{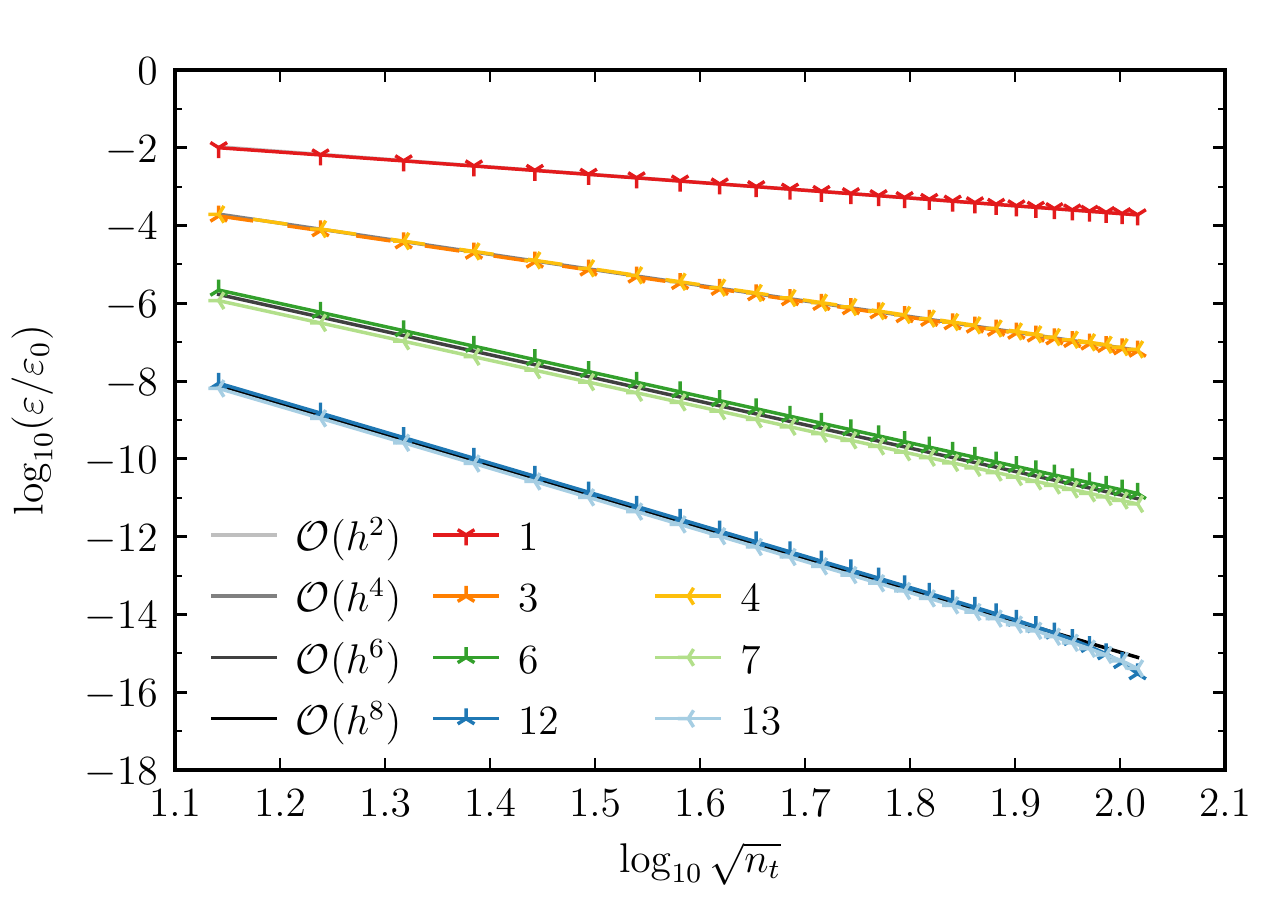}
\caption{Cube, $d=3$\vpad}
\end{subfigure}
\hspace{0.25em}
\begin{subfigure}[b]{.49\textwidth}
\includegraphics[scale=.64,clip=true,trim=2.3in 0in 2.8in 0in]{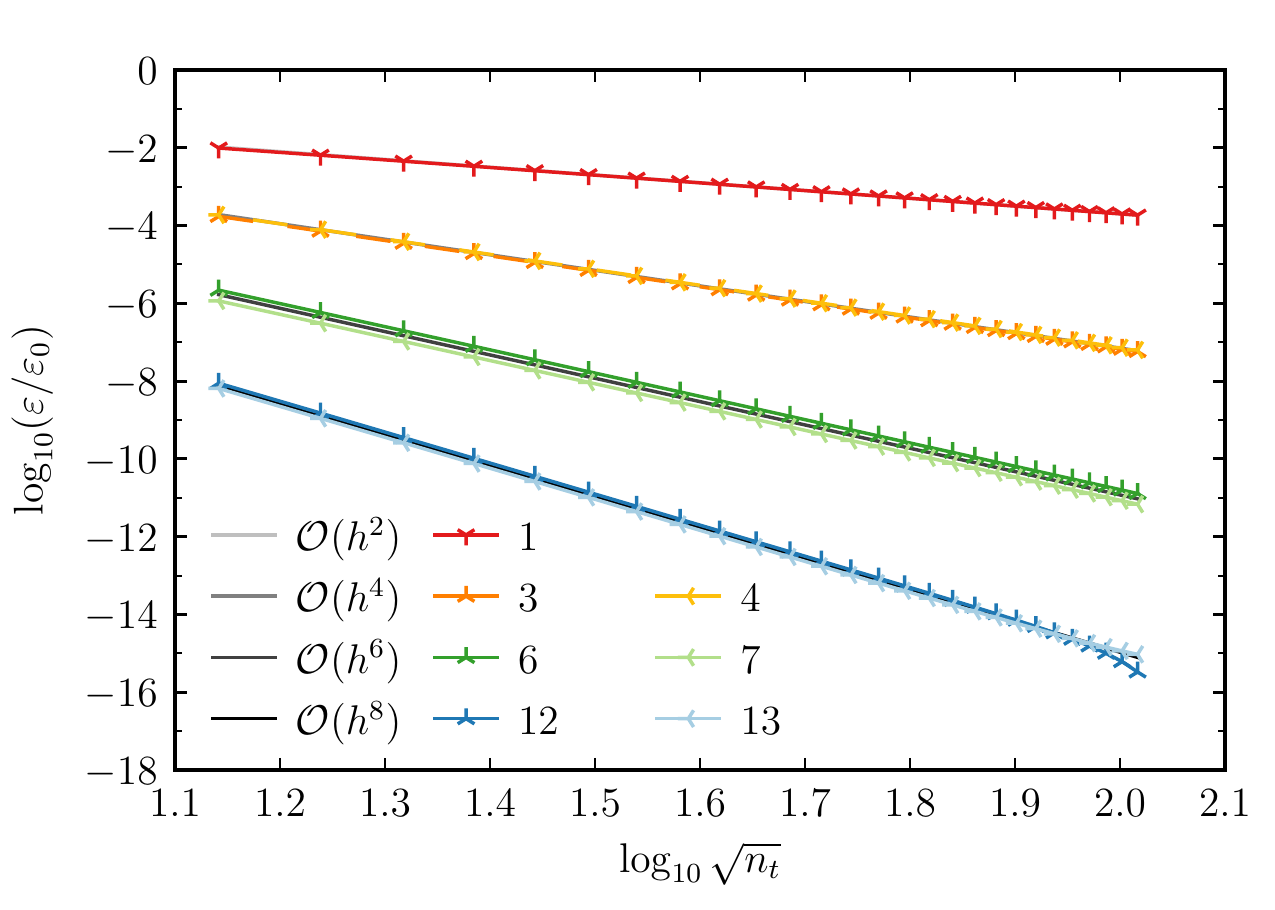}
\caption{Rhombic Prism, $d=3$\vpad}
\end{subfigure}
\caption{Numerical-integration error: $\varepsilon=|e_b(\mathbf{J}_{h_\text{MS}})|$~\eqref{eq:b_error_cancel} for different amounts of quadrature points.}
\vskip-\dp\strutbox
\label{fig:part6}
\end{figure}

\begin{figure}
\centering
\begin{subfigure}[b]{.49\textwidth}
\includegraphics[scale=.64,clip=true,trim=2.3in 0in 2.8in 0in]{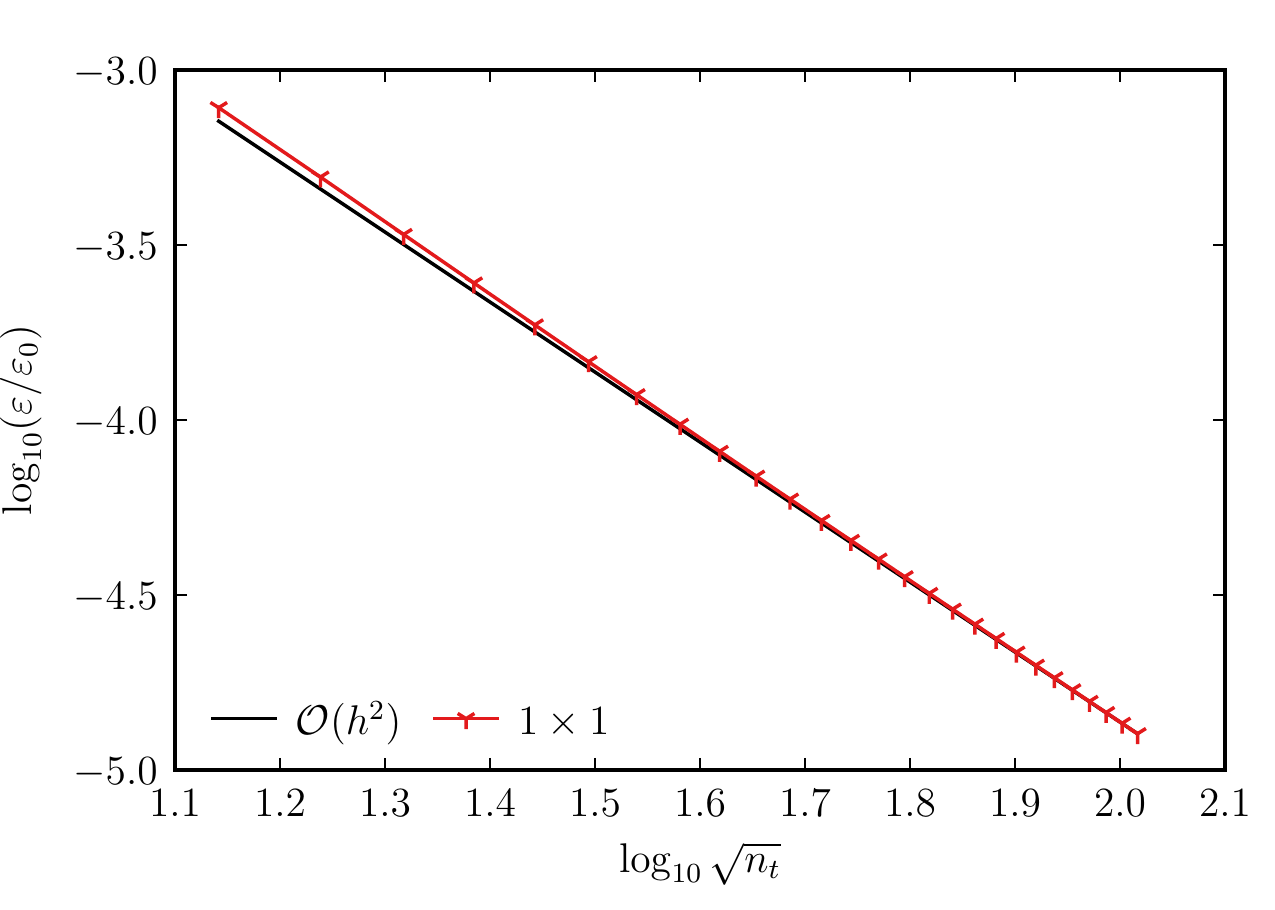}
\caption{Cube, $d=1$\vpad}
\end{subfigure}
\hspace{0.25em}
\begin{subfigure}[b]{.49\textwidth}
\includegraphics[scale=.64,clip=true,trim=2.3in 0in 2.8in 0in]{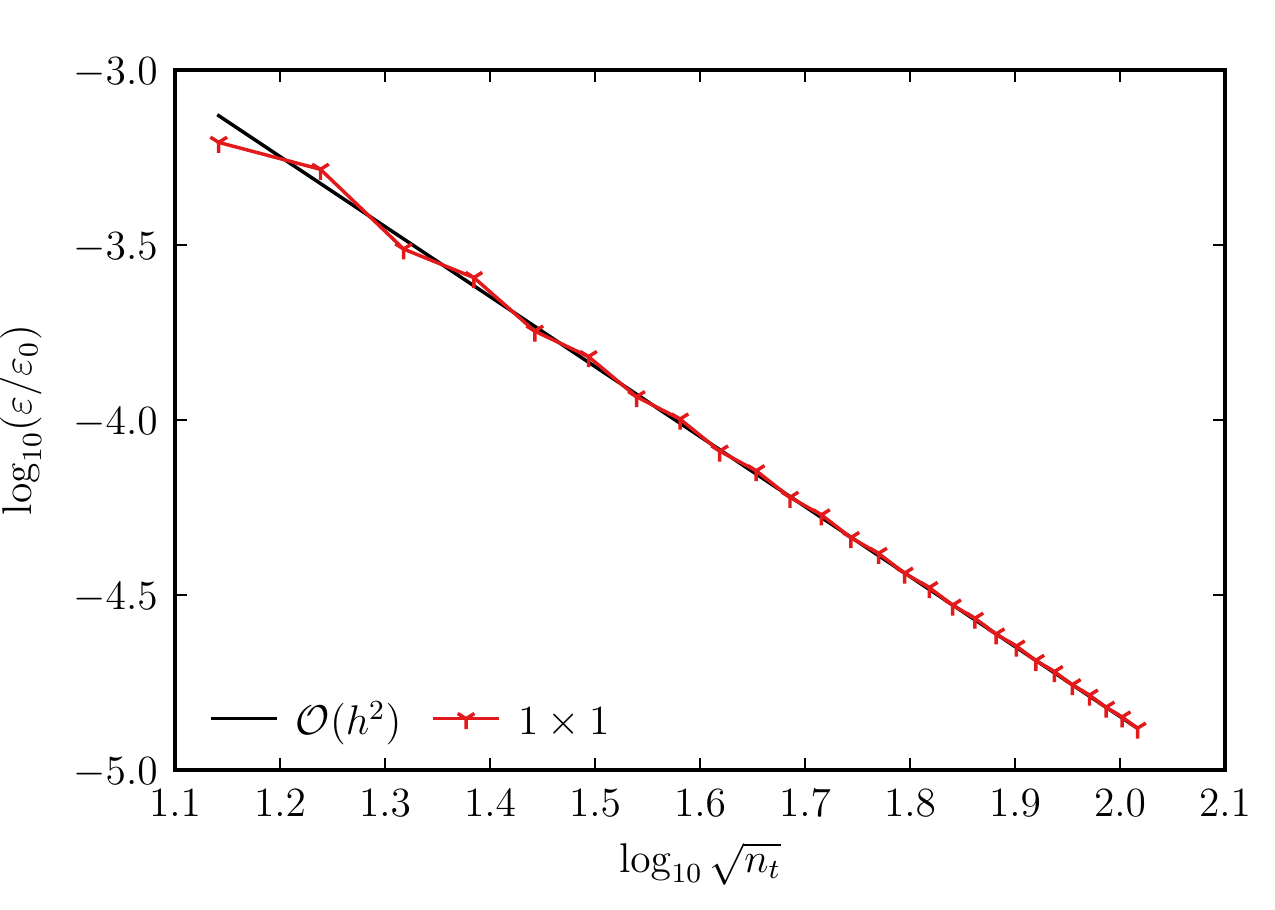}
\caption{Rhombic Prism, $d=1$\vpad}
\end{subfigure}
\\
\begin{subfigure}[b]{.49\textwidth}
\includegraphics[scale=.64,clip=true,trim=2.3in 0in 2.8in 0in]{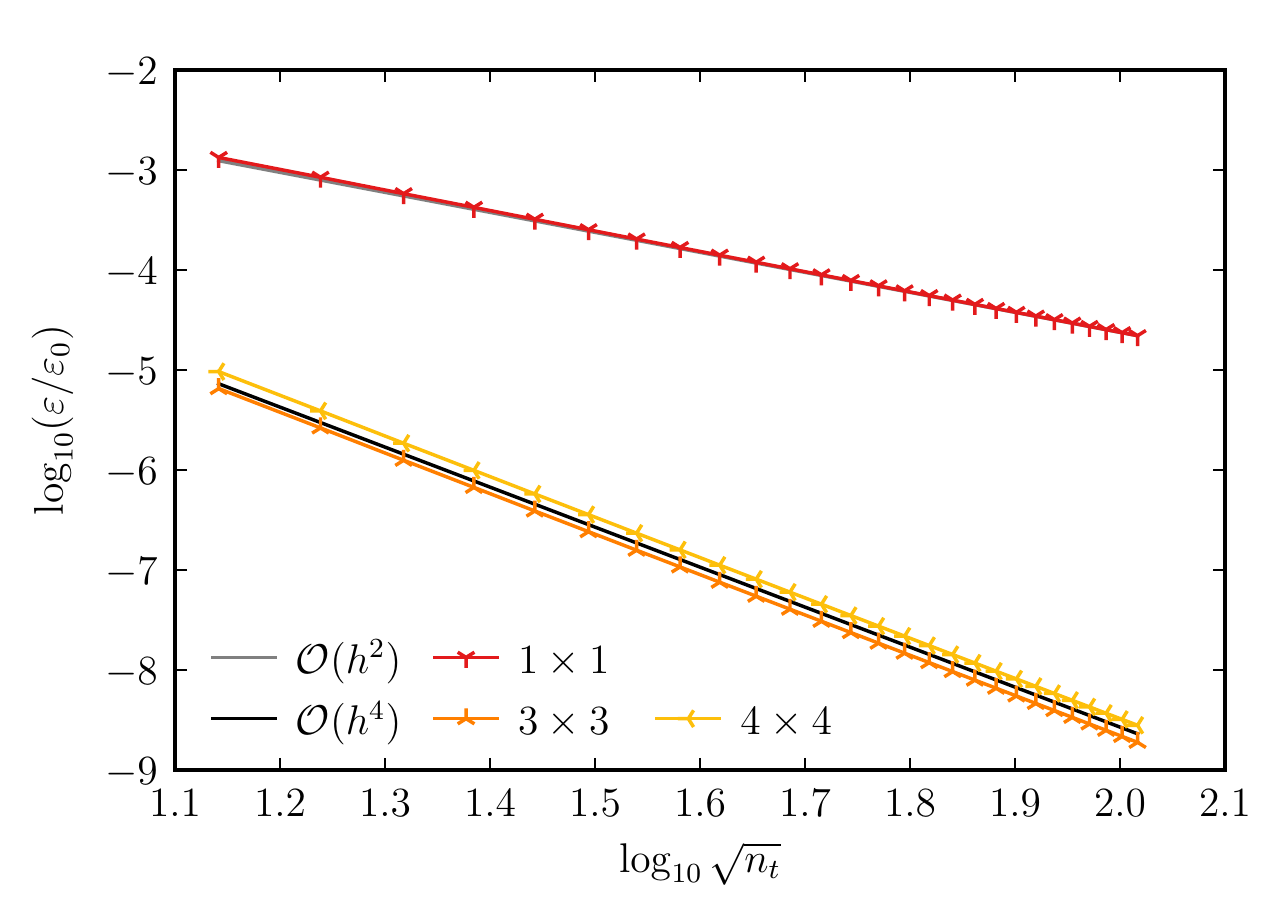}
\caption{Cube, $d=2$\vpad}
\end{subfigure}
\hspace{0.25em}
\begin{subfigure}[b]{.49\textwidth}
\includegraphics[scale=.64,clip=true,trim=2.3in 0in 2.8in 0in]{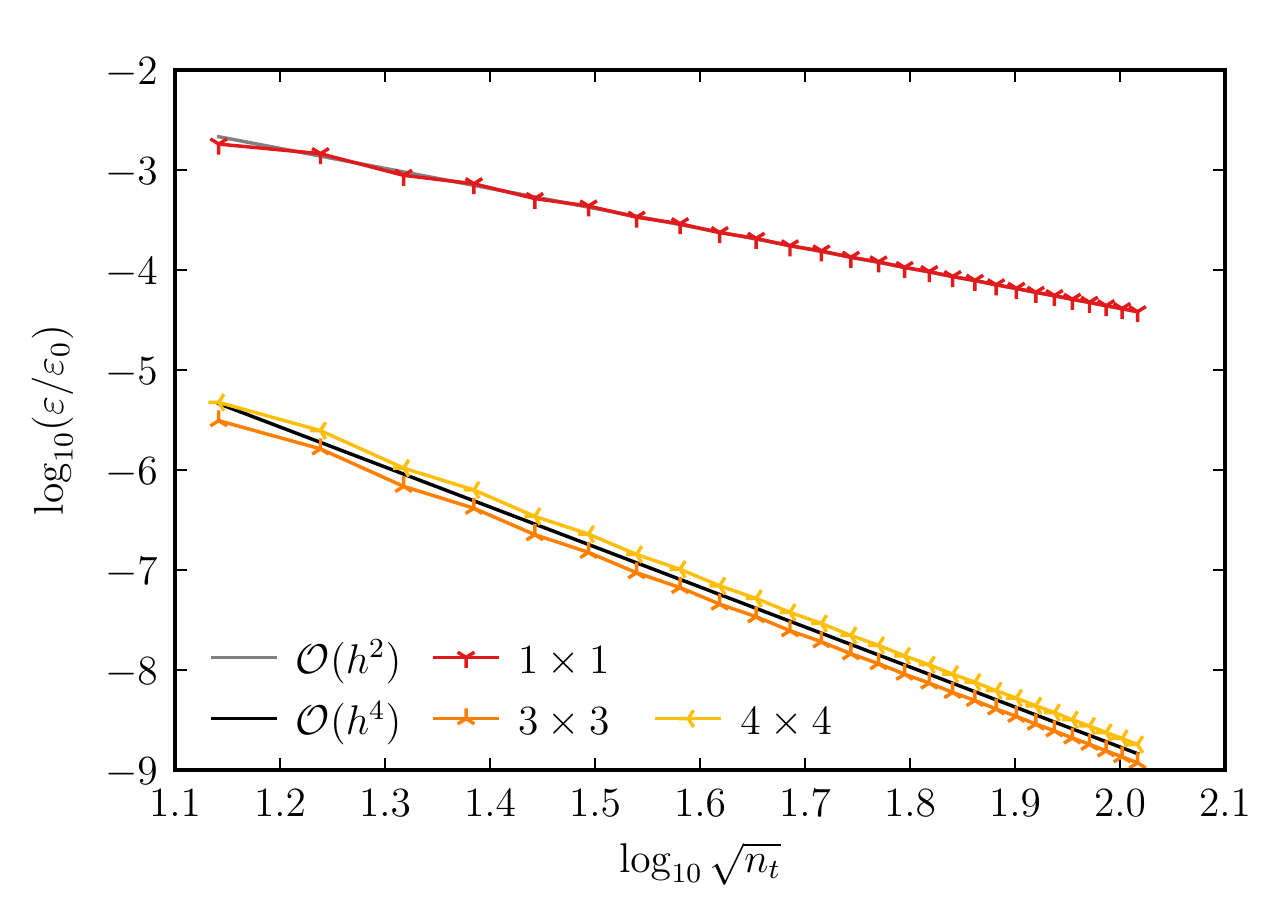}
\caption{Rhombic Prism, $d=2$\vpad}
\end{subfigure}
\\
\begin{subfigure}[b]{.49\textwidth}
\includegraphics[scale=.64,clip=true,trim=2.3in 0in 2.8in 0in]{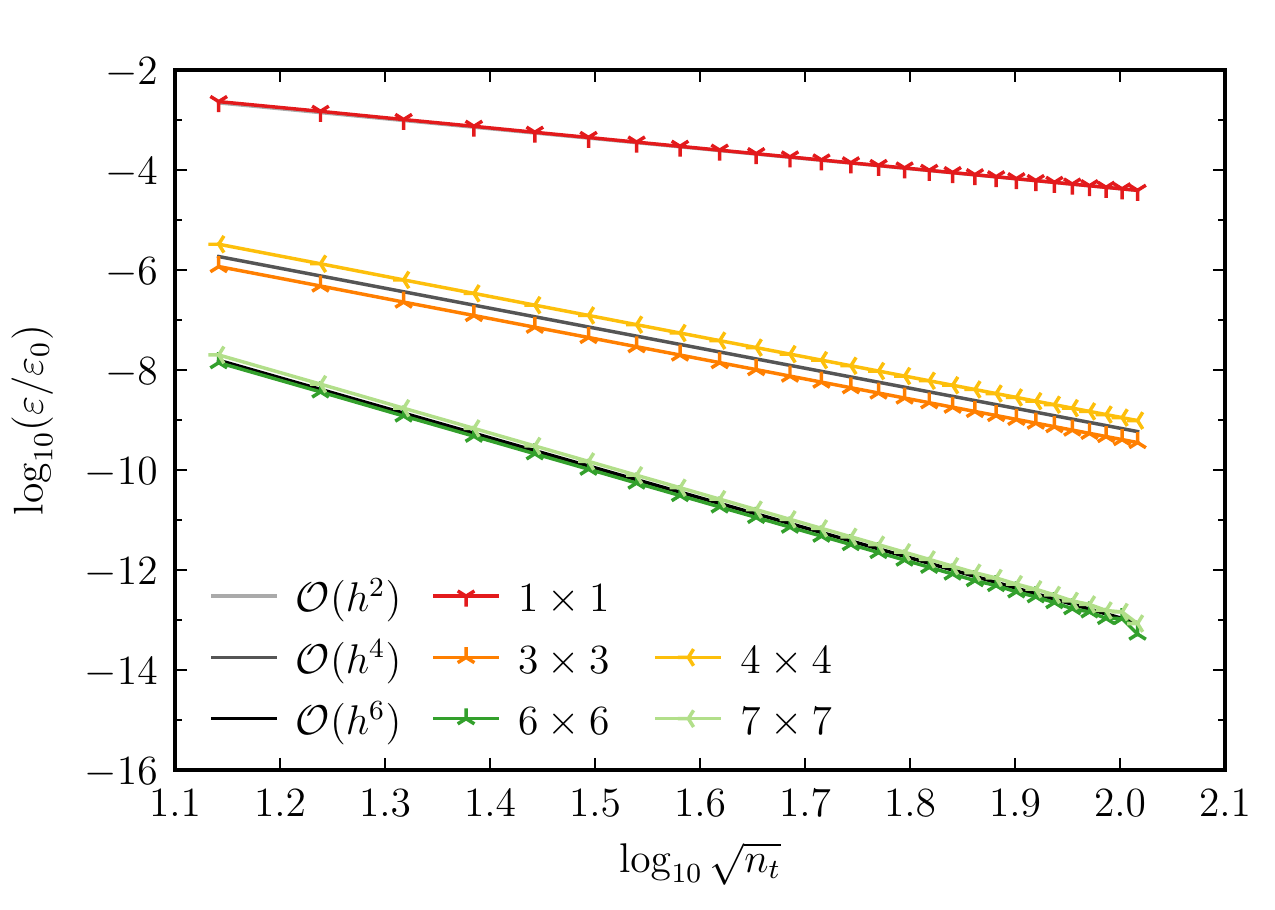}
\caption{Cube, $d=3$\vpad}
\end{subfigure}
\hspace{0.25em}
\begin{subfigure}[b]{.49\textwidth}
\includegraphics[scale=.64,clip=true,trim=2.3in 0in 2.8in 0in]{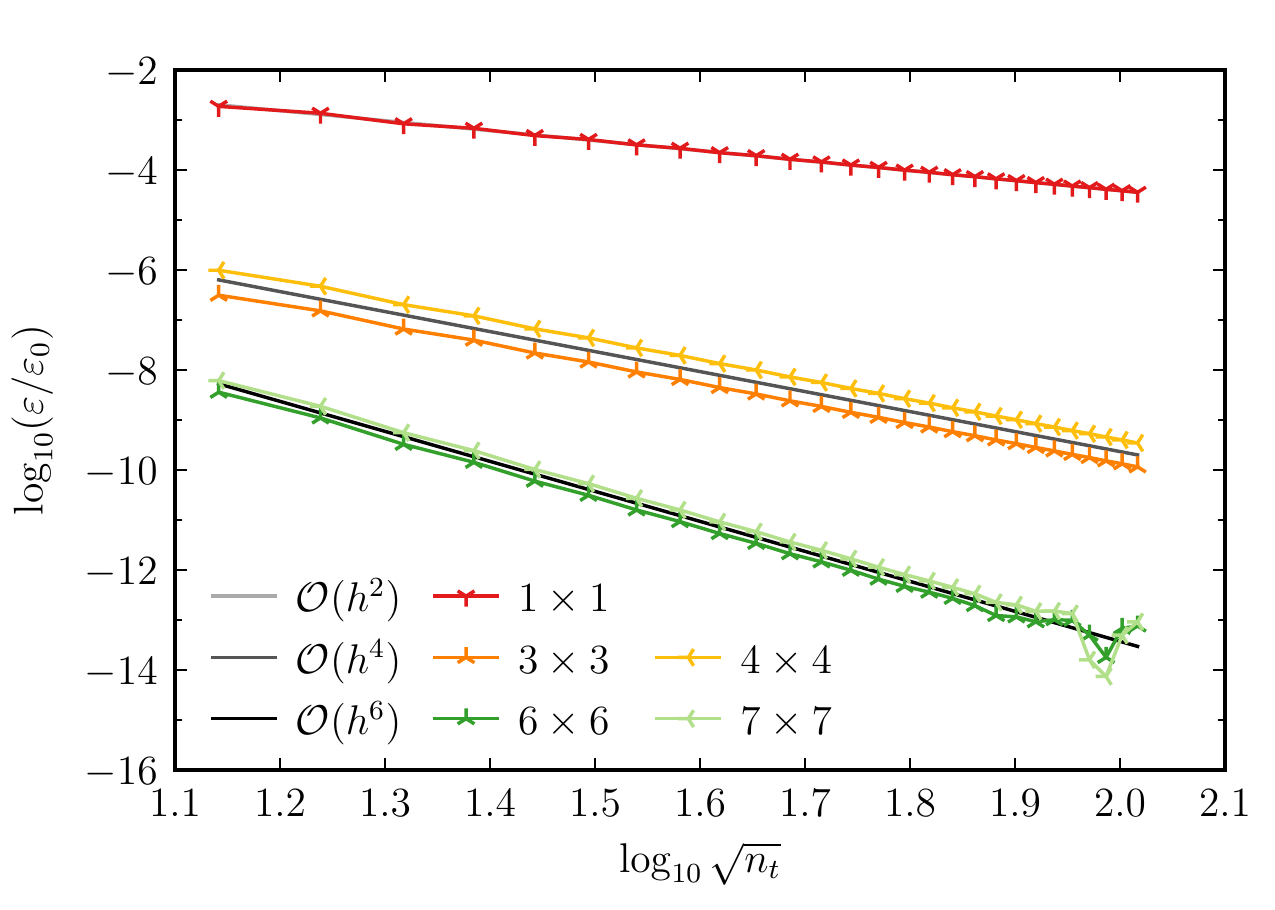}
\caption{Rhombic Prism, $d=3$\vpad}
\end{subfigure}
\caption{Numerical-integration error: $\varepsilon={\|\mathbf{e}_\mathbf{J}\|}_\infty$ from~\eqref{eq:opt_a} for different amounts of quadrature points.}
\vskip-\dp\strutbox
\label{fig:part7}
\end{figure}

\begin{figure}
\centering
\begin{subfigure}[b]{.49\textwidth}
\includegraphics[scale=.64,clip=true,trim=2.3in 0in 2.8in 0in]{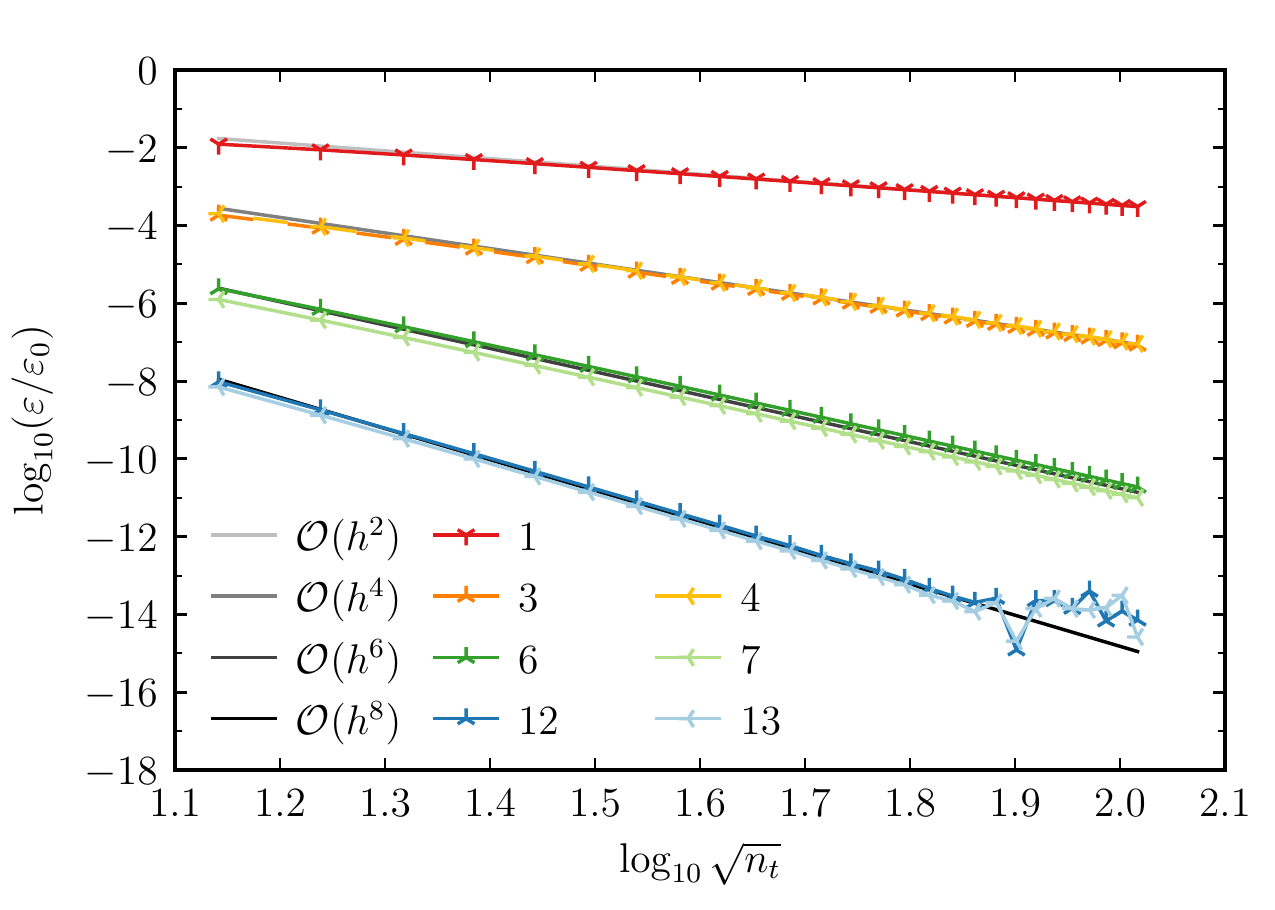}
\caption{Cube, $d=1$\vpad}
\end{subfigure}
\hspace{0.25em}
\begin{subfigure}[b]{.49\textwidth}
\includegraphics[scale=.64,clip=true,trim=2.3in 0in 2.8in 0in]{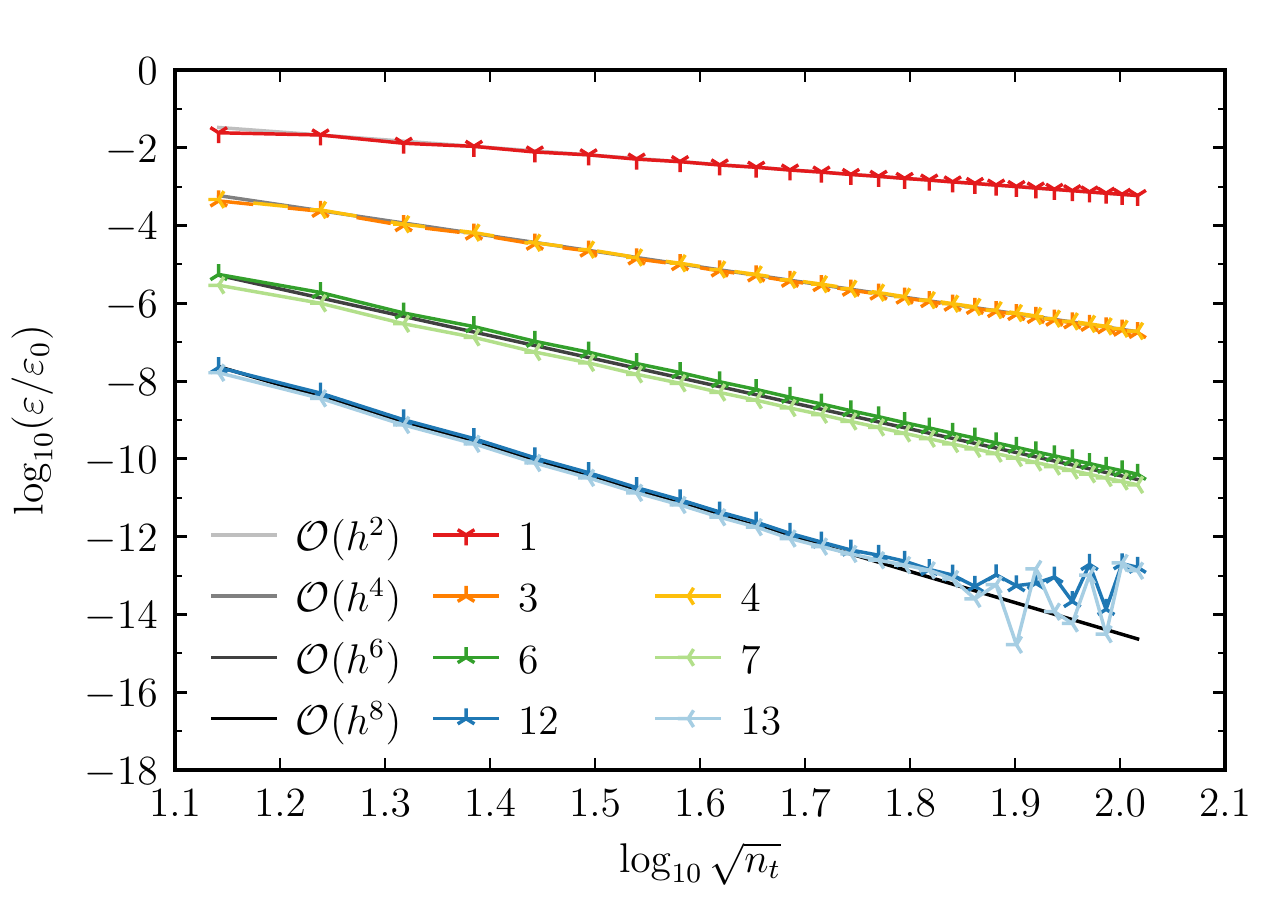}
\caption{Rhombic Prism, $d=1$\vpad}
\end{subfigure}
\\
\begin{subfigure}[b]{.49\textwidth}
\includegraphics[scale=.64,clip=true,trim=2.3in 0in 2.8in 0in]{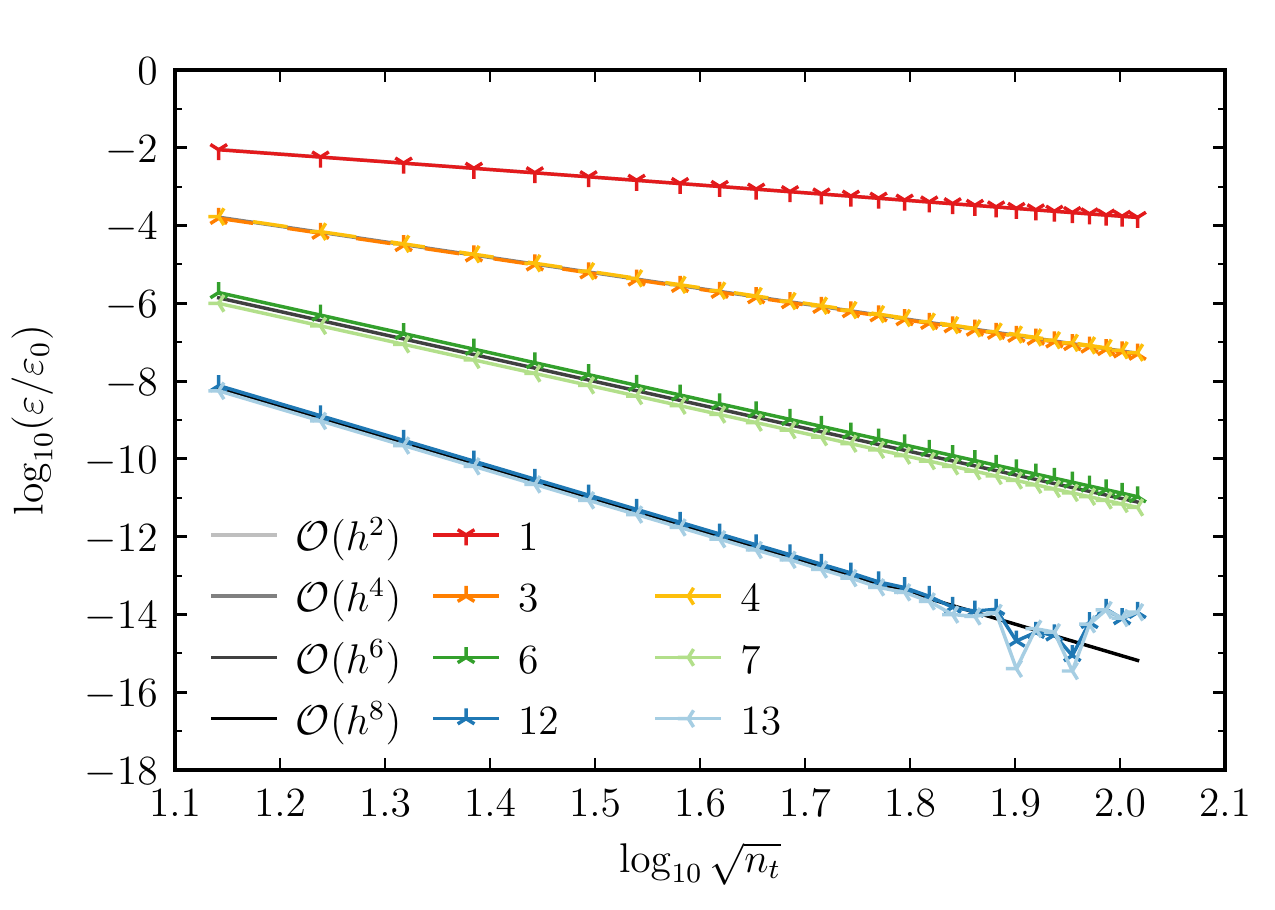}
\caption{Cube, $d=2$\vpad}
\end{subfigure}
\hspace{0.25em}
\begin{subfigure}[b]{.49\textwidth}
\includegraphics[scale=.64,clip=true,trim=2.3in 0in 2.8in 0in]{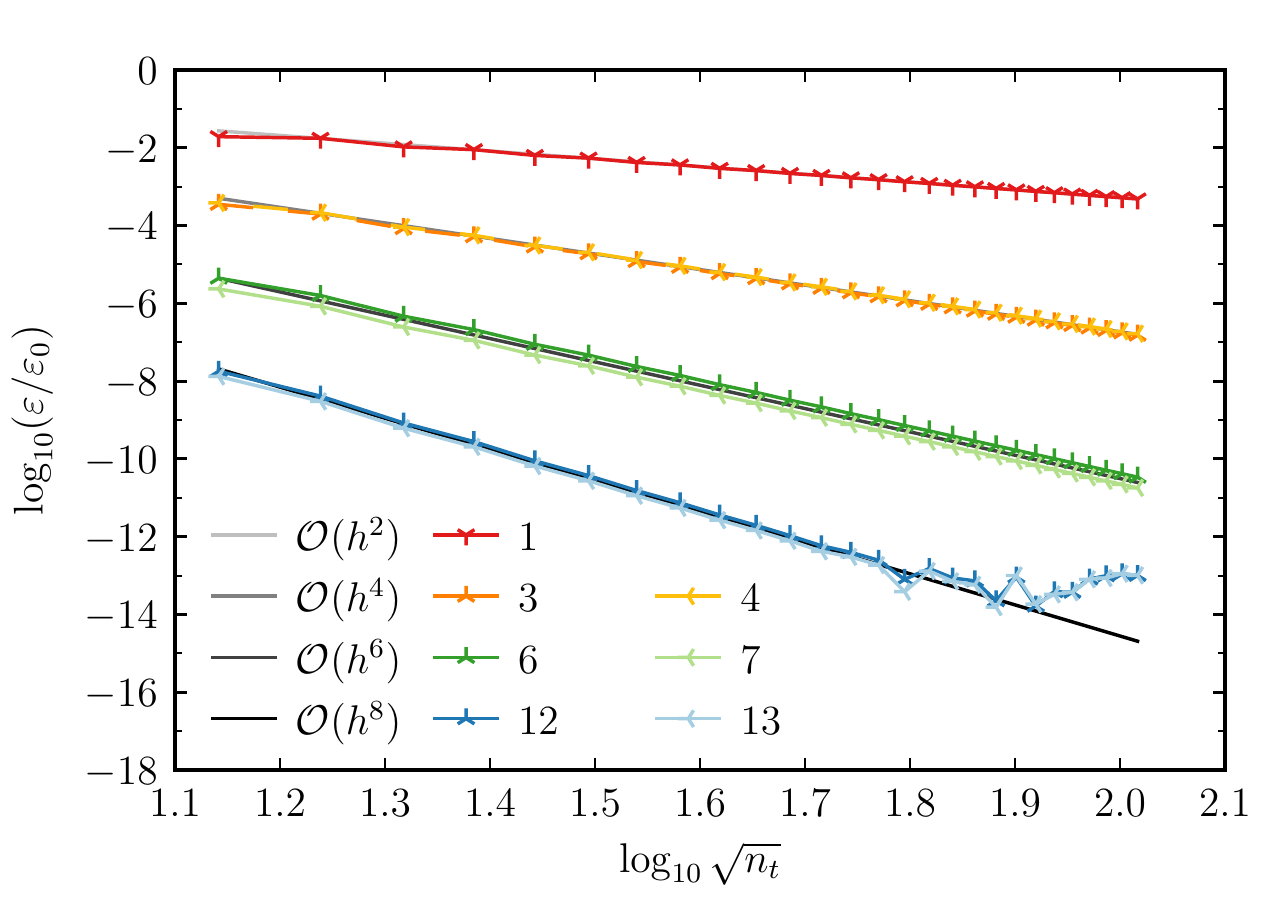}
\caption{Rhombic Prism, $d=2$\vpad}
\end{subfigure}
\\
\begin{subfigure}[b]{.49\textwidth}
\includegraphics[scale=.64,clip=true,trim=2.3in 0in 2.8in 0in]{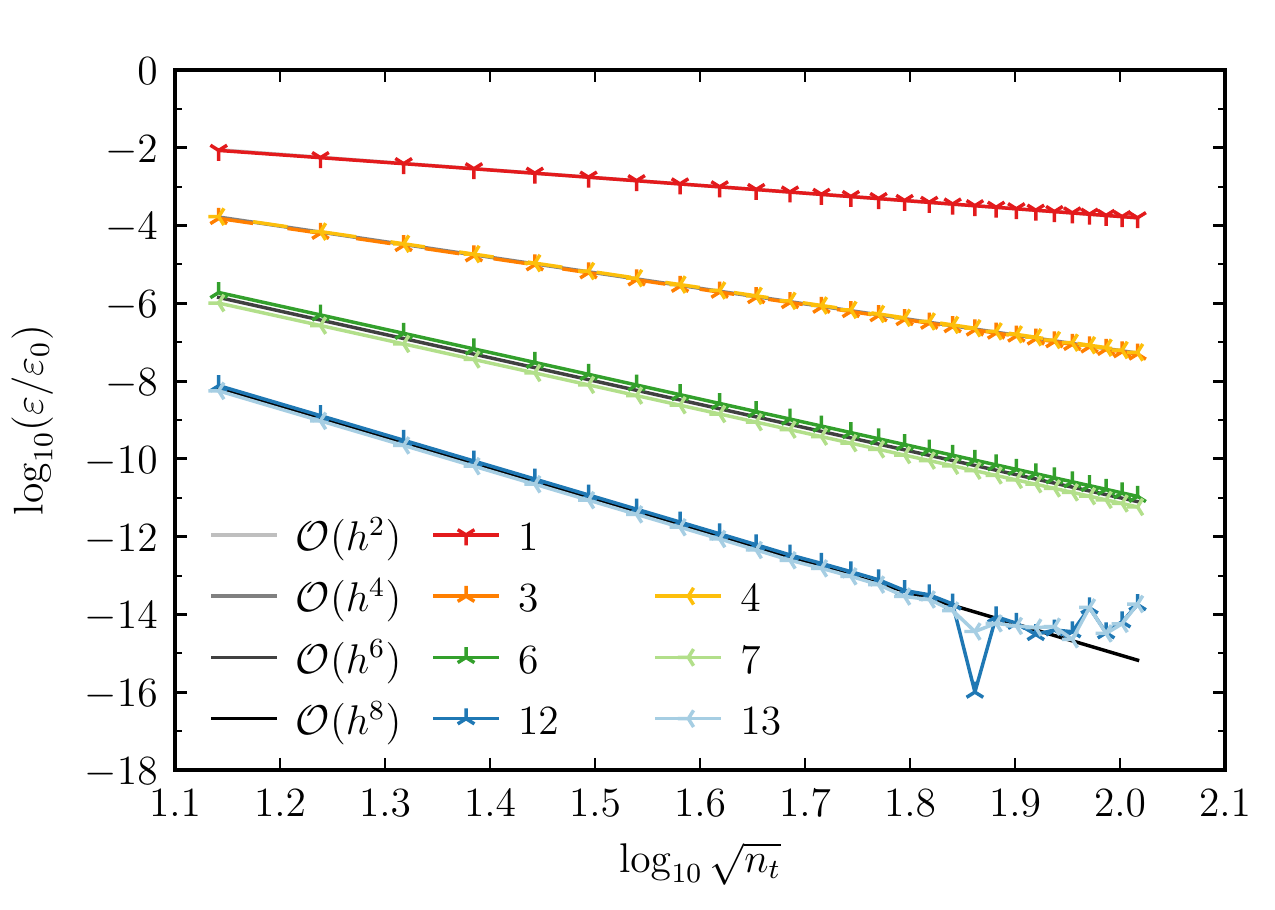}
\caption{Cube, $d=3$\vpad}
\end{subfigure}
\hspace{0.25em}
\begin{subfigure}[b]{.49\textwidth}
\includegraphics[scale=.64,clip=true,trim=2.3in 0in 2.8in 0in]{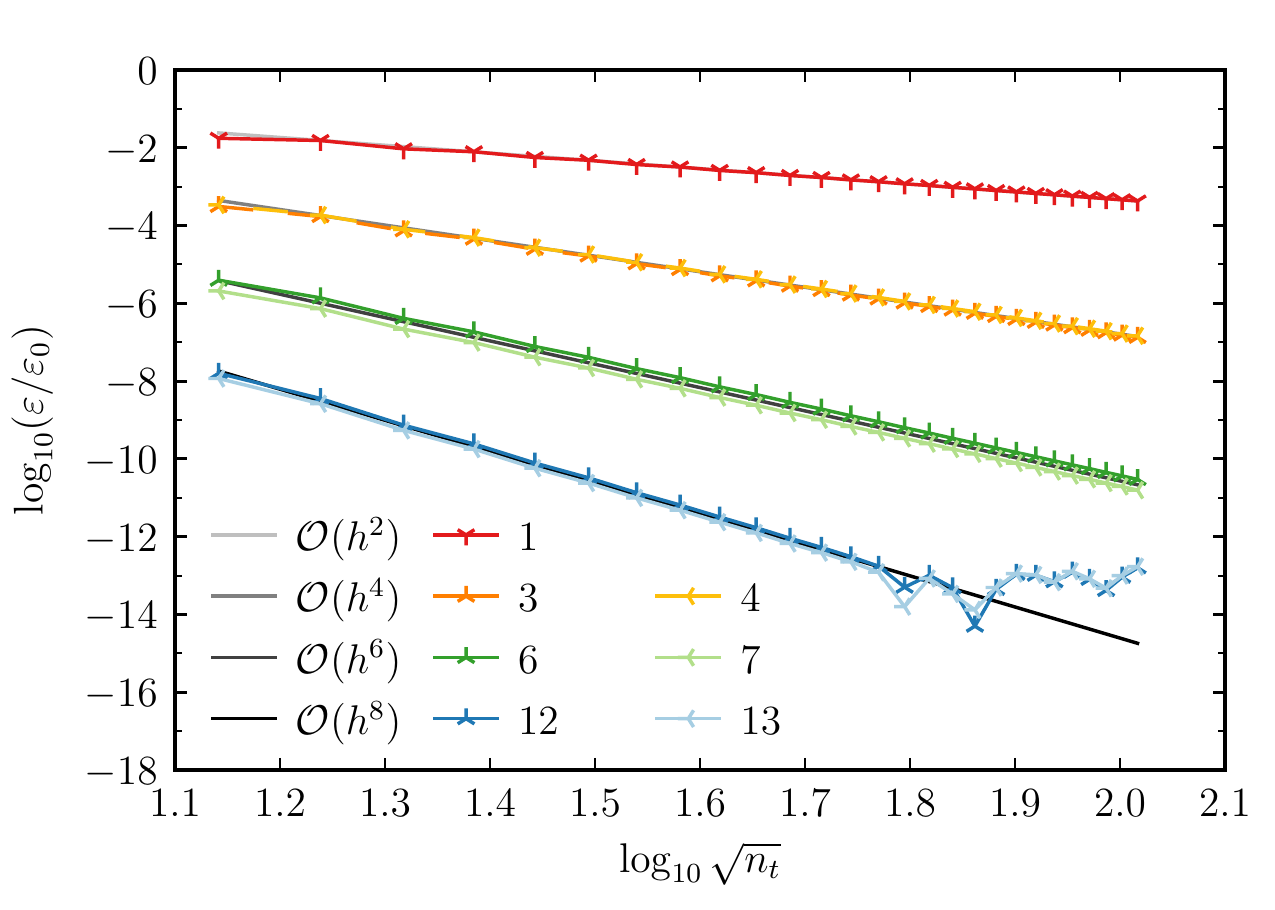}
\caption{Rhombic Prism, $d=3$\vpad}
\end{subfigure}
\caption{Numerical-integration error: $\varepsilon={\|\mathbf{e}_\mathbf{J}\|}_\infty$ from~\eqref{eq:opt_b} for different amounts of quadrature points.}
\vskip-\dp\strutbox
\label{fig:part8}
\end{figure}

\begin{figure}
\centering
\begin{subfigure}[b]{.49\textwidth}
\includegraphics[scale=.64,clip=true,trim=2.3in 0in 2.8in 0in]{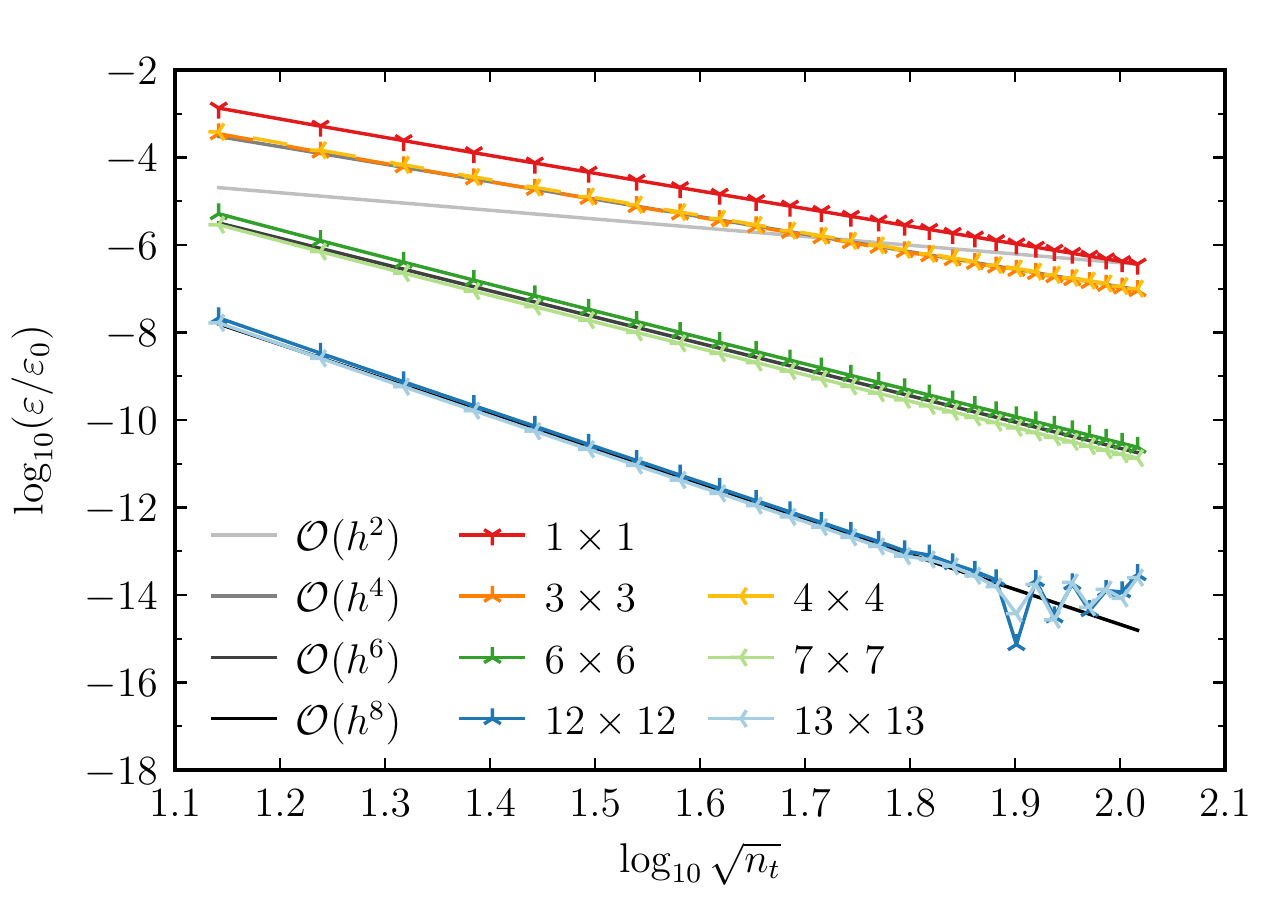}
\caption{Cube, $d=1$\vpad}
\label{fig:p3_c1_G1}
\end{subfigure}
\hspace{0.25em}
\begin{subfigure}[b]{.49\textwidth}
\includegraphics[scale=.64,clip=true,trim=2.3in 0in 2.8in 0in]{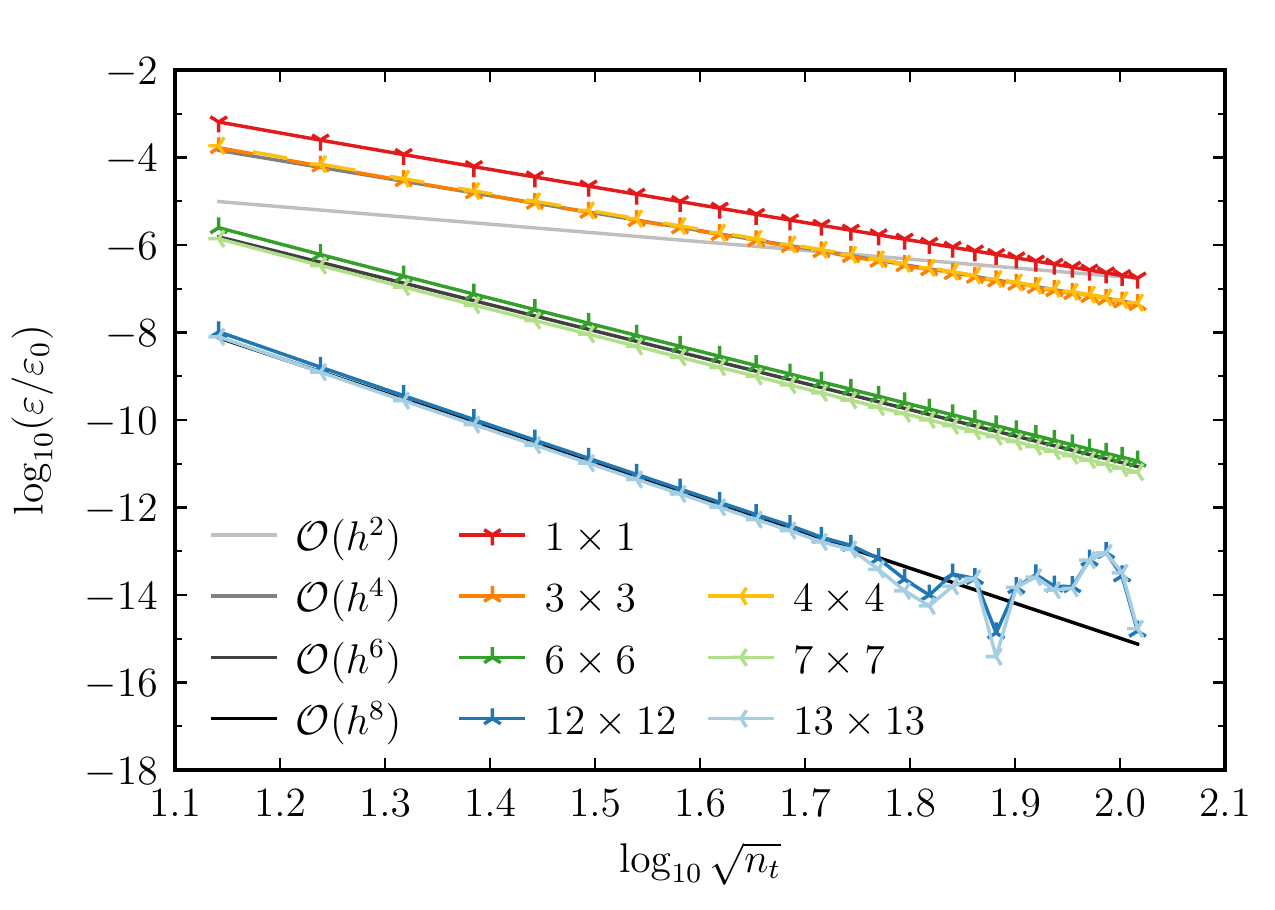}
\caption{Rhombic Prism, $d=1$\vpad}
\label{fig:p3_c2_G1}
\end{subfigure}
\\
\begin{subfigure}[b]{.49\textwidth}
\includegraphics[scale=.64,clip=true,trim=2.3in 0in 2.8in 0in]{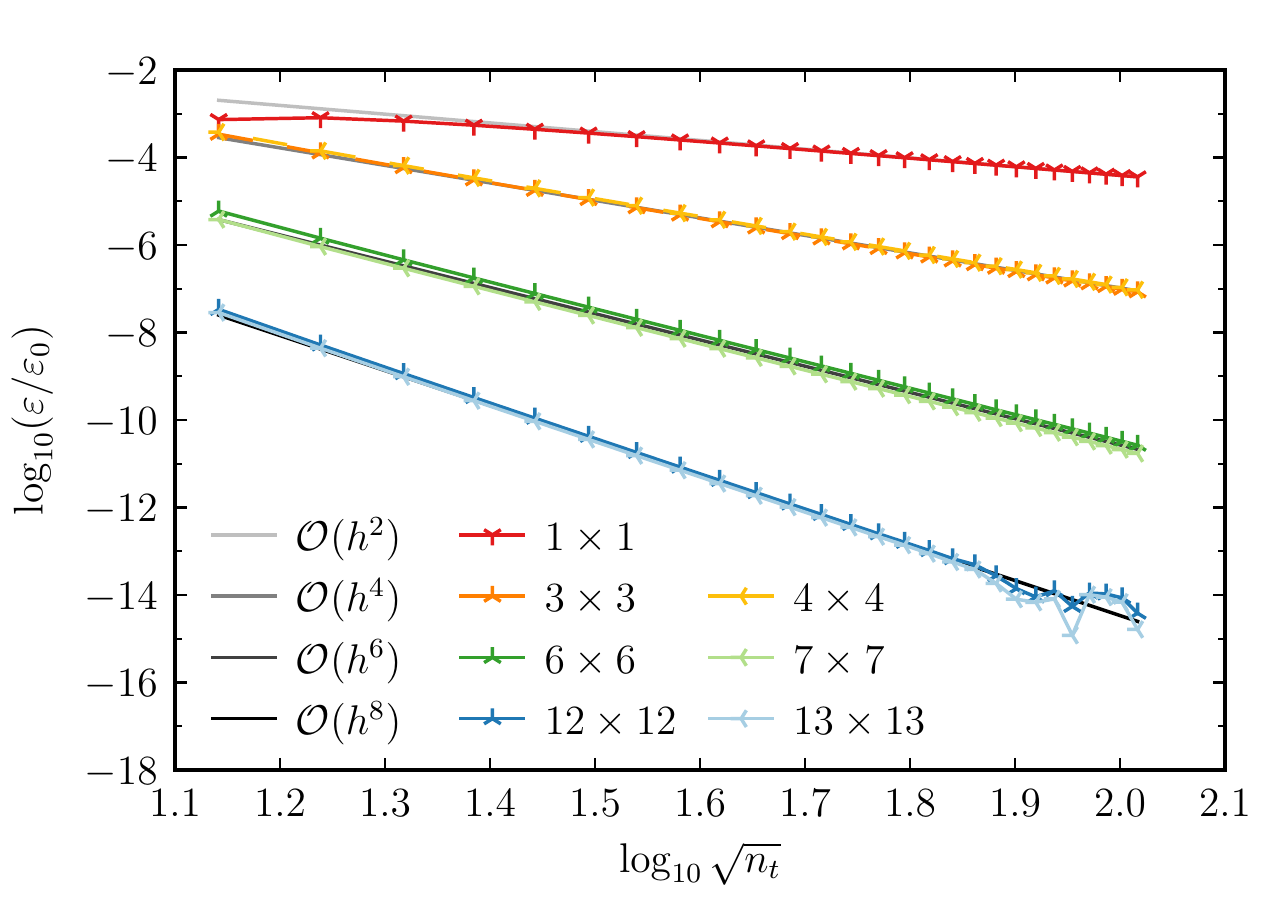}
\caption{Cube, $d=2$\vpad}
\end{subfigure}
\hspace{0.25em}
\begin{subfigure}[b]{.49\textwidth}
\includegraphics[scale=.64,clip=true,trim=2.3in 0in 2.8in 0in]{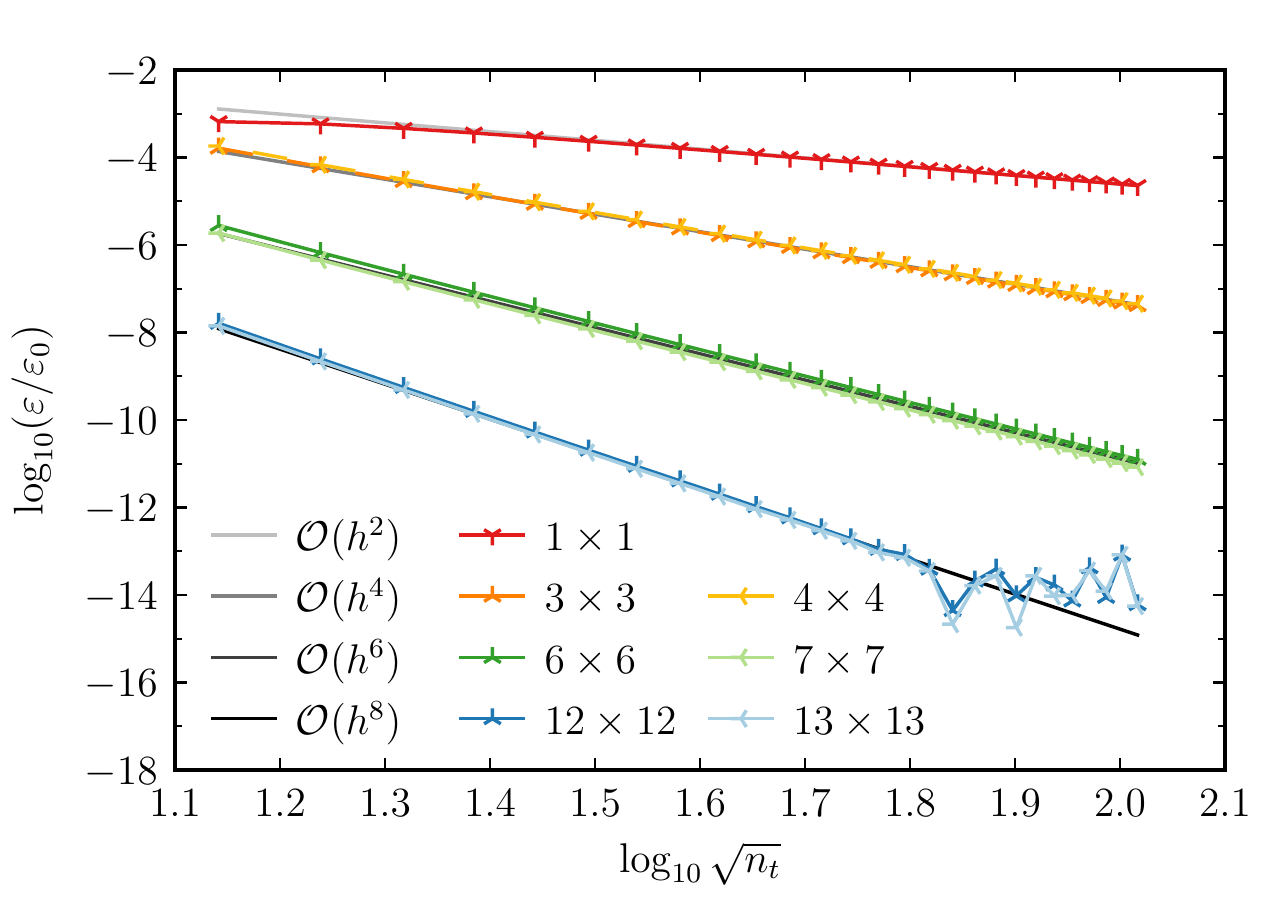}
\caption{Rhombic Prism, $d=2$\vpad}
\end{subfigure}
\\
\begin{subfigure}[b]{.49\textwidth}
\includegraphics[scale=.64,clip=true,trim=2.3in 0in 2.8in 0in]{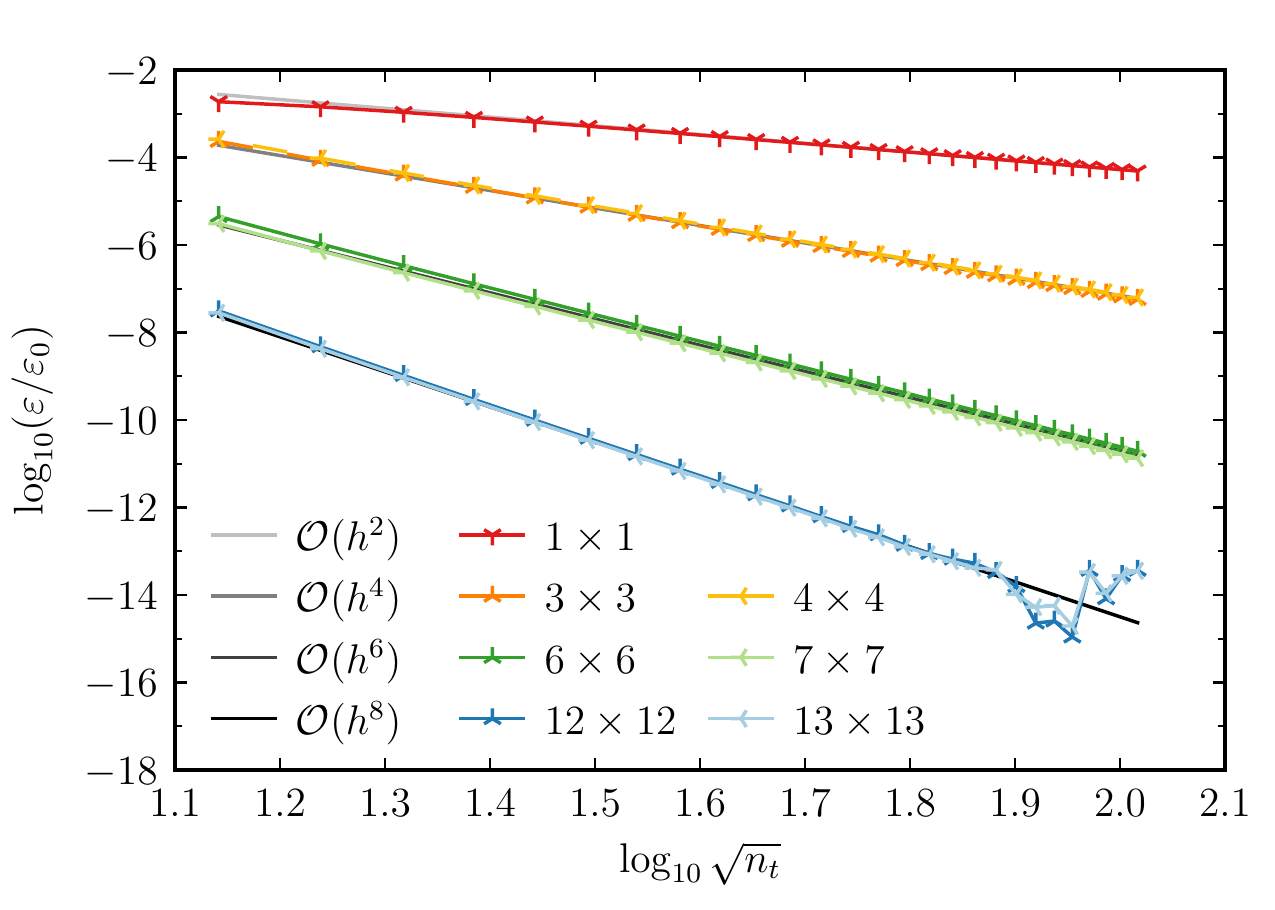}
\caption{Cube, $d=3$\vpad}
\end{subfigure}
\hspace{0.25em}
\begin{subfigure}[b]{.49\textwidth}
\includegraphics[scale=.64,clip=true,trim=2.3in 0in 2.8in 0in]{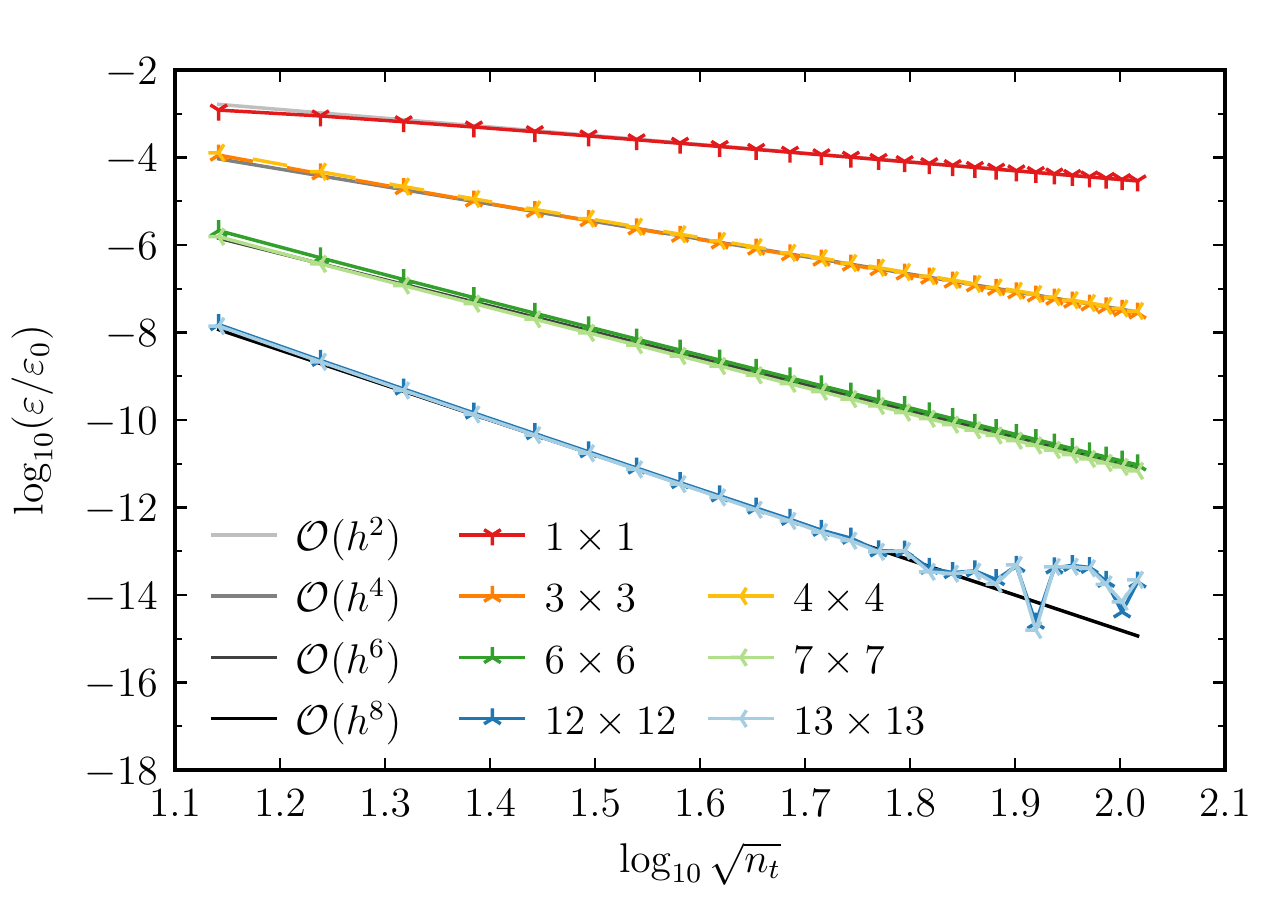}
\caption{Rhombic Prism, $d=3$\vpad}
\end{subfigure}
\caption{Numerical-integration error: $\varepsilon=|e_a(\mathbf{J}_\text{MS})|$~\eqref{eq:a_error_eliminate} for different amounts of quadrature points.}
\vskip-\dp\strutbox
\label{fig:part3}
\end{figure}

\begin{figure}
\centering
\begin{subfigure}[b]{.49\textwidth}
\includegraphics[scale=.64,clip=true,trim=2.3in 0in 2.8in 0in]{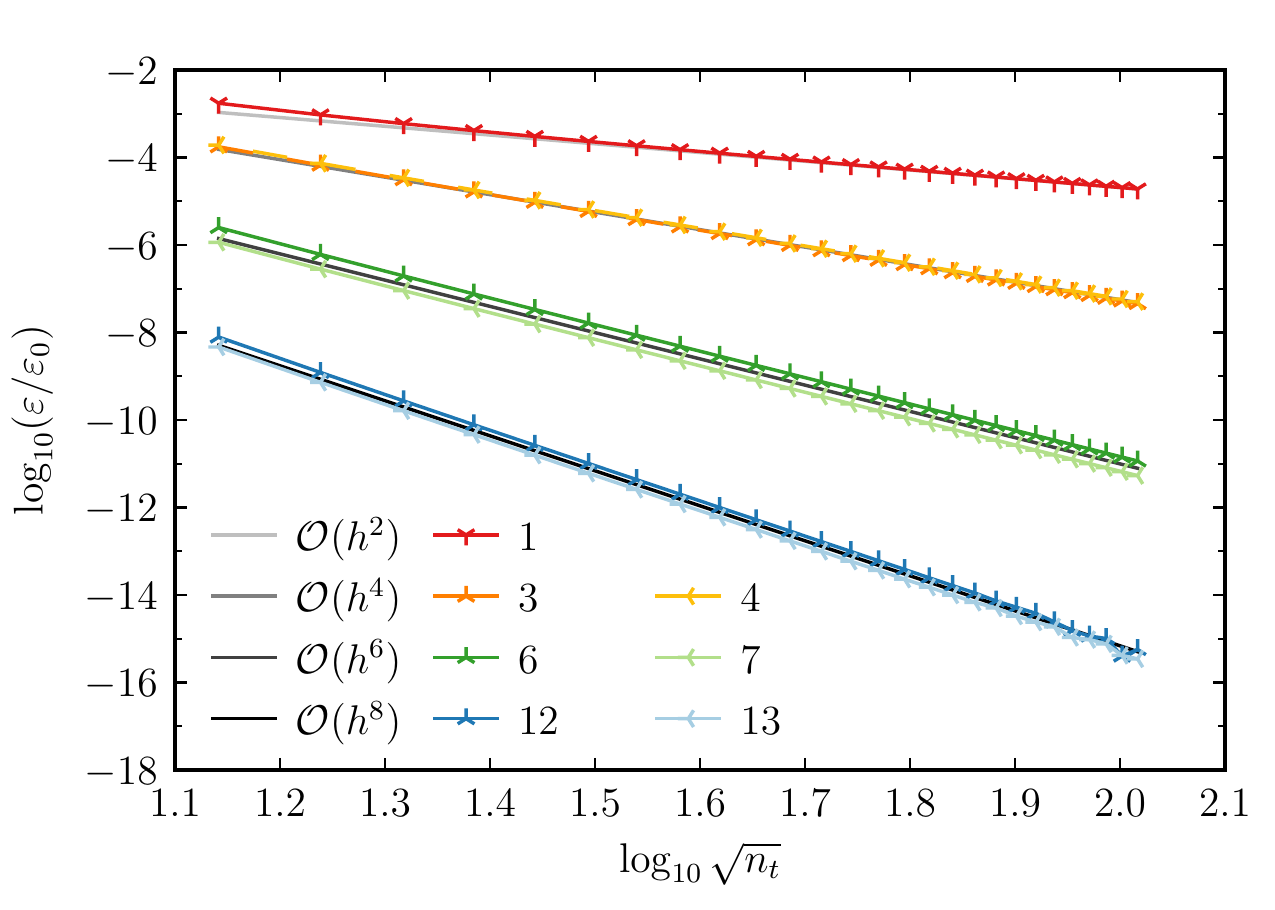}

\caption{Cube, $d=1$\vpad}
\end{subfigure}
\hspace{0.25em}
\begin{subfigure}[b]{.49\textwidth}
\includegraphics[scale=.64,clip=true,trim=2.3in 0in 2.8in 0in]{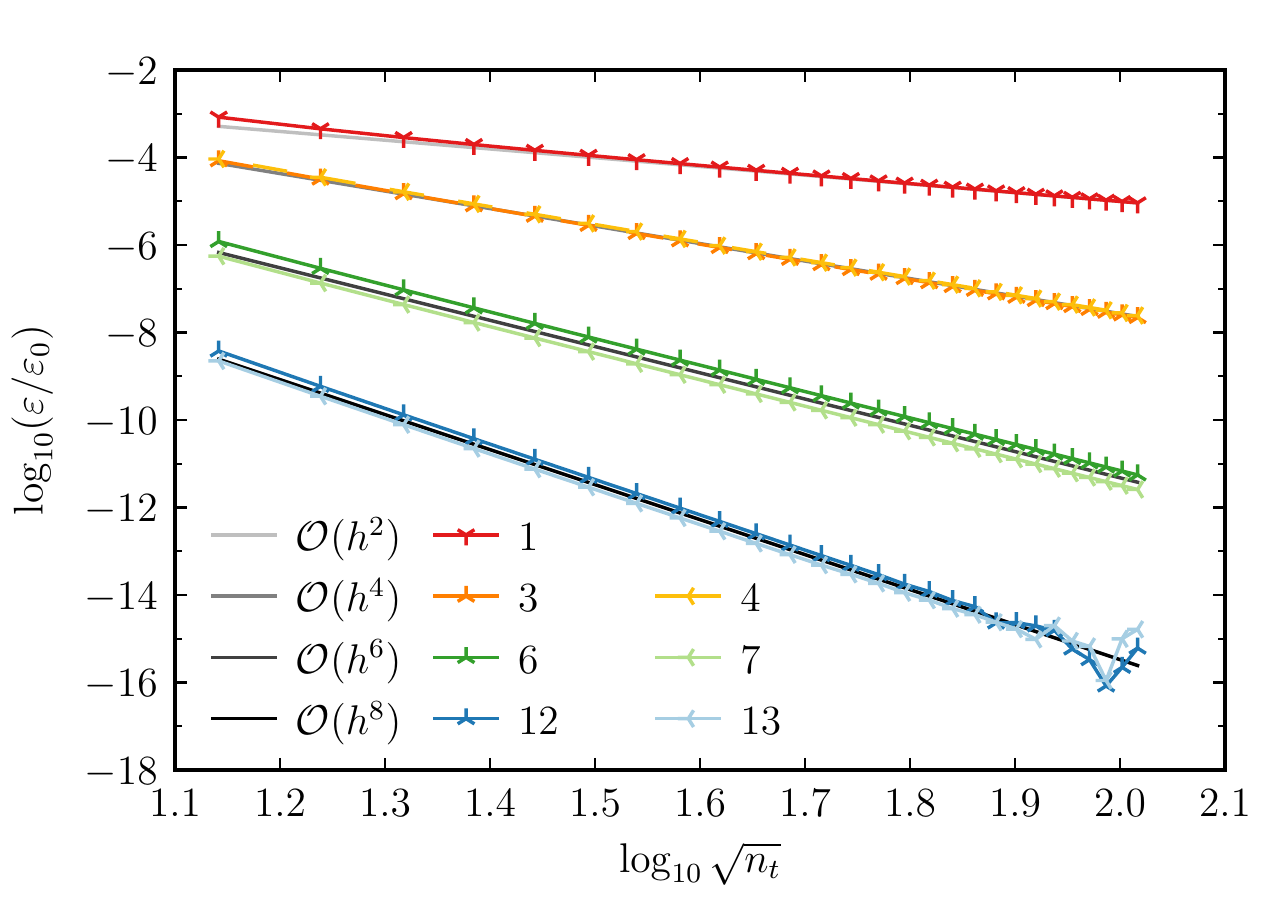}
\caption{Rhombic Prism, $d=1$\vpad}
\end{subfigure}
\\
\begin{subfigure}[b]{.49\textwidth}
\includegraphics[scale=.64,clip=true,trim=2.3in 0in 2.8in 0in]{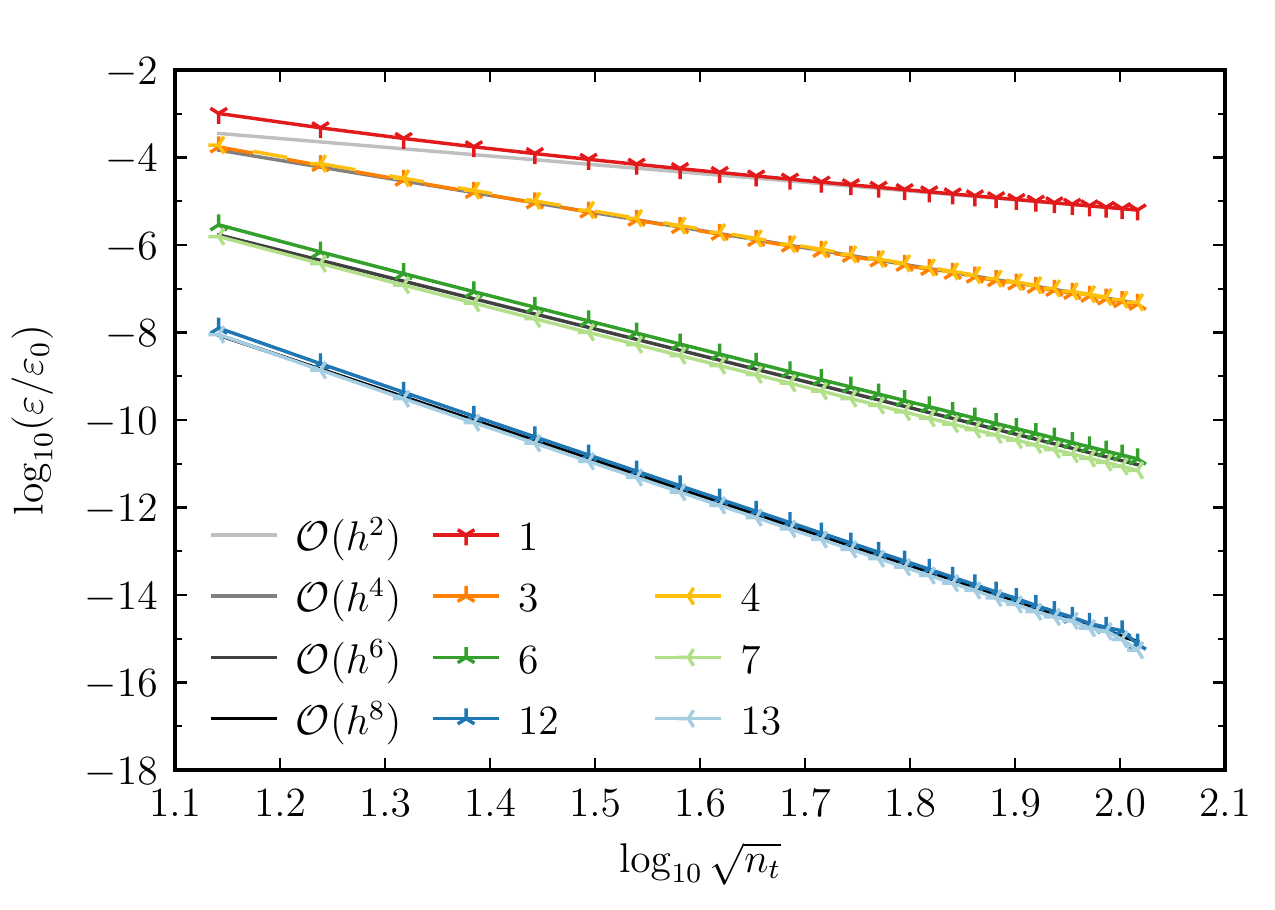}
\caption{Cube, $d=2$\vpad}
\end{subfigure}
\hspace{0.25em}
\begin{subfigure}[b]{.49\textwidth}
\includegraphics[scale=.64,clip=true,trim=2.3in 0in 2.8in 0in]{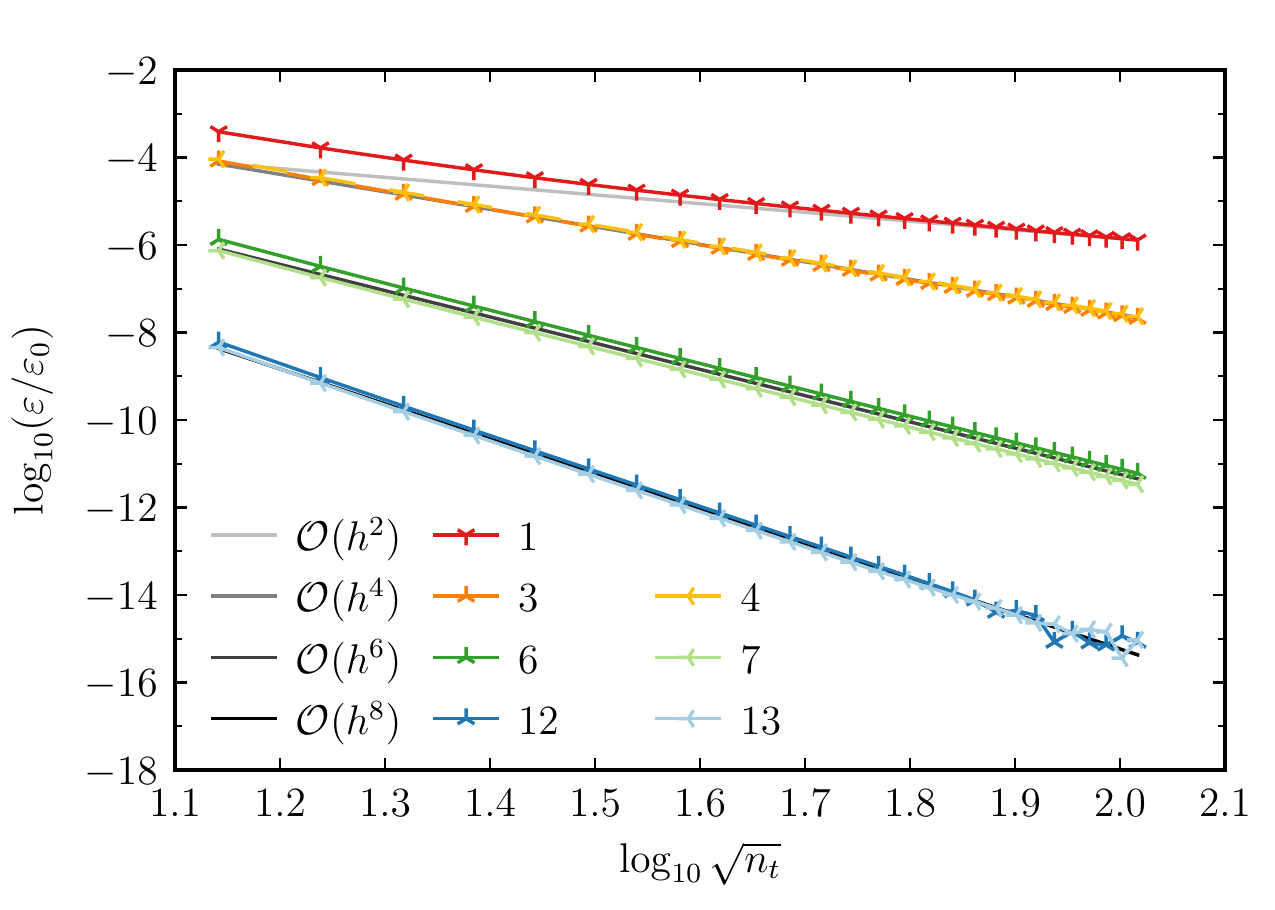}
\caption{Rhombic Prism, $d=2$\vpad}
\end{subfigure}
\\
\begin{subfigure}[b]{.49\textwidth}
\includegraphics[scale=.64,clip=true,trim=2.3in 0in 2.8in 0in]{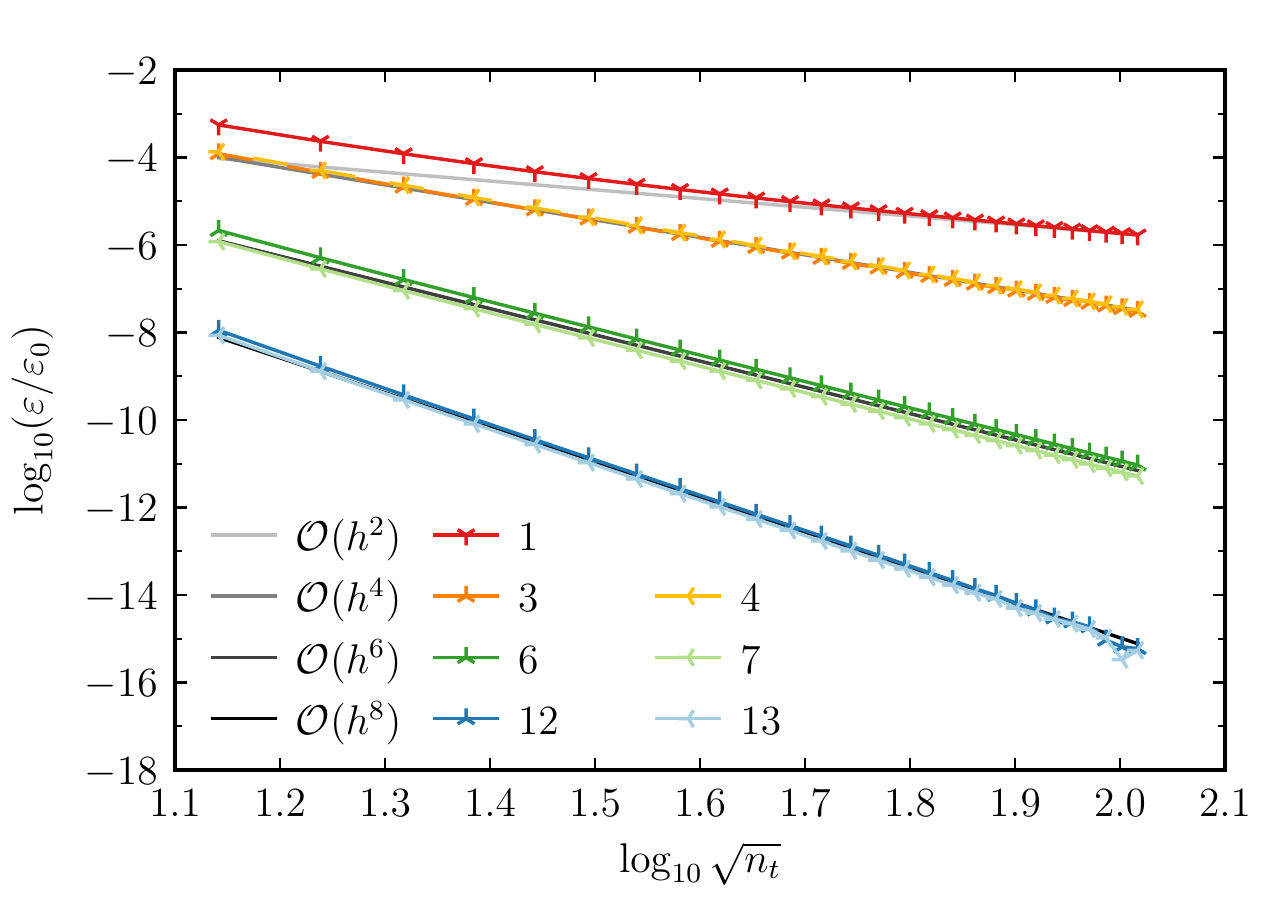}
\caption{Cube, $d=3$\vpad}
\end{subfigure}
\hspace{0.25em}
\begin{subfigure}[b]{.49\textwidth}
\includegraphics[scale=.64,clip=true,trim=2.3in 0in 2.8in 0in]{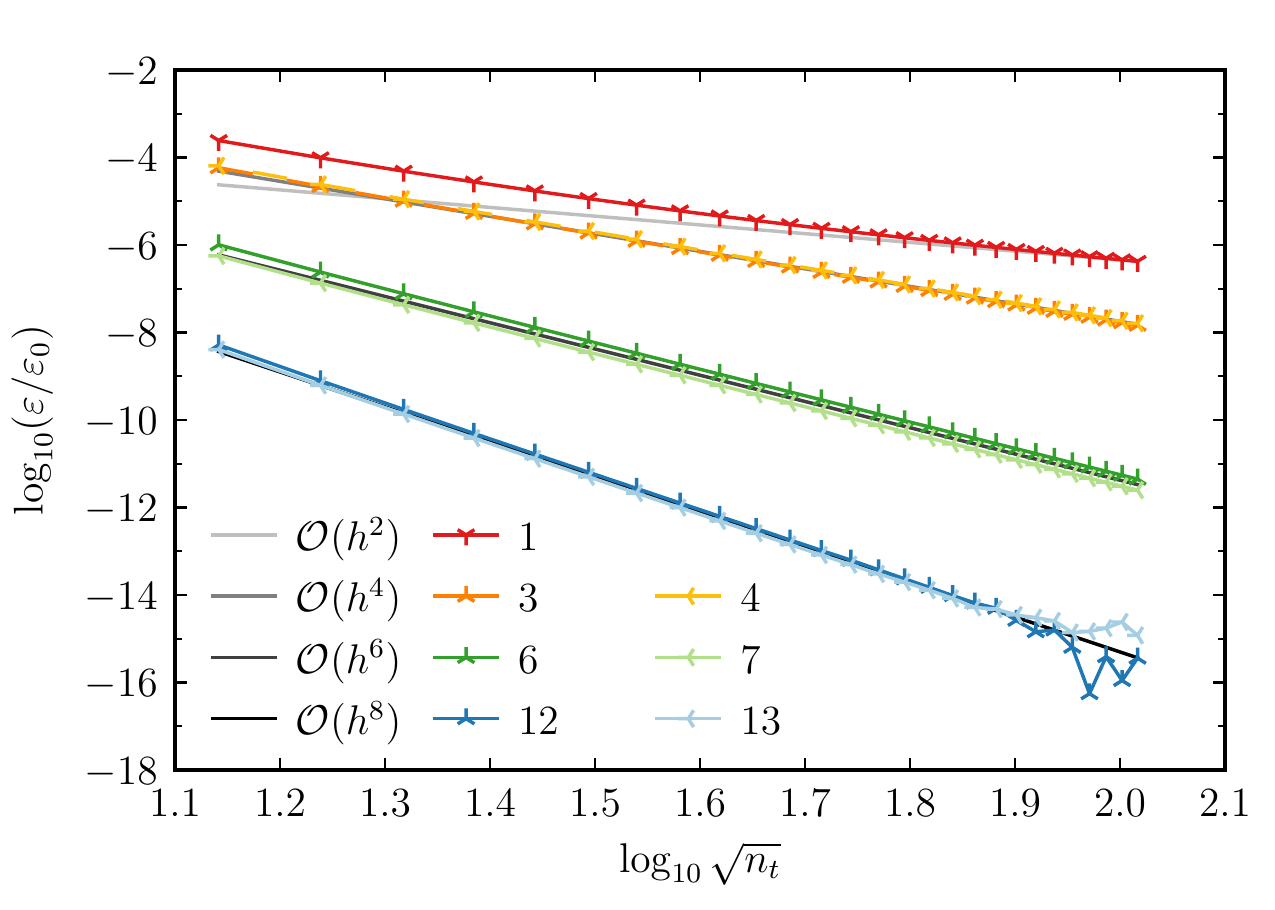}
\caption{Rhombic Prism, $d=3$\vpad}
\end{subfigure}
\caption{Numerical-integration error: $\varepsilon=|e_b(\mathbf{J}_\text{MS})|$~\eqref{eq:b_error_eliminate} for different amounts of quadrature points.}
\vskip-\dp\strutbox
\label{fig:part4}
\end{figure}

\begin{figure}
\centering
\begin{subfigure}[b]{.49\textwidth}
\includegraphics[scale=.64,clip=true,trim=2.3in 0in 2.8in 0in]{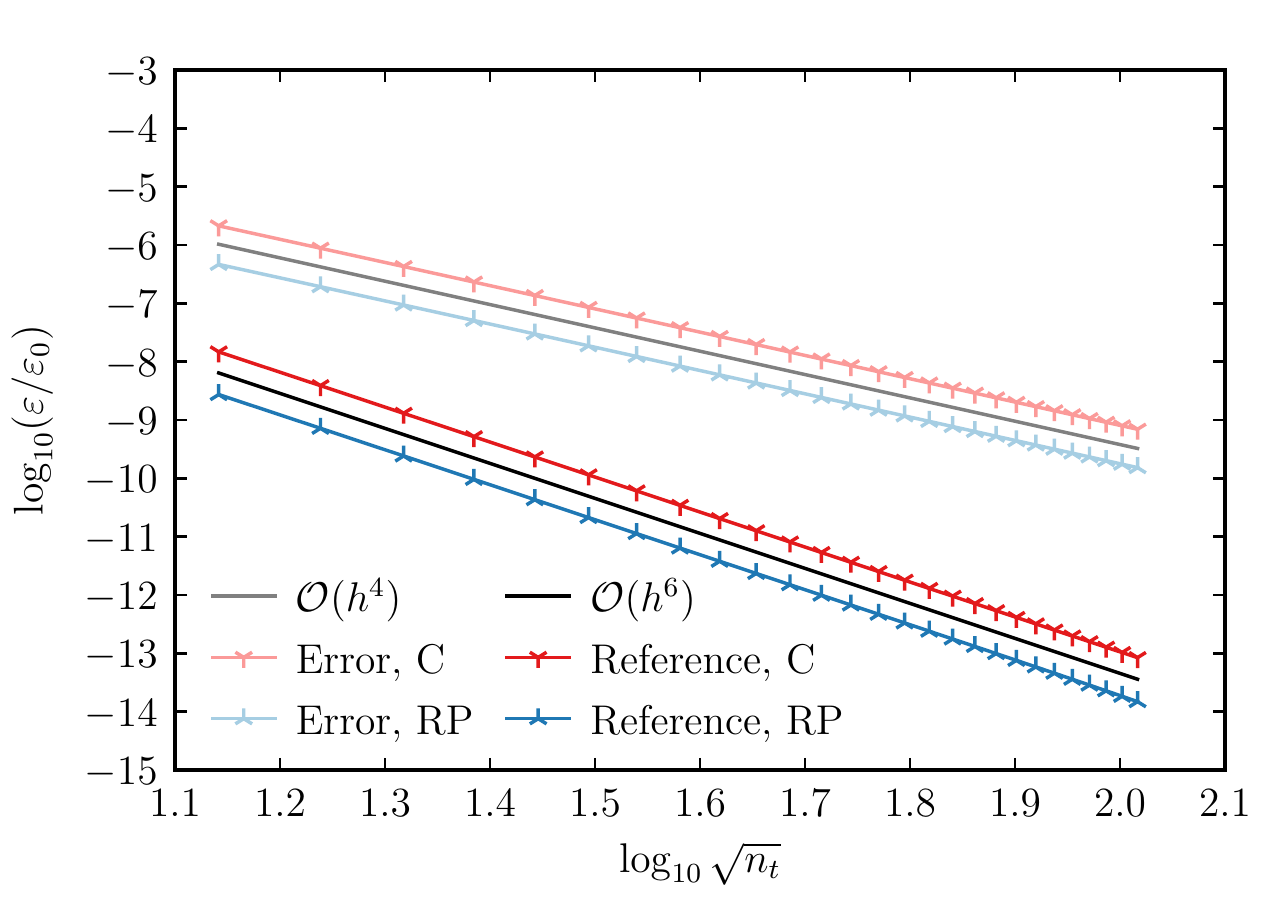}
\caption{$\varepsilon=\big|e_a\big(\mathbf{J}_{h_\text{MS}}\big)\big|$\vpad}
\label{fig:part5_bug}
\end{subfigure}
\hspace{0.25em}
\begin{subfigure}[b]{.49\textwidth}
\includegraphics[scale=.64,clip=true,trim=2.3in 0in 2.8in 0in]{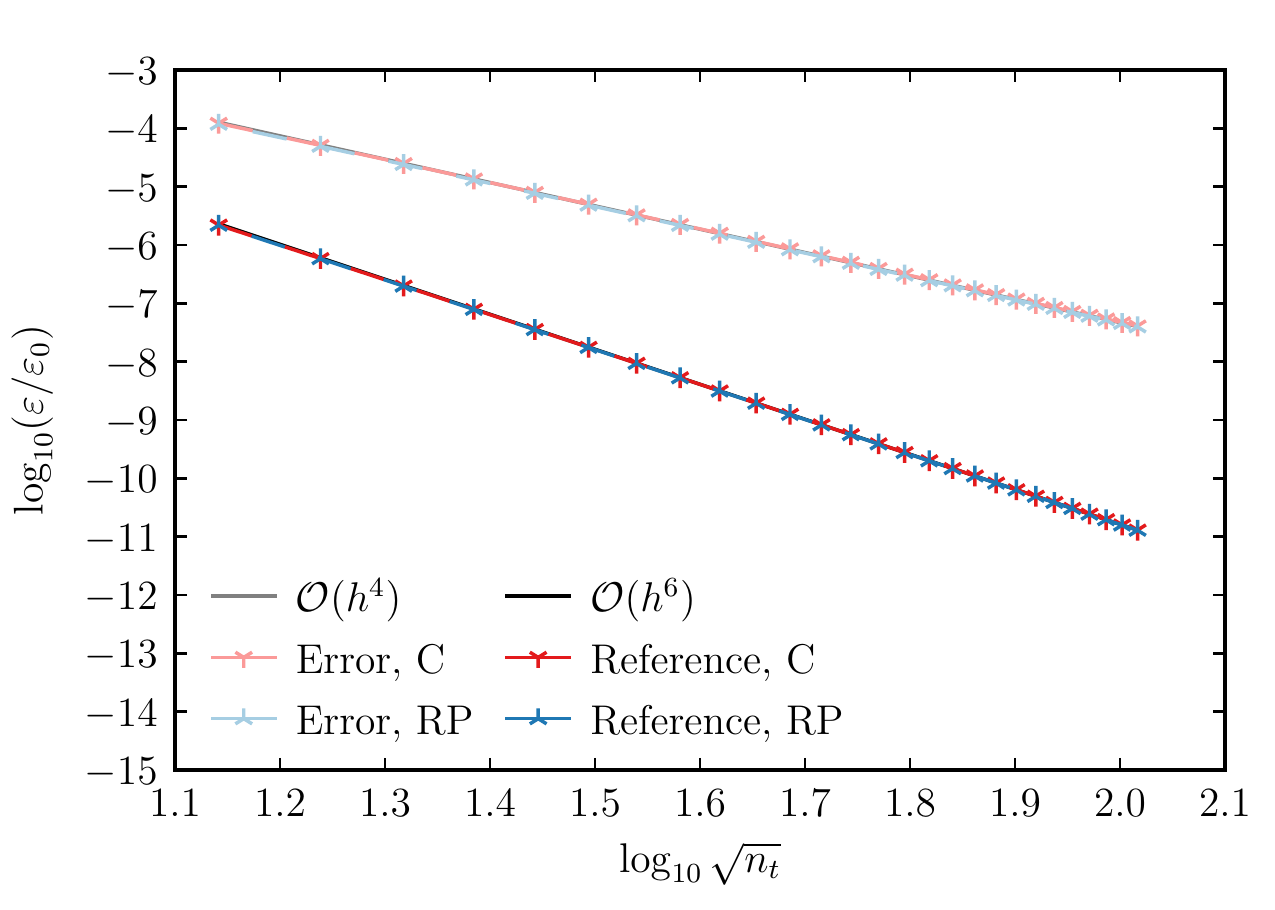}
\caption{$\varepsilon=\big|e_b\big(\mathbf{J}_{h_\text{MS}}\big)\big|$\vpad}
\label{fig:part6_bug}
\end{subfigure}
\\
\begin{subfigure}[b]{.49\textwidth}
\includegraphics[scale=.64,clip=true,trim=2.3in 0in 2.8in 0in]{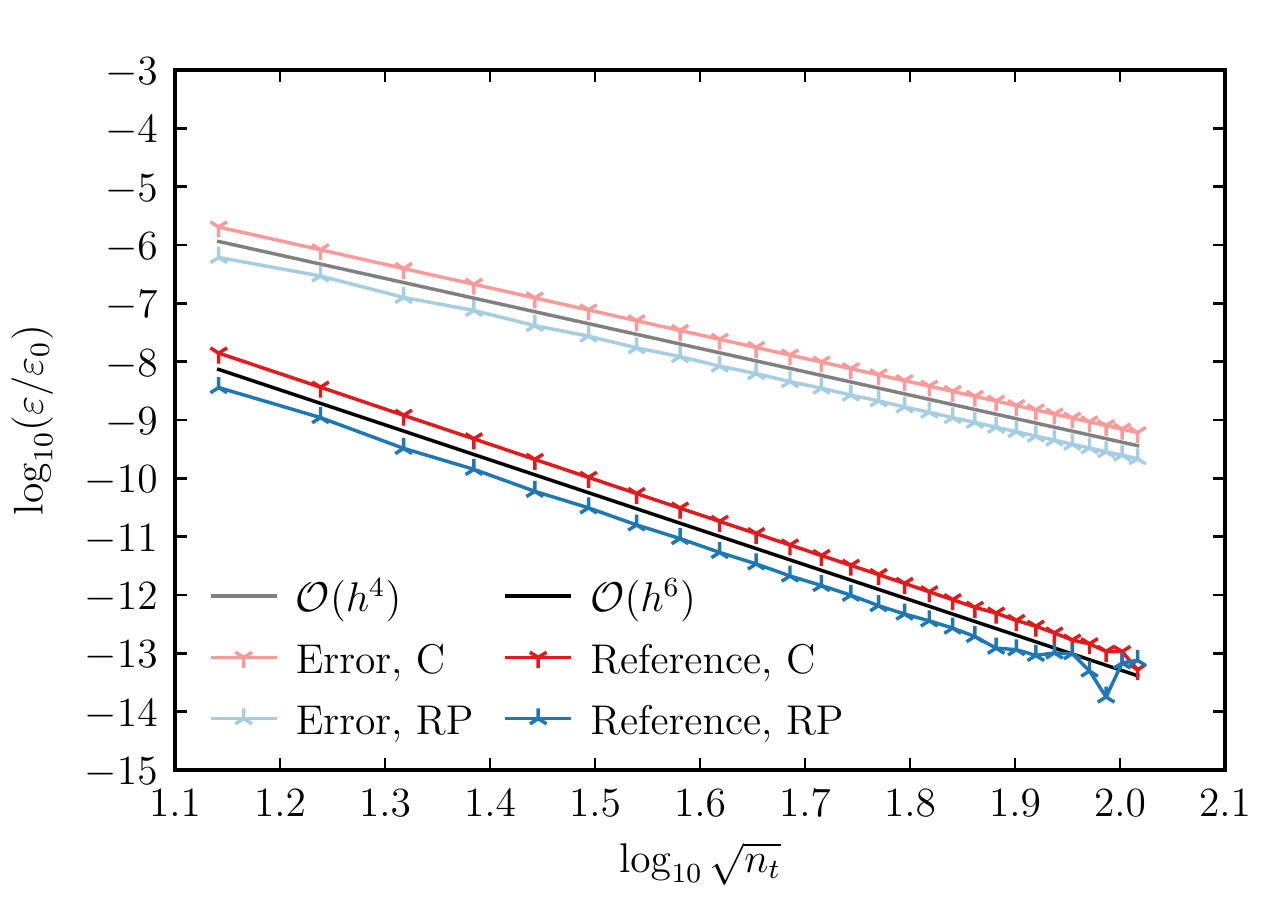}
\caption{$\varepsilon={\|\mathbf{e}_\mathbf{J}\|}_\infty$, $a^q\big(\mathbf{J}_h,\mathbf{J}_{h_\text{MS}}\big) = a\big(\mathbf{J}_{h_\text{MS}},\mathbf{J}_{h_\text{MS}}\big)$\vpad}
\label{fig:part7_bug}
\end{subfigure}
\hspace{0.25em}
\begin{subfigure}[b]{.49\textwidth}
\includegraphics[scale=.64,clip=true,trim=2.3in 0in 2.8in 0in]{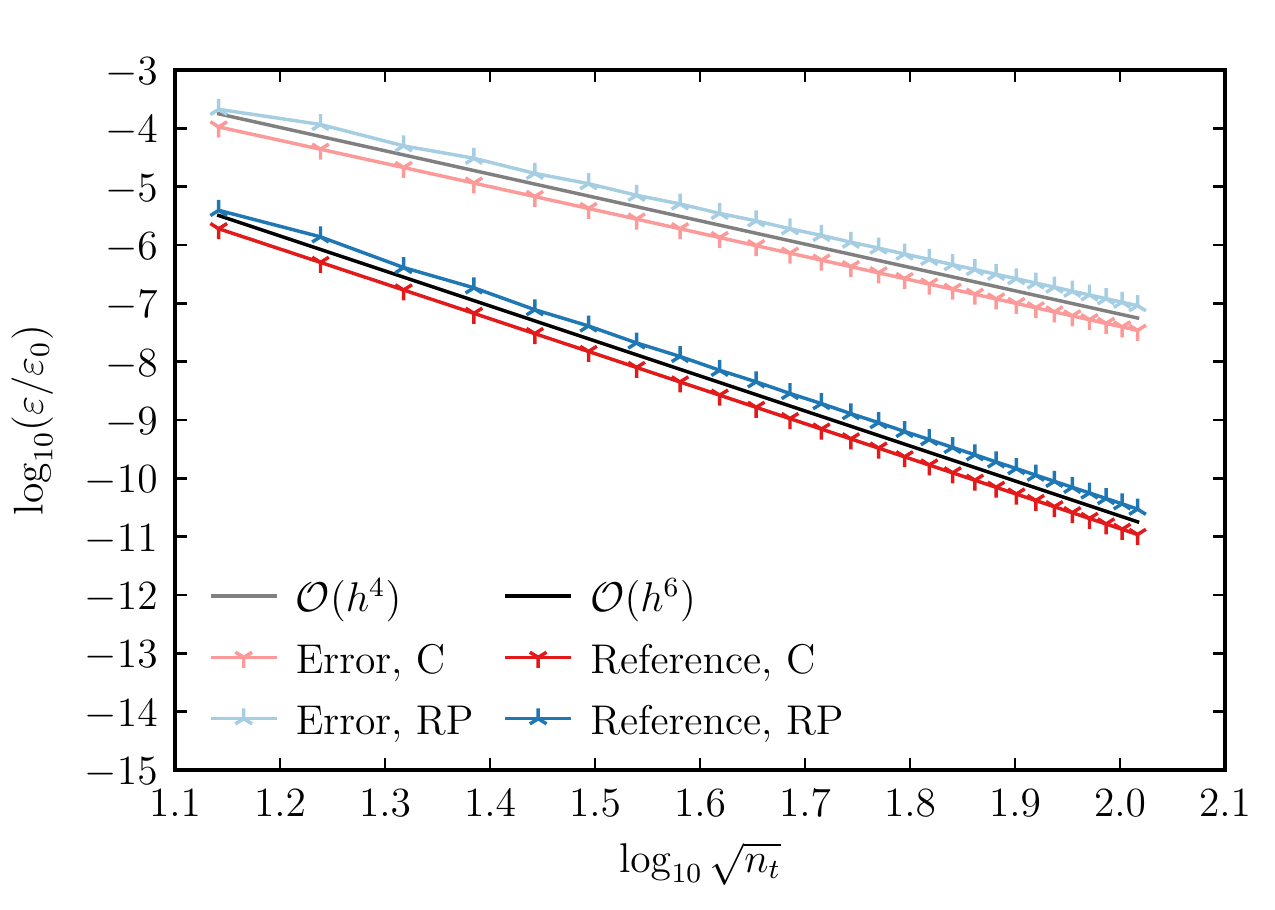}
\caption{$\varepsilon={\|\mathbf{e}_\mathbf{J}\|}_\infty$, $b^q\big(\mathbf{H}^\mathcal{I}_\text{MS},\mathbf{J}_h\big) = b\big(\mathbf{H}^\mathcal{I}_\text{MS},\mathbf{J}_{h_\text{MS}}\big)$\vpad}
\label{fig:part8_bug}
\end{subfigure}
\\
\begin{subfigure}[b]{.49\textwidth}
\includegraphics[scale=.64,clip=true,trim=2.3in 0in 2.8in 0in]{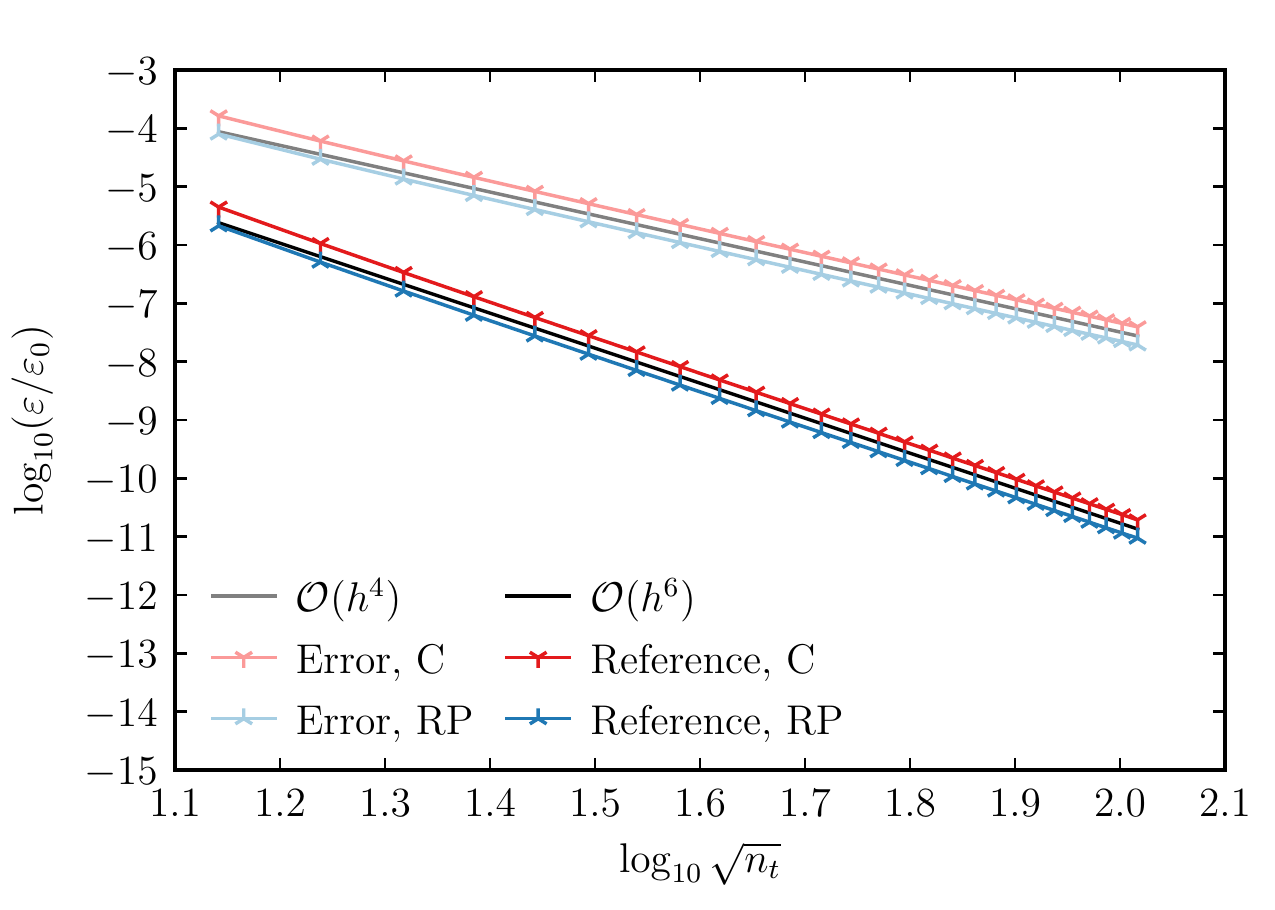}
\caption{$\varepsilon=|e_a(\mathbf{J}_\text{MS})|$\vpad}
\label{fig:part3_bug}
\end{subfigure}
\hspace{0.25em}
\begin{subfigure}[b]{.49\textwidth}
\includegraphics[scale=.64,clip=true,trim=2.3in 0in 2.8in 0in]{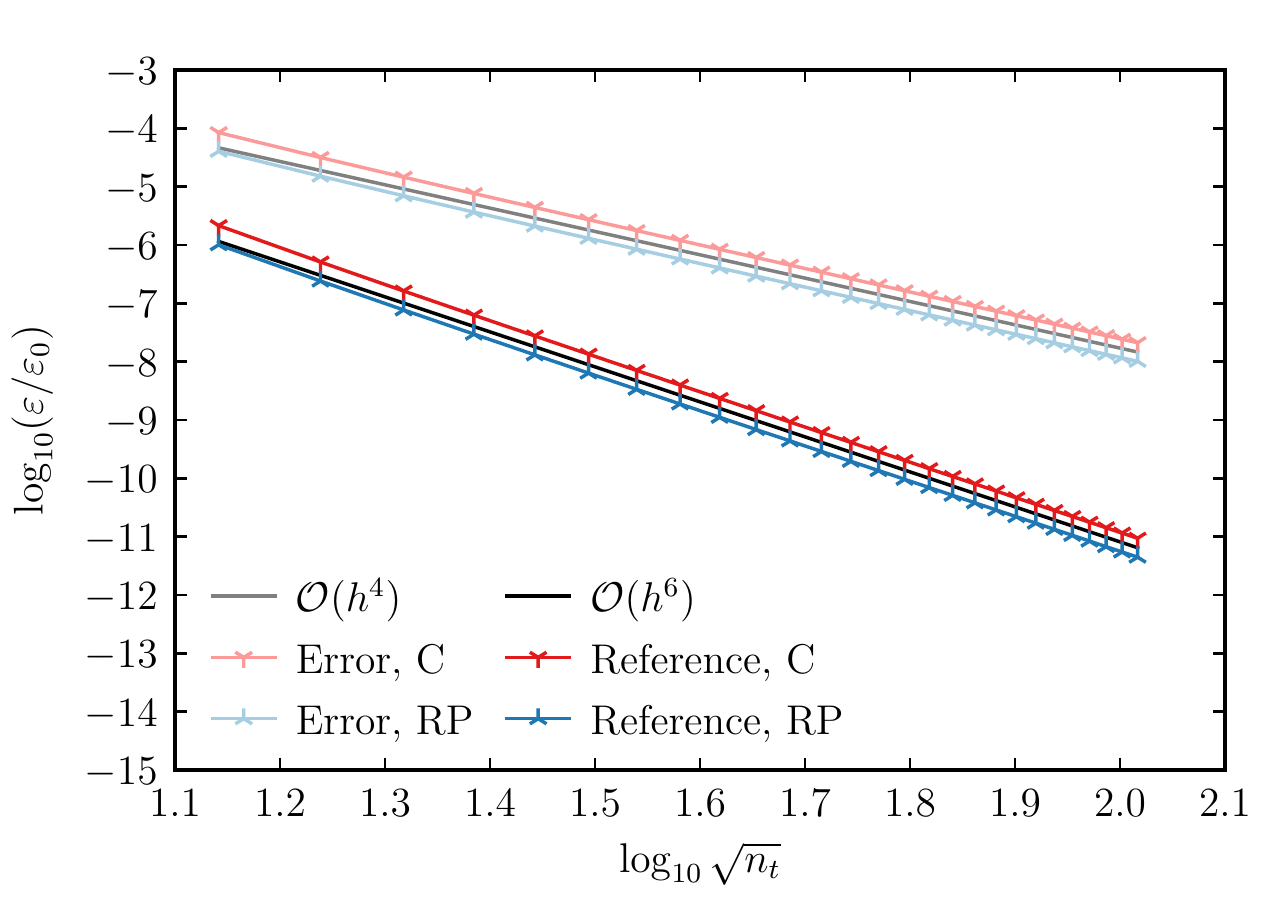}
\caption{$\varepsilon=|e_b(\mathbf{J}_\text{MS})|$\vpad}
\label{fig:part4_bug}
\end{subfigure}
\caption{Numerical-integration error: $\varepsilon$ in the presence of a coding error for $d=3$.}
\vskip-\dp\strutbox
\label{fig:part38_bug}
\end{figure}

Figure~\ref{fig:part5} shows the numerical-integration error $e_a(\mathbf{J}_{h_\text{MS}})$~\eqref{eq:a_error_cancel} when the solution-discretization error is canceled for $d=\{1,\,2,\,3\}$ in~\eqref{eq:G_mms}, both geometries, and both terms.  In the legend entries, the first number is the amount of quadrature points used to compute the integral over $S$, whereas the second is the amount used to compute the integral over $S'$.  
Each of the solutions converges at the expected rate listed in Table~\ref{tab:dunavant_properties}.

Figure~\ref{fig:part6} shows the numerical-integration error $e_b(\mathbf{J}_{h_\text{MS}})$~\eqref{eq:b_error_cancel} when the solution-discretization error is canceled.  In the legend entries, the number is the amount of quadrature points used to compute the integral.  Each of the solutions converges at the expected rate.  \reviewerTwo{For the finest meshes considered, the round-off error arising from the double-precision calculations exceeds the numerical-integration error.}

To relate the $e_a(\mathbf{J}_{h_\text{MS}})$~\eqref{eq:a_error_cancel} and $e_b(\mathbf{J}_{h_\text{MS}})$~\eqref{eq:b_error_cancel} to the discretization error $\mathbf{e}_\mathbf{J}$~\eqref{eq:solution_error}, we can solve the optimization problems~\eqref{eq:opt_a} and~\eqref{eq:opt_b}.  These discretization errors are respectively shown in Figures~\ref{fig:part7} and~\ref{fig:part8} and converge at the same rates as those in Figures~\ref{fig:part5} and~\ref{fig:part6}, \reviewerTwo{until the round-off error exceeds the discretization error}.

Figures~\ref{fig:part3} and~\ref{fig:part4} show the numerical-integration errors $e_a(\mathbf{J}_\text{MS})$~\eqref{eq:a_error_eliminate} and $e_b(\mathbf{J}_\text{MS})$~\eqref{eq:b_error_eliminate} when the solution-discretization error is eliminated for $d=\{1,\,2,\,3\}$ in~\eqref{eq:G_mms}, both geometries, and both terms.  Each of the solutions converges at the expected rate listed in Table~\ref{tab:dunavant_properties}, except for $d=1$ in Figures~\ref{fig:p3_c1_G1} and~\ref{fig:p3_c2_G1}, where the convergence rates for the $1\times 1$ quadrature combination are $\mathcal{O}(h^4)$ instead of $\mathcal{O}(h^2)$.  \reviewerTwo{Once more, for the finer meshes, the round-off error exceeds the discretization error}.

To test the ability to detect a coding error, we replace the optimal 6-point quadrature rule that can exactly integrate polynomials up to degree 4 (Figure~\ref{fig:quad_optimal_6}) with a suboptimal rule (Figure~\ref{fig:quad_suboptimal_6}) that can integrate polynomials up to degree 3.  
With this coding error, for $d=3$, Figures~\ref{fig:part5_bug}--\ref{fig:part4_bug} respectively show the approaches presented in Figures~\ref{fig:part5}--\ref{fig:part4}.  The convergence rates are $\mathcal{O}(h^4)$ for the cases with the coding errors, compared to the expected $\mathcal{O}(h^6)$ rates without.  Therefore, each of these methods detects the coding error.
\subsection{With Curvature} 
\label{sec:with}

\begin{figure}
\centering
\includegraphics[scale=.28,clip=true,trim=.7in 0in 0in 0in]{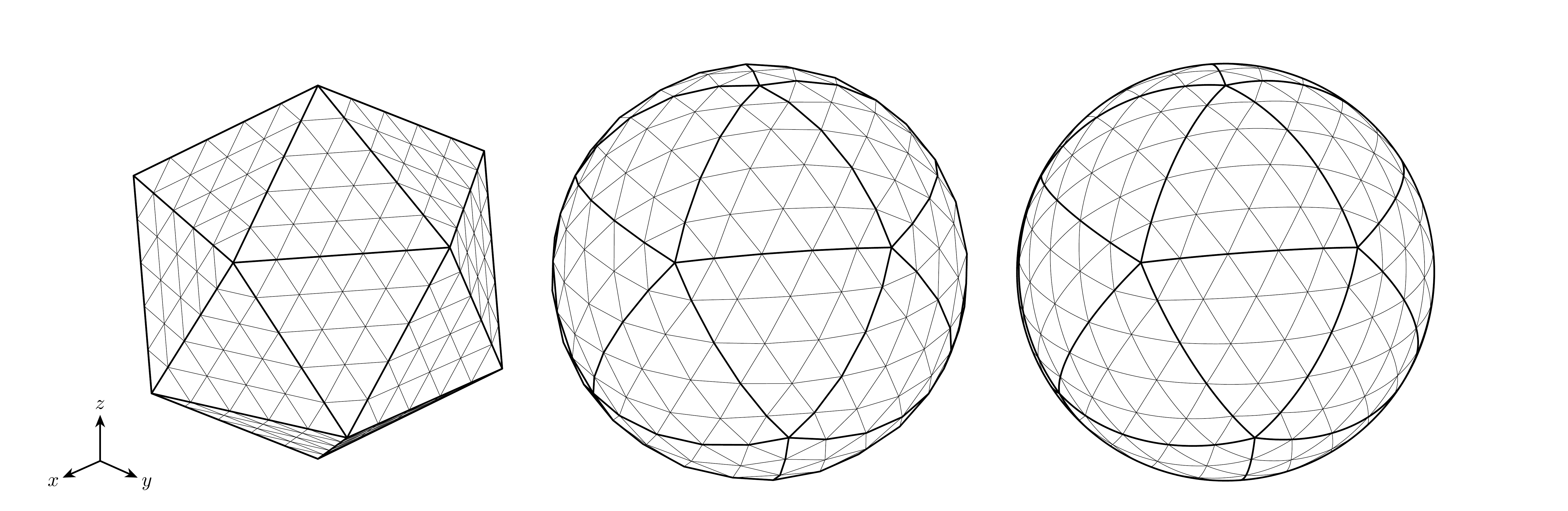}
\caption{Meshes for the regular icosahedron (left), the faceted sphere (center), and the smooth sphere (right), with $\ntriangles=500$.}
\vskip-\dp\strutbox
\label{fig:sphere}
\end{figure}

\begin{figure}
\centering
\includegraphics[scale=.28,clip=true,trim=0in 0in 0in 0in]{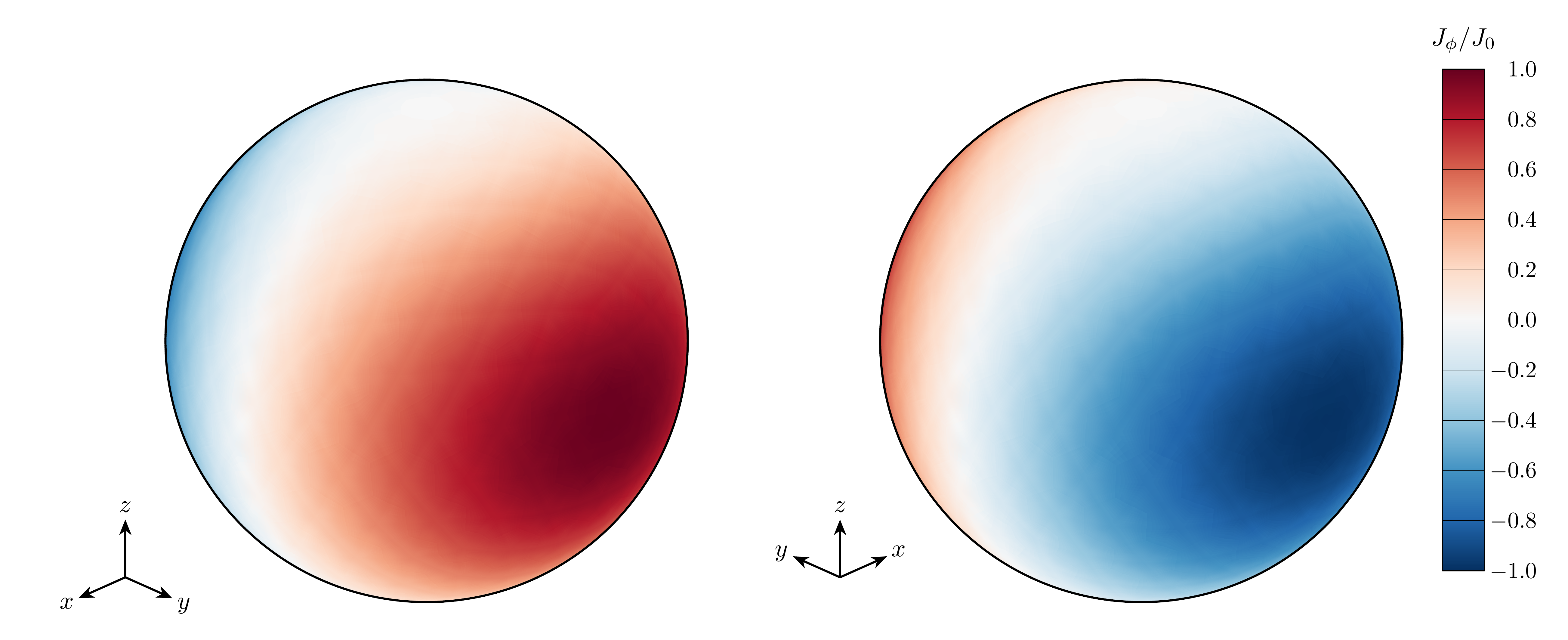}
\caption{Manufactured surface current density $\mathbf{J}_\text{MS}$ for the sphere.}
\vskip-\dp\strutbox
\label{fig:sphere_solution}
\end{figure}

To account for curvature in the domain, we consider a spherical domain with a radius of $1$~m.  The mesh for the sphere is obtained by generating a regular icosahedron inscribed in the sphere, subdividing the twenty triangular faces, and projecting the nodes radially onto the sphere.  To account for the curvature of the sphere, instead of discretizing it with planar triangles, we can discretize it with spherical triangles.  These three phases -- the icosahedron, the faceted sphere, and the smooth sphere -- are depicted in Figure~\ref{fig:sphere} with the total number of triangles $\ntriangles=500$.  The icosahedron is oriented in a manner such it has vertices at $z=\pm 1$~m.

We manufacture the surface current density $\mathbf{J}_\text{MS}(\mathbf{x}) = J_\phi(\theta,\phi)\mathbf{e}_\phi$, where 
\begin{align}
J_\phi(\theta,\phi)=J_0\sin^2\theta\sin\phi,
\label{eq:jphi}
\end{align}
and $x=r\sin\theta\cos\phi$, $y=r\sin\theta\sin\phi$, $z=r\cos\theta$, and $r=\sqrt{x^2+y^2+z^2}$.  Equation~\eqref{eq:jphi} is of class $C^\infty$.  Figure~\ref{fig:sphere_solution} shows plots of~\eqref{eq:jphi}.
%

\subsubsection{Domain-Discretization Error} 

\begin{figure}
\centering
\begin{subfigure}[b]{.49\textwidth}
\includegraphics[scale=.64,clip=true,trim=2.3in 0in 2.8in 0in]{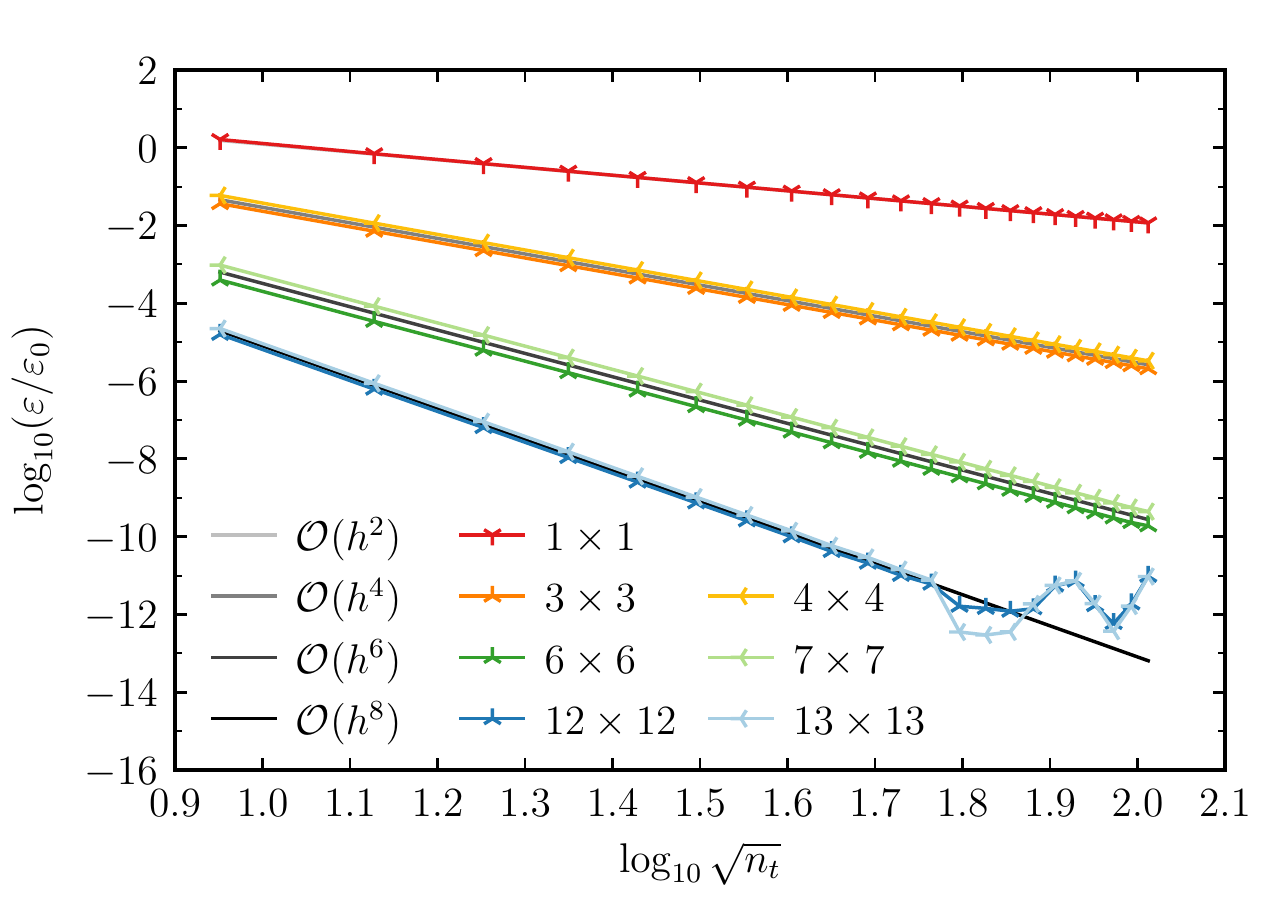}
\caption{$a^q(\mathbf{J}_\text{MS},\mathbf{J}_\text{MS})$, $d=1$\vpad}
\label{fig:p23_d1}
\end{subfigure}
\hspace{0.25em}
\begin{subfigure}[b]{.49\textwidth}
\includegraphics[scale=.64,clip=true,trim=2.3in 0in 2.8in 0in]{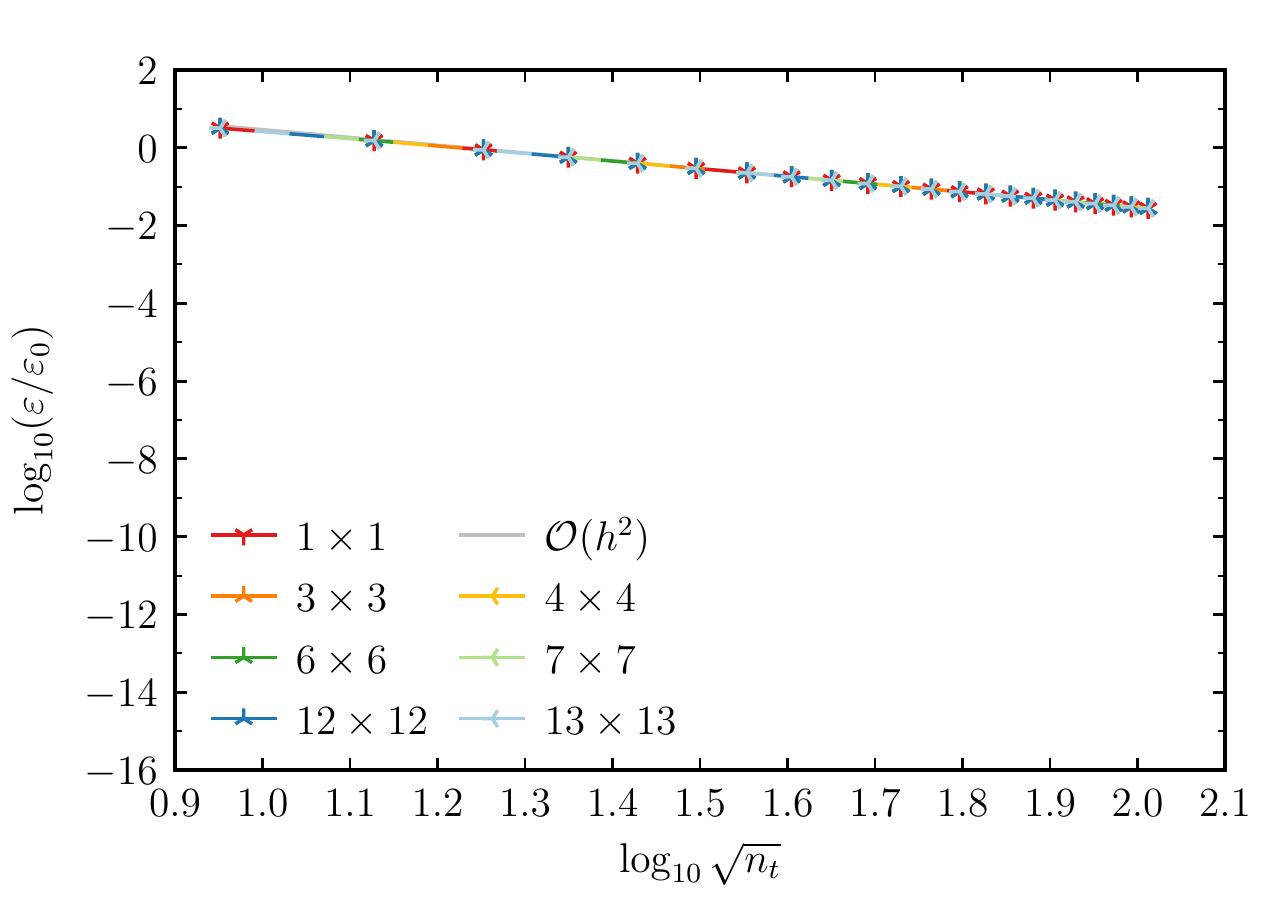}
\caption{$a_h^q(\mathbf{J}_\text{MS},\mathbf{J}_\text{MS})$, $d=1$\vpad}
\label{fig:p13_d1}
\end{subfigure}
\\
\begin{subfigure}[b]{.49\textwidth}
\includegraphics[scale=.64,clip=true,trim=2.3in 0in 2.8in 0in]{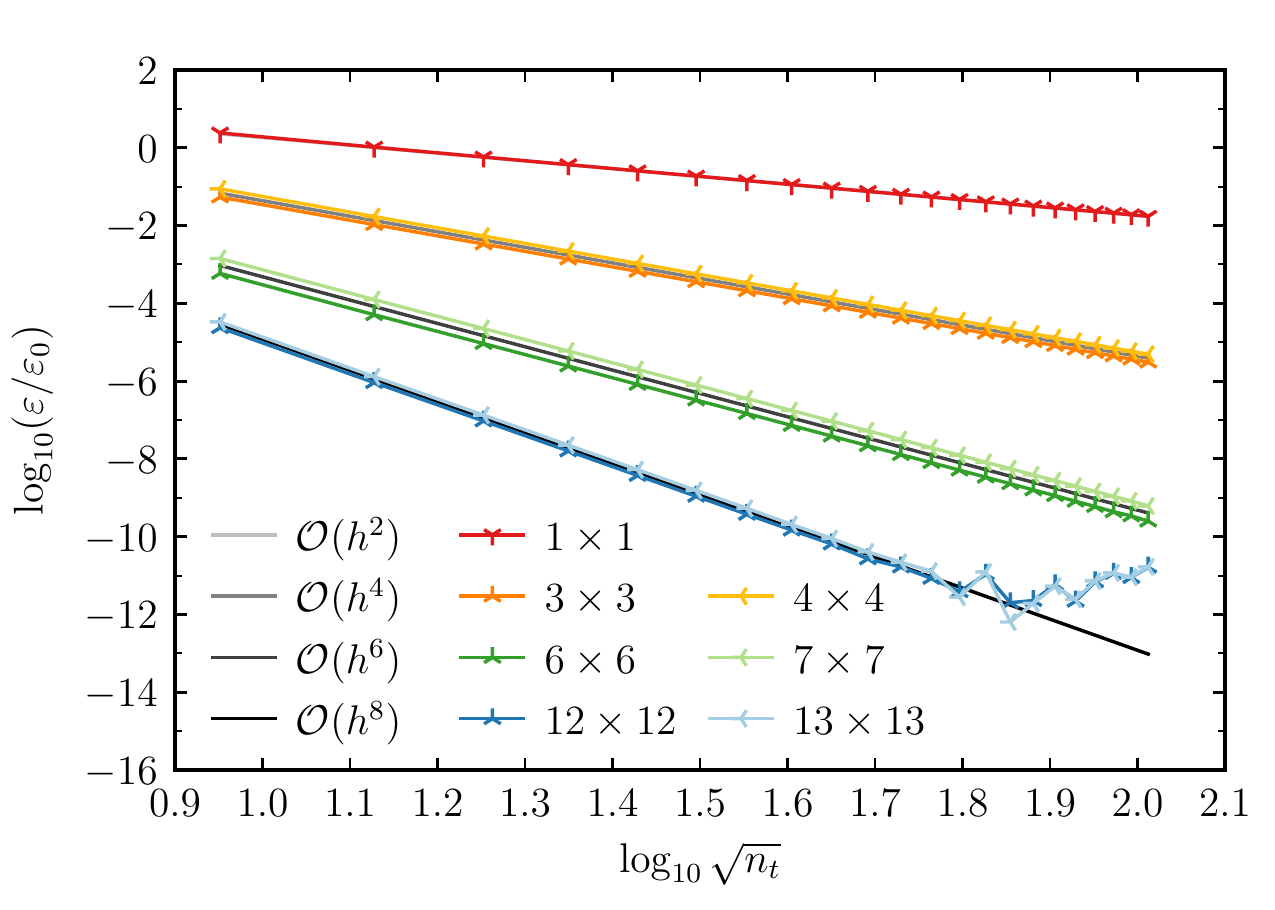}
\caption{$a^q(\mathbf{J}_\text{MS},\mathbf{J}_\text{MS})$, $d=2$\vpad}
\label{fig:p23_d2}
\end{subfigure}
\hspace{0.25em}
\begin{subfigure}[b]{.49\textwidth}
\includegraphics[scale=.64,clip=true,trim=2.3in 0in 2.8in 0in]{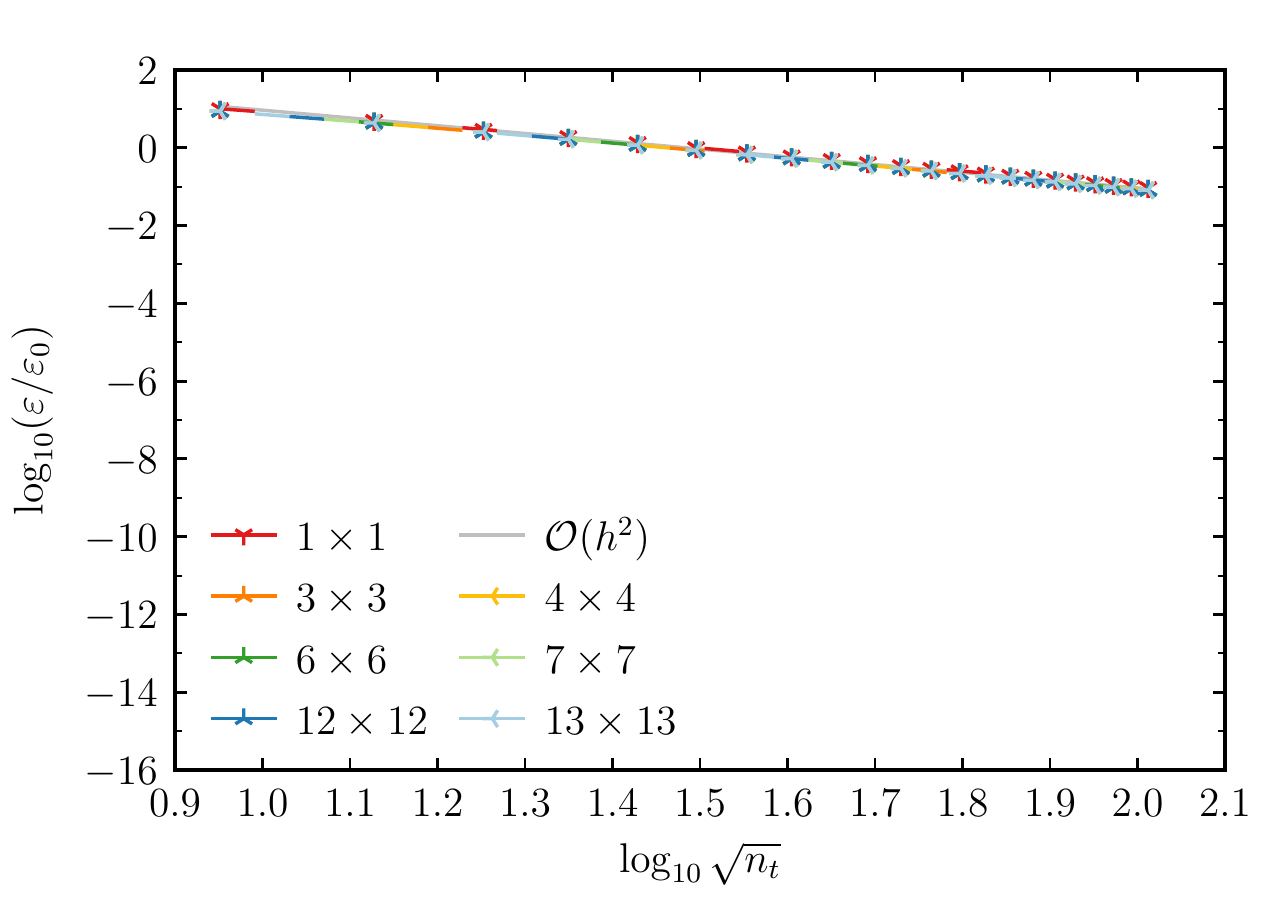}
\caption{$a_h^q(\mathbf{J}_\text{MS},\mathbf{J}_\text{MS})$, $d=2$\vpad}
\label{fig:p13_d2}
\end{subfigure}
\\
\begin{subfigure}[b]{.49\textwidth}
\includegraphics[scale=.64,clip=true,trim=2.3in 0in 2.8in 0in]{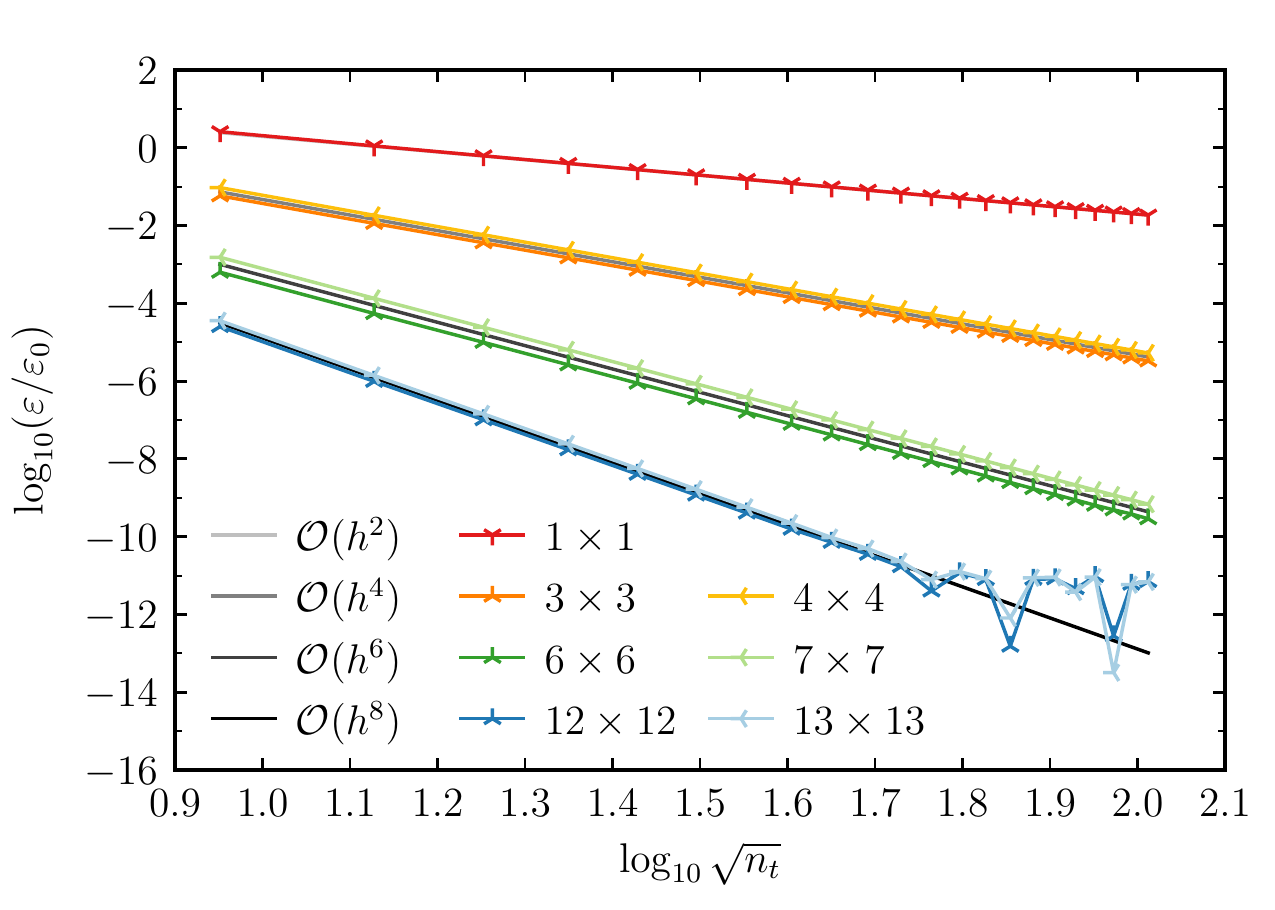}
\caption{$a^q(\mathbf{J}_\text{MS},\mathbf{J}_\text{MS})$, $d=3$\vpad}
\label{fig:p23_d3}
\end{subfigure}
\hspace{0.25em}
\begin{subfigure}[b]{.49\textwidth}
\includegraphics[scale=.64,clip=true,trim=2.3in 0in 2.8in 0in]{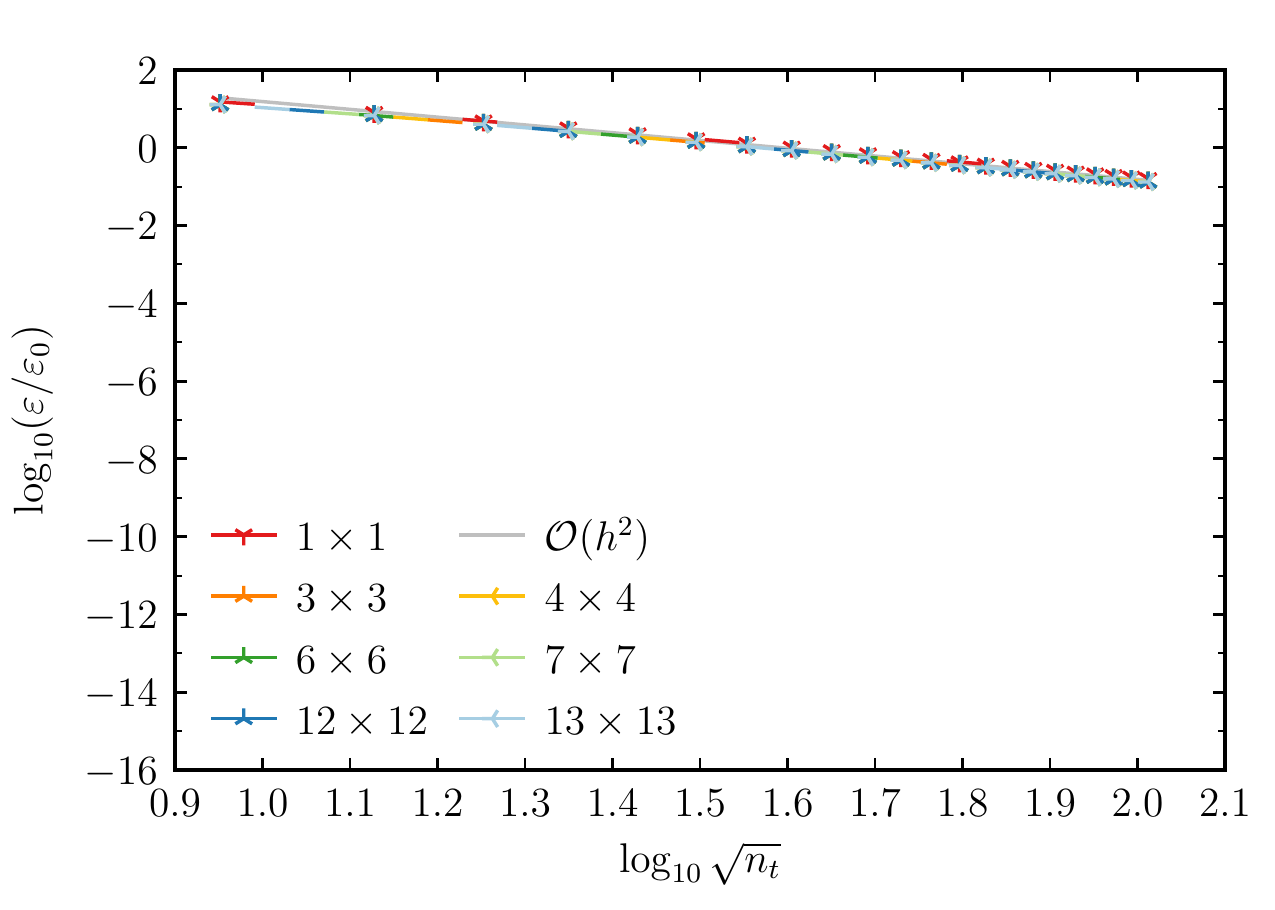}
\caption{$a_h^q(\mathbf{J}_\text{MS},\mathbf{J}_\text{MS})$, $d=3$\vpad}
\label{fig:p13_d3}
\end{subfigure}
\caption{Domain-Discretization Error: $\varepsilon=|e_a(\mathbf{J}_\text{MS})|$ for different amounts of quadrature points.}
\vskip-\dp\strutbox
\label{fig:part23_sphere}
\end{figure}

To address the domain-discretization error, we perform the assessments described in Section~\ref{sec:dde}.
For $d=\{1,\,2,\,3\}$ in~\eqref{eq:G_mms}, Figure~\ref{fig:part23_sphere} shows $e_a(\mathbf{J}_\text{MS})$~\eqref{eq:a_error_eliminate}, as described in Section~\ref{sec:ac}.  For the left column (\ref{fig:p23_d1}, \ref{fig:p23_d2}, \ref{fig:p23_d3}), $a^q(\mathbf{J}_\text{MS},\mathbf{J}_\text{MS})$ in~\eqref{eq:a_error_eliminate} is computed on meshes composed of spherical triangles.  For the right column (\ref{fig:p13_d1}, \ref{fig:p13_d2}, \ref{fig:p13_d3}), $a^q(\mathbf{J}_\text{MS},\mathbf{J}_\text{MS})$ is approximated by $a_h^q(\mathbf{J}_\text{MS},\mathbf{J}_\text{MS})$ and computed on meshes composed of planar triangles.
By accounting for the curvature, $e_a(\mathbf{J}_\text{MS})$ converges at the expected rates listed in Table~\ref{tab:dunavant_properties} \reviewerTwo{(until the round-off error exceeds the domain-discretization error)} for $a^q(\mathbf{J}_\text{MS},\mathbf{J}_\text{MS})$, whereas its rate is limited to $\mathcal{O}(h^2)$ for $a_h^q(\mathbf{J}_\text{MS},\mathbf{J}_\text{MS})$.

\begin{figure}
\centering
\begin{subfigure}[b]{.49\textwidth}
\includegraphics[scale=.64,clip=true,trim=2.3in 0in 2.8in 0in]{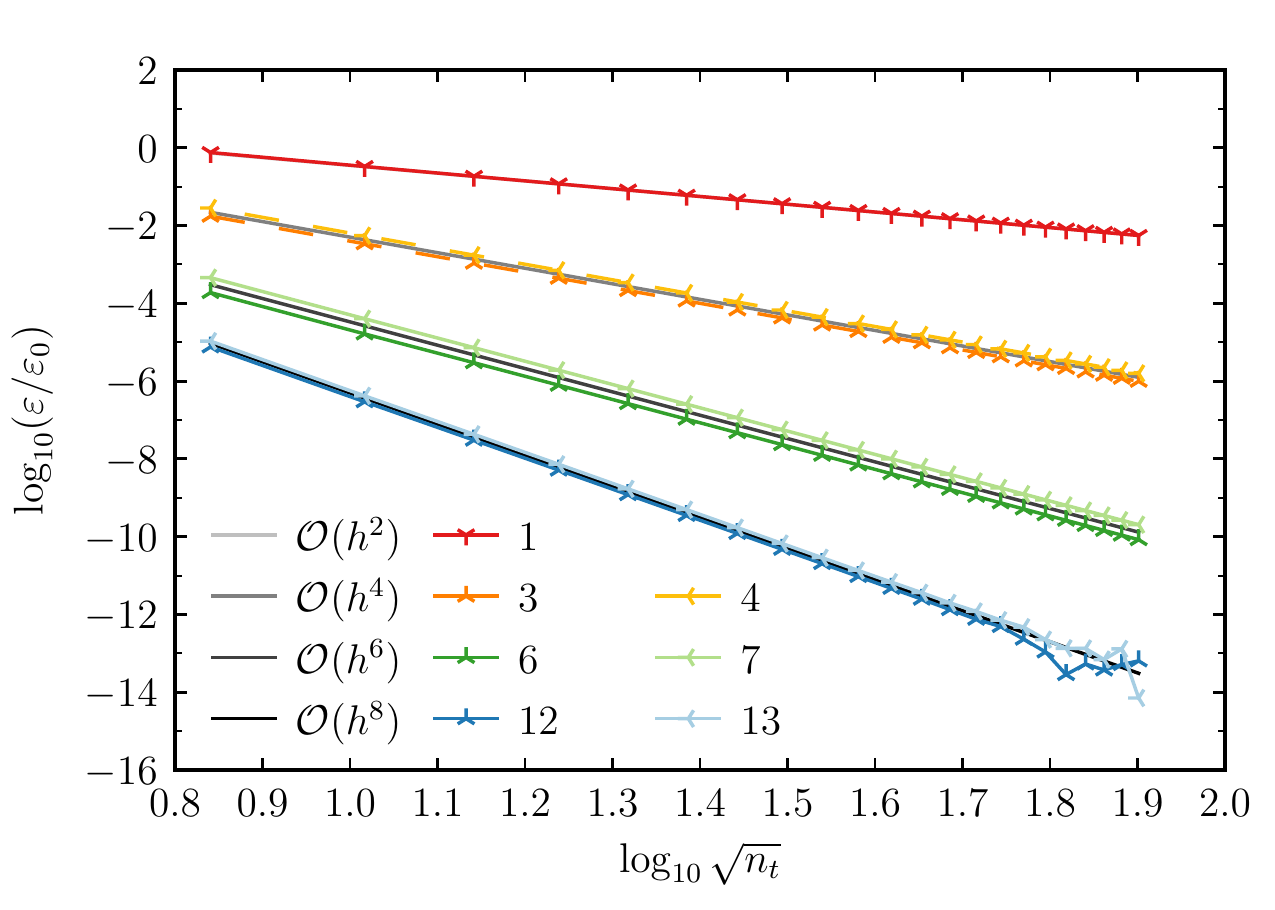}
\caption{$b^q\big(\mathbf{H}^\mathcal{I}_\text{MS},\mathbf{J}_\text{MS}\big)$, $d=1$\vpad}
\label{fig:p24_d1}
\end{subfigure}
\hspace{0.25em}
\begin{subfigure}[b]{.49\textwidth}
\includegraphics[scale=.64,clip=true,trim=2.3in 0in 2.8in 0in]{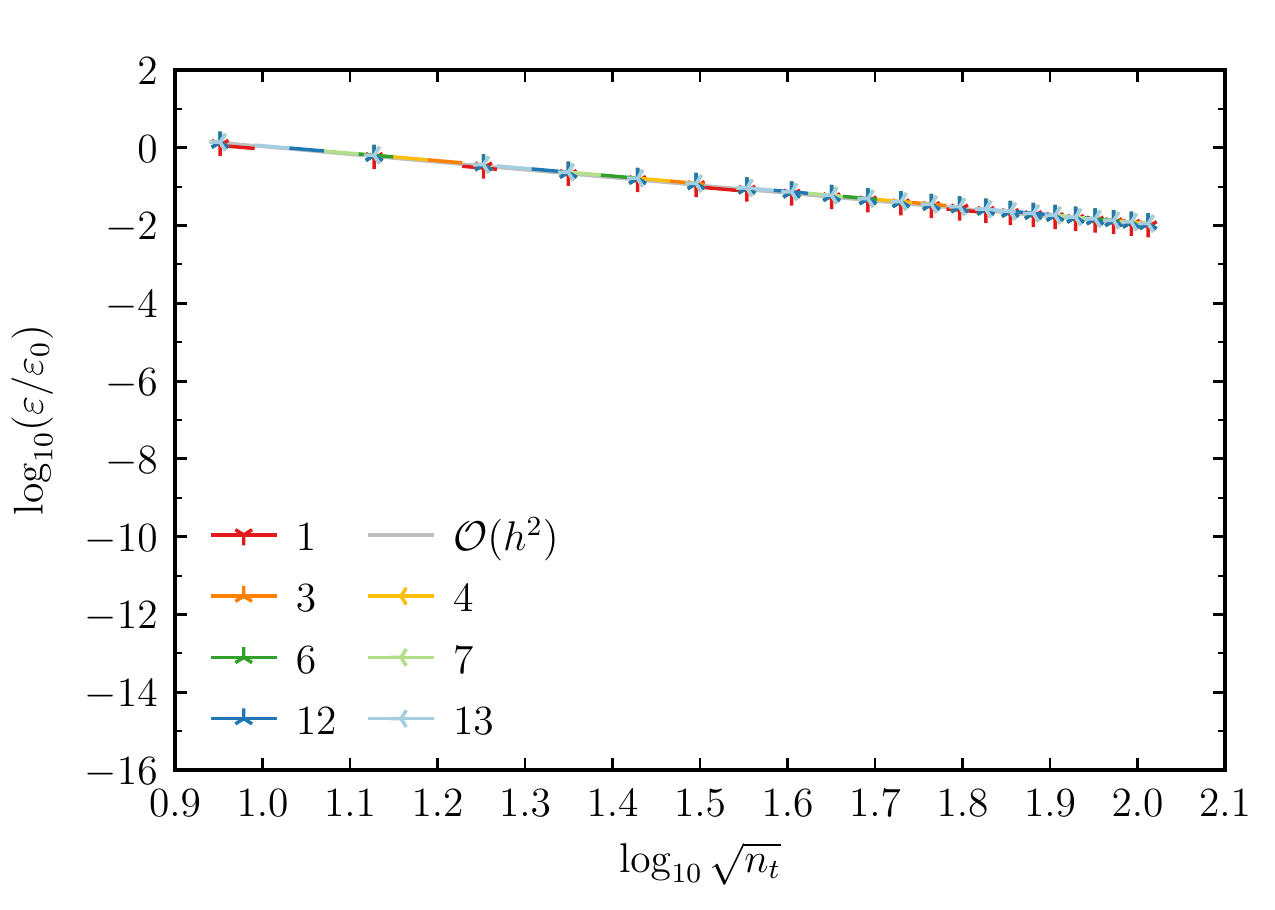}
\caption{$b_h^q\big(\mathbf{H}^\mathcal{I}_\text{MS},\mathbf{J}_\text{MS}\big)$, $d=1$\vpad}
\label{fig:p14_d1}
\end{subfigure}
\\
\begin{subfigure}[b]{.49\textwidth}
\includegraphics[scale=.64,clip=true,trim=2.3in 0in 2.8in 0in]{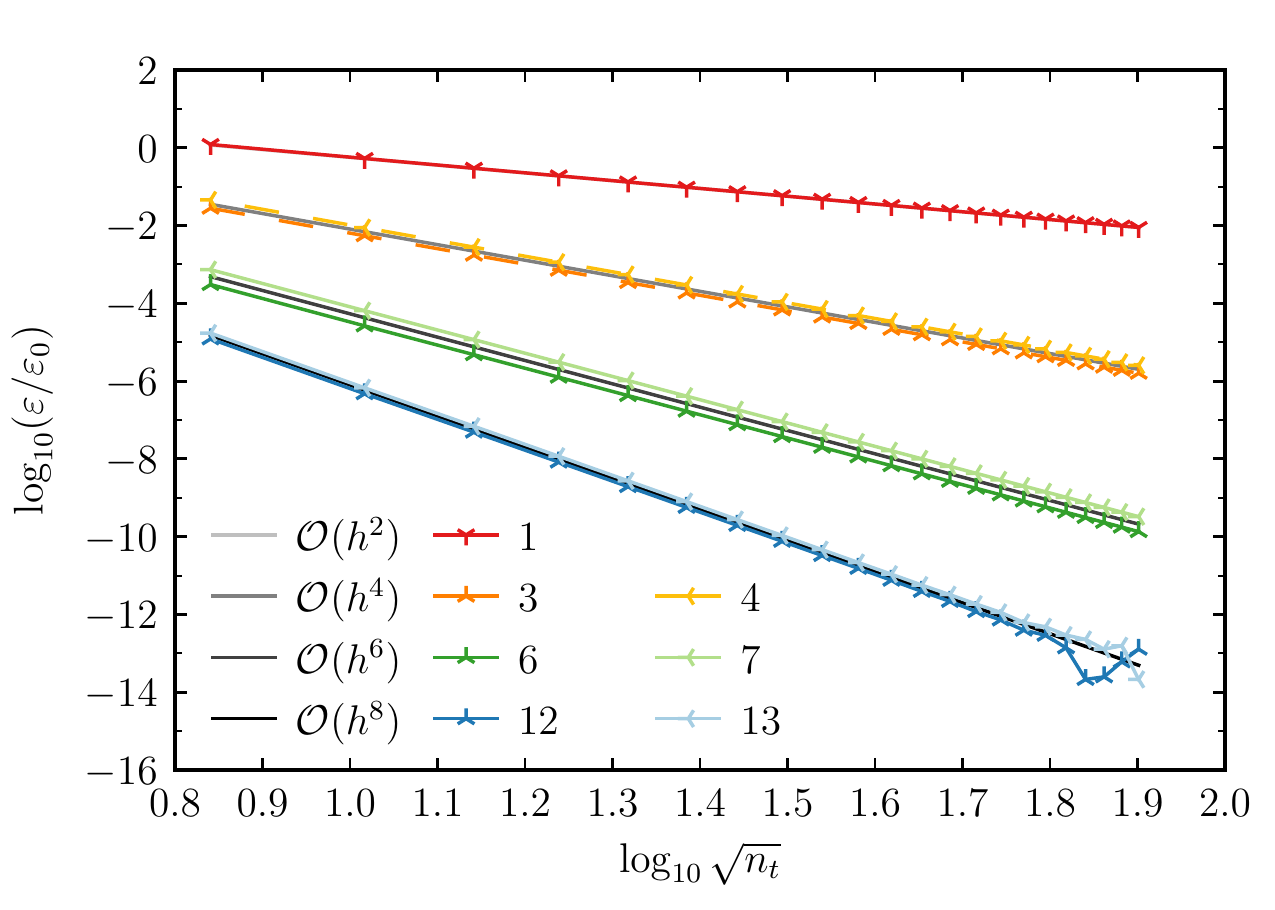}
\caption{$b^q\big(\mathbf{H}^\mathcal{I}_\text{MS},\mathbf{J}_\text{MS}\big)$, $d=2$\vpad}
\label{fig:p24_d2}
\end{subfigure}
\hspace{0.25em}
\begin{subfigure}[b]{.49\textwidth}
\includegraphics[scale=.64,clip=true,trim=2.3in 0in 2.8in 0in]{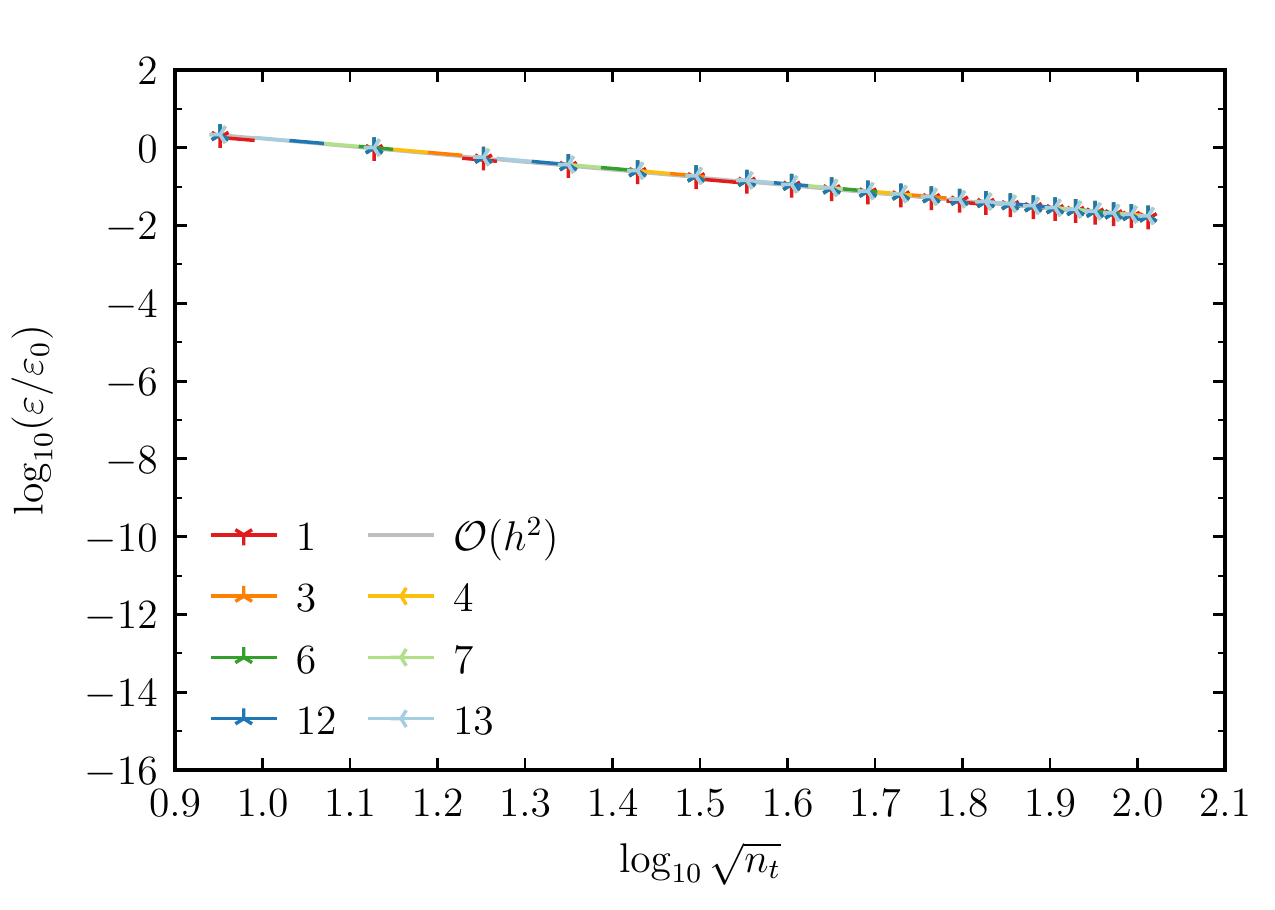}
\caption{$b_h^q\big(\mathbf{H}^\mathcal{I}_\text{MS},\mathbf{J}_\text{MS}\big)$, $d=2$\vpad}
\label{fig:p14_d2}
\end{subfigure}
\\
\begin{subfigure}[b]{.49\textwidth}
\includegraphics[scale=.64,clip=true,trim=2.3in 0in 2.8in 0in]{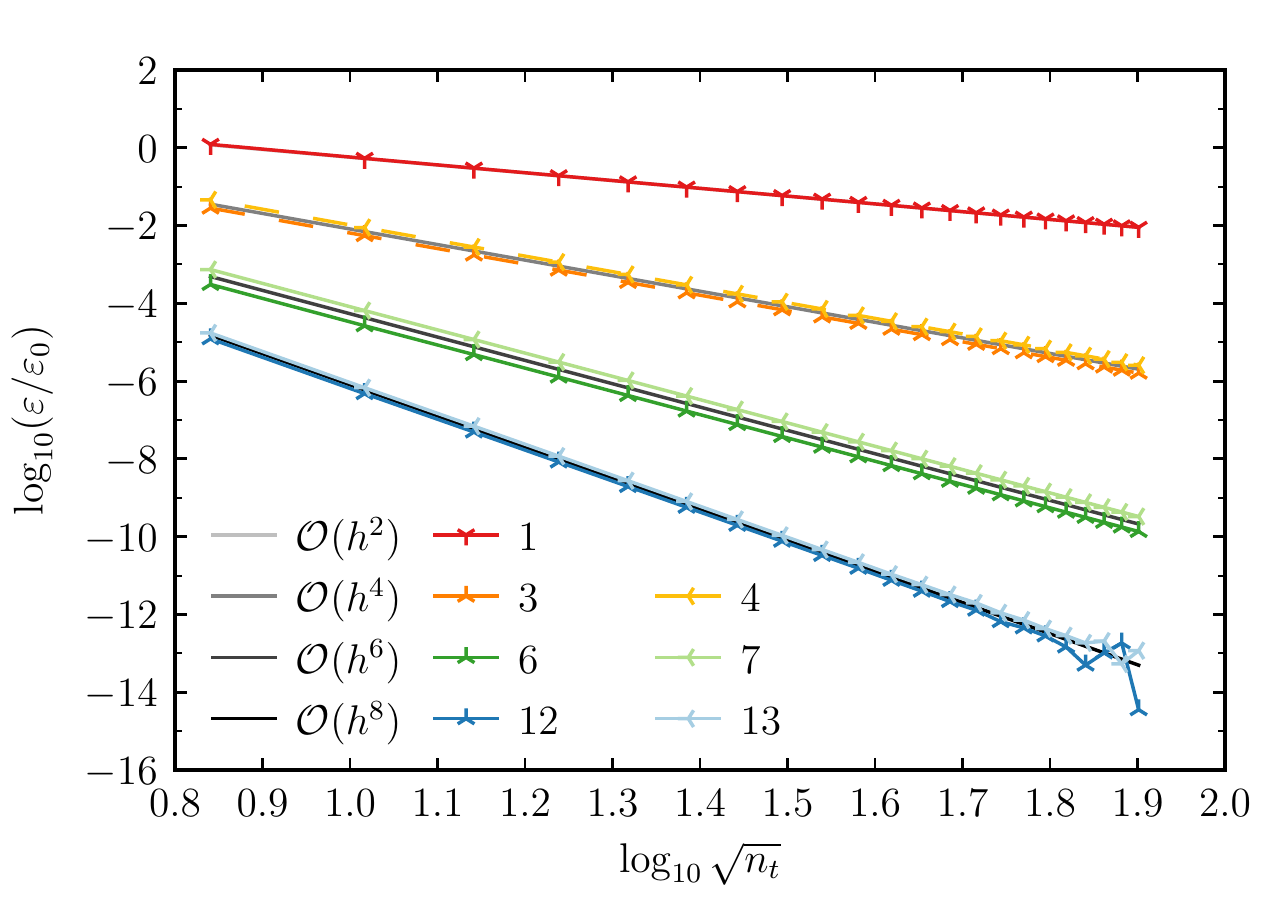}
\caption{$b^q\big(\mathbf{H}^\mathcal{I}_\text{MS},\mathbf{J}_\text{MS}\big)$, $d=3$\vpad}
\label{fig:p24_d3}
\end{subfigure}
\hspace{0.25em}
\begin{subfigure}[b]{.49\textwidth}
\includegraphics[scale=.64,clip=true,trim=2.3in 0in 2.8in 0in]{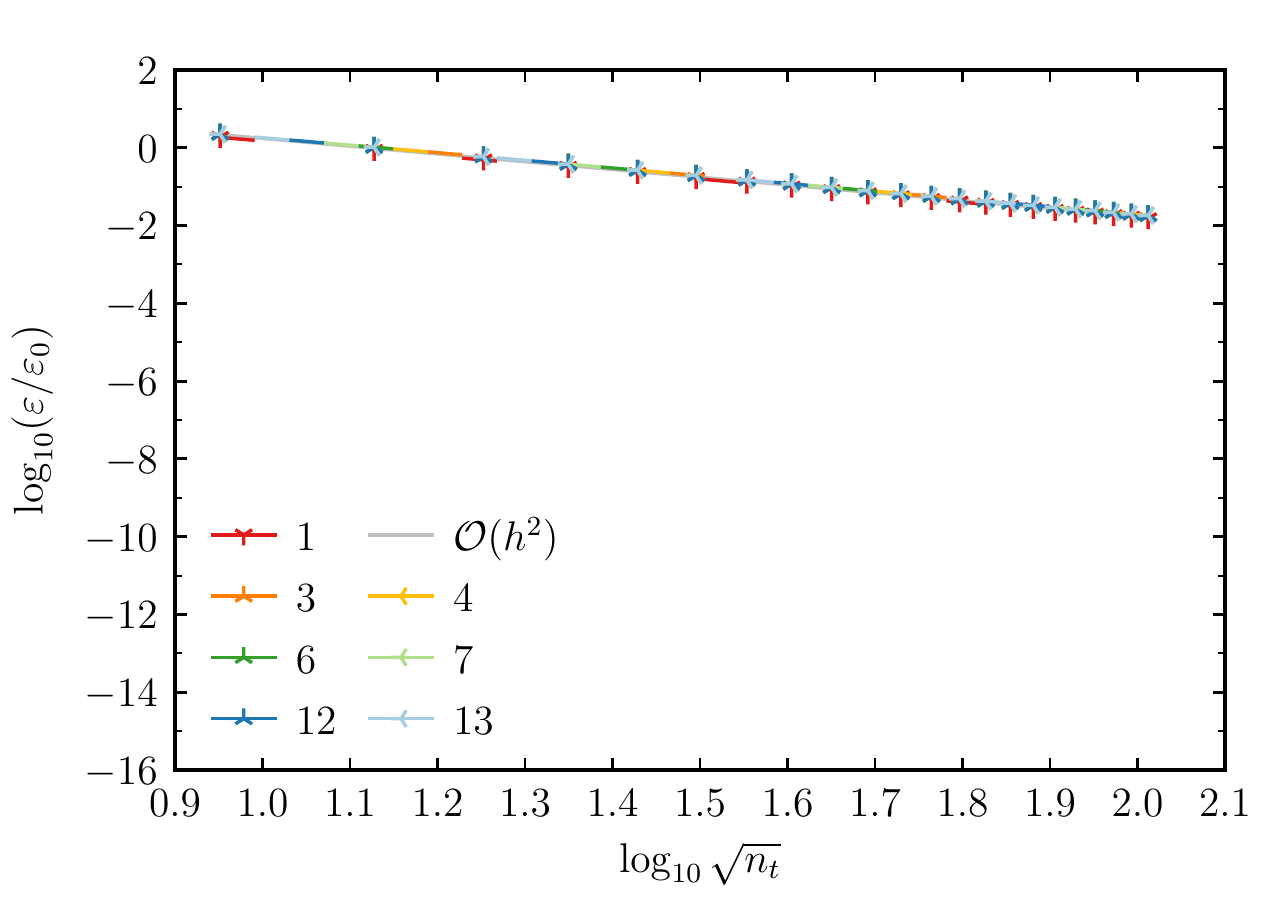}
\caption{$b_h^q\big(\mathbf{H}^\mathcal{I}_\text{MS},\mathbf{J}_\text{MS}\big)$, $d=3$\vpad}
\label{fig:p14_d3}
\end{subfigure}
\caption{Domain-Discretization Error: $\varepsilon=|e_b(\mathbf{J}_\text{MS})|$ for different amounts of quadrature points.}
\vskip-\dp\strutbox
\label{fig:part24_sphere}
\end{figure}

\begin{figure}
\centering
\begin{subfigure}[b]{.49\textwidth}
\includegraphics[scale=.64,clip=true,trim=2.3in 0in 2.8in 0in]{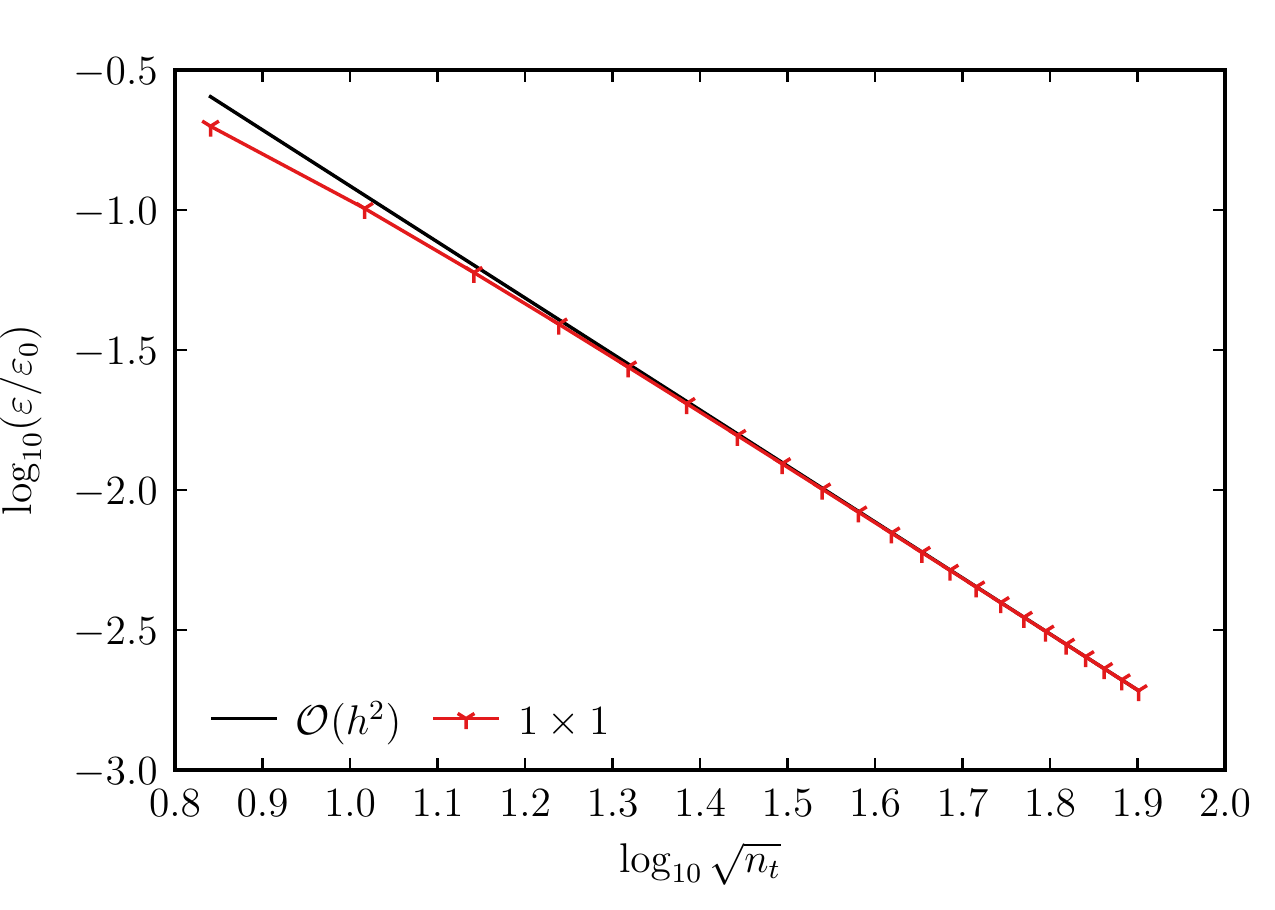}
\caption{$\varepsilon=e_a\big(\mathbf{J}_{h_\text{MS}}\big)$, $d=1$\vpad}
\label{fig:p5_d1}
\end{subfigure}
\hspace{0.25em}
\begin{subfigure}[b]{.49\textwidth}
\includegraphics[scale=.64,clip=true,trim=2.3in 0in 2.8in 0in]{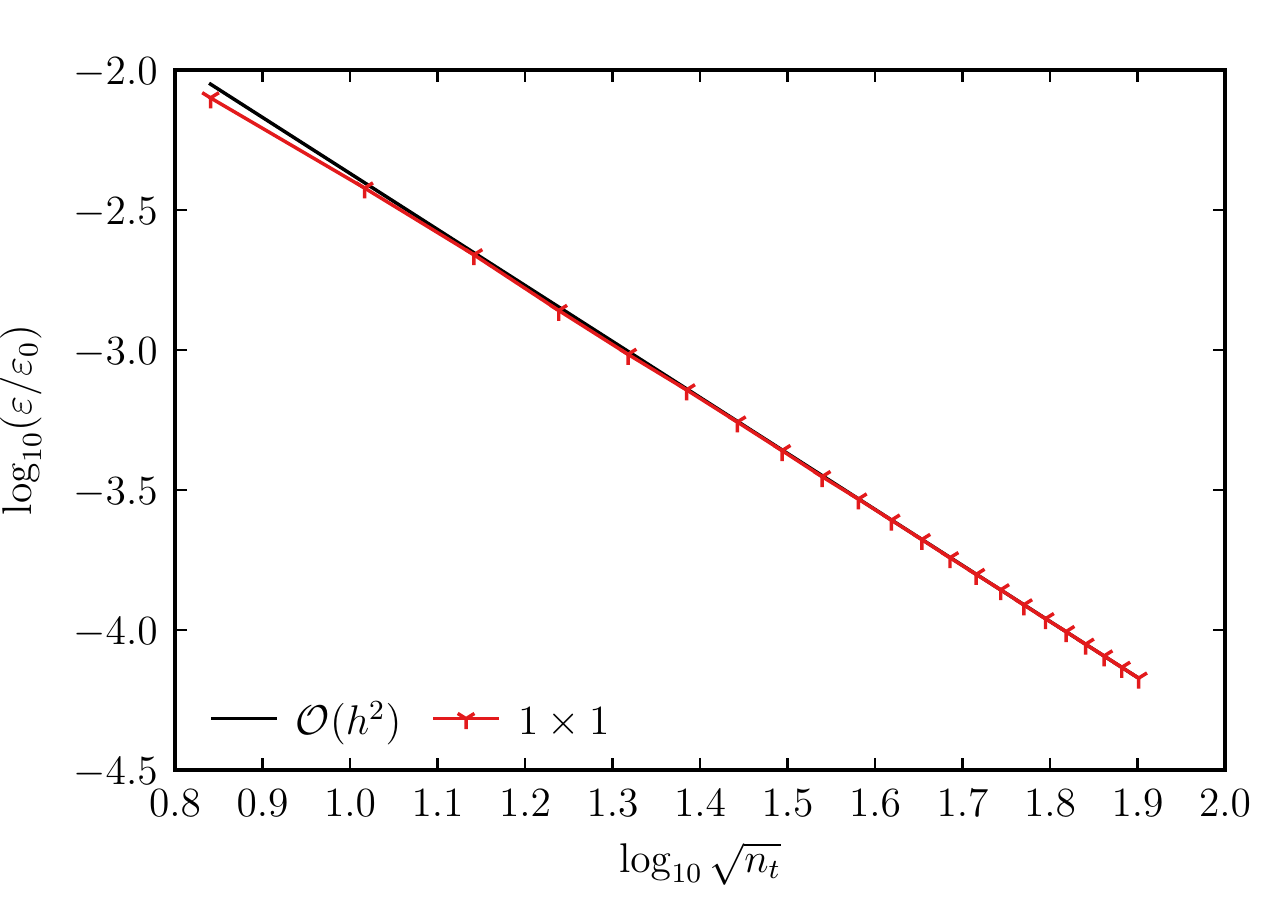}
\caption{$\varepsilon={\|\mathbf{e}_\mathbf{J}\|}_\infty$ from~\eqref{eq:opt_a}, $d=1$\vpad}
\label{fig:p7_d1}
\end{subfigure}
\\
\begin{subfigure}[b]{.49\textwidth}
\includegraphics[scale=.64,clip=true,trim=2.3in 0in 2.8in 0in]{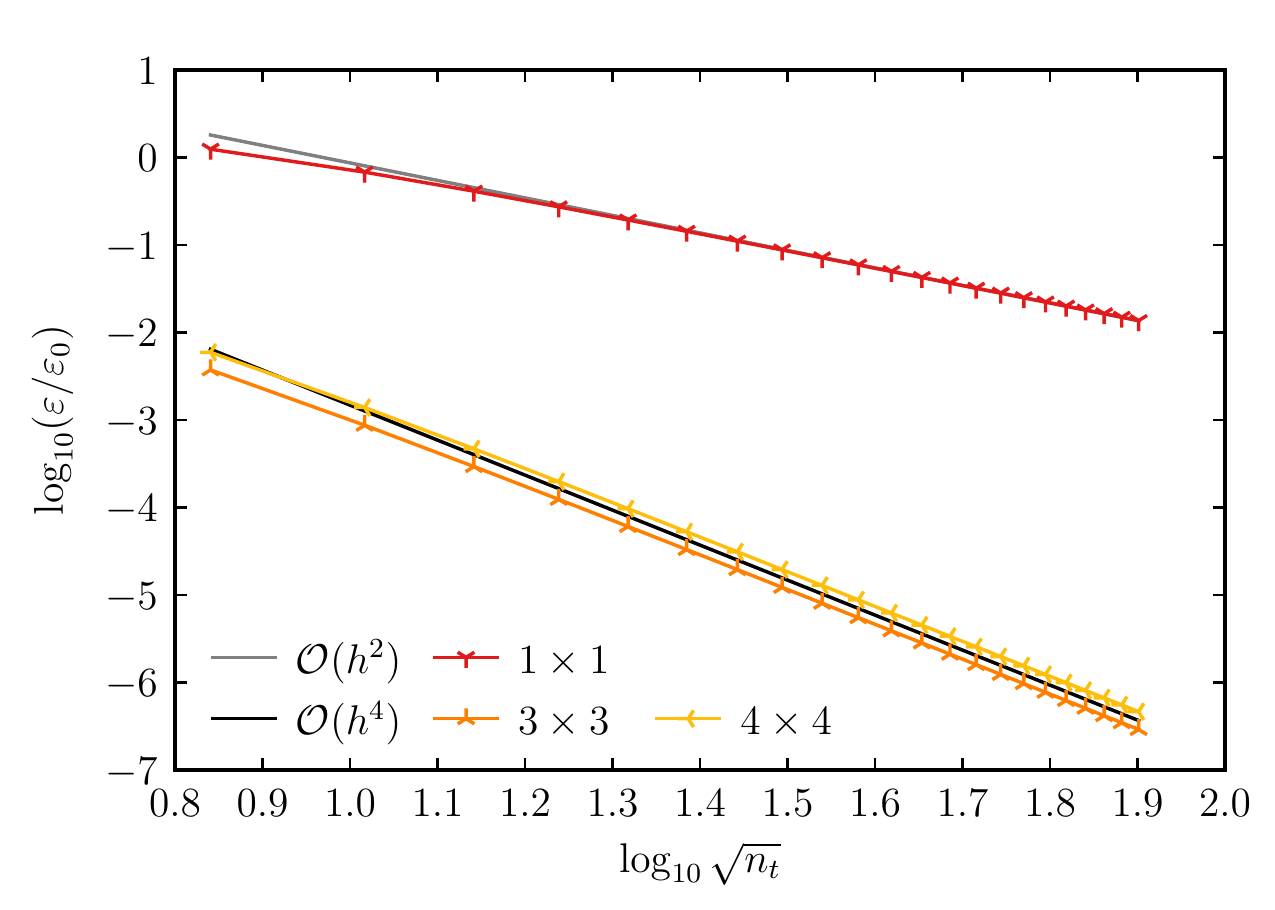}
\caption{$\varepsilon=e_a\big(\mathbf{J}_{h_\text{MS}}\big)$, $d=2$\vpad}
\label{fig:p5_d2}
\end{subfigure}
\hspace{0.25em}
\begin{subfigure}[b]{.49\textwidth}
\includegraphics[scale=.64,clip=true,trim=2.3in 0in 2.8in 0in]{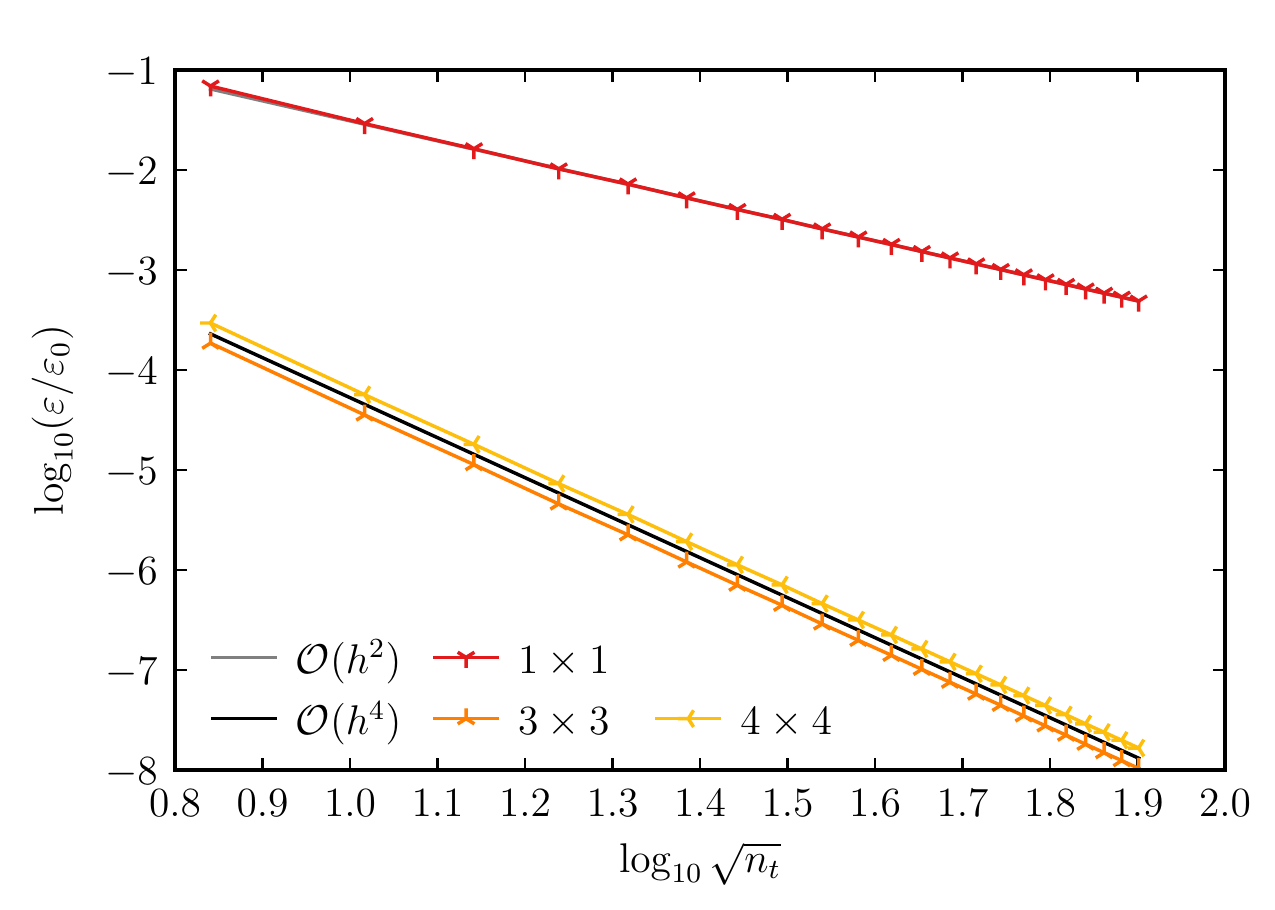}
\caption{$\varepsilon={\|\mathbf{e}_\mathbf{J}\|}_\infty$ from~\eqref{eq:opt_a}, $d=2$\vpad}
\label{fig:p7_d2}
\end{subfigure}
\\
\begin{subfigure}[b]{.49\textwidth}
\includegraphics[scale=.64,clip=true,trim=2.3in 0in 2.8in 0in]{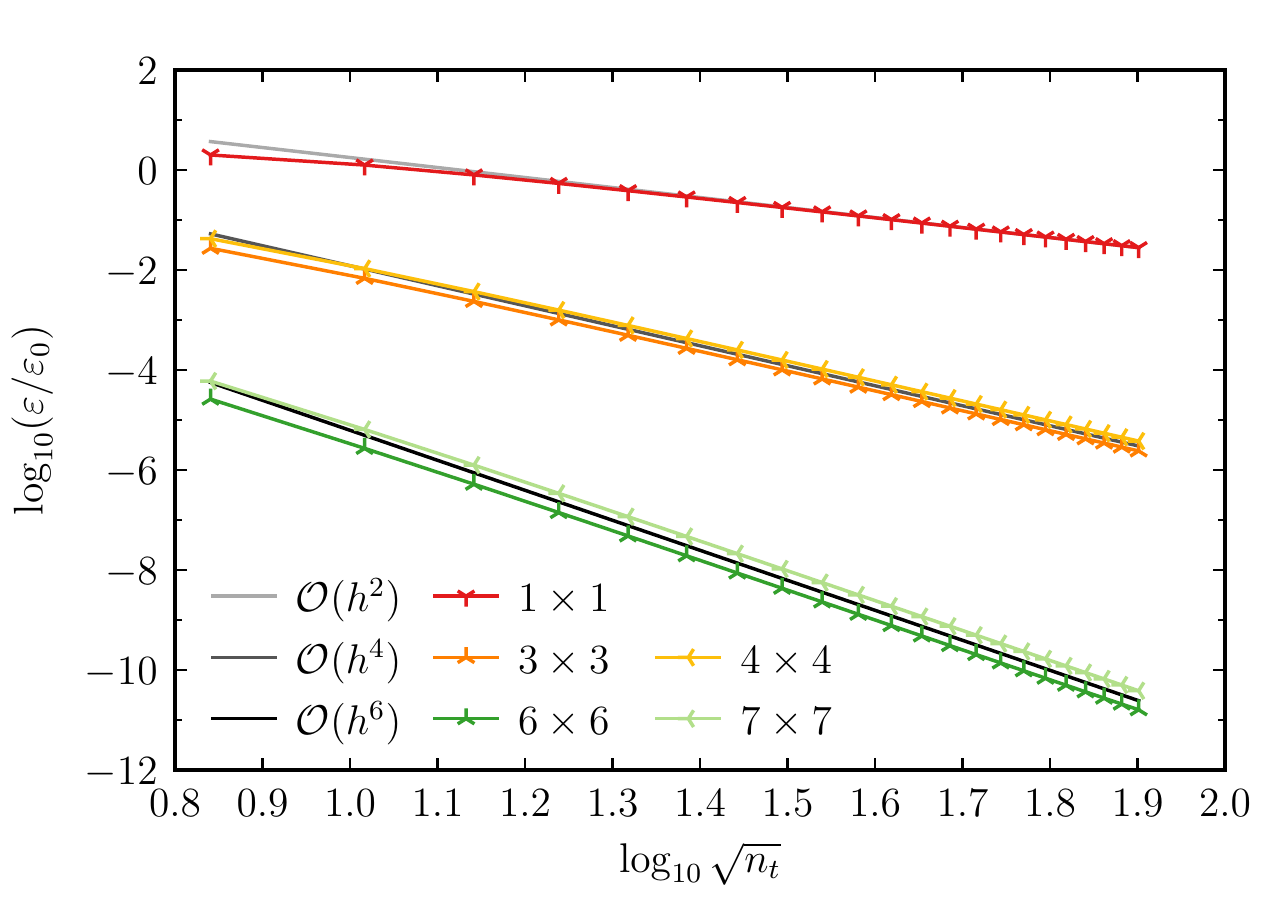}
\caption{$\varepsilon=e_a\big(\mathbf{J}_{h_\text{MS}}\big)$, $d=3$\vpad}
\label{fig:p5_d3}
\end{subfigure}
\hspace{0.25em}
\begin{subfigure}[b]{.49\textwidth}
\includegraphics[scale=.64,clip=true,trim=2.3in 0in 2.8in 0in]{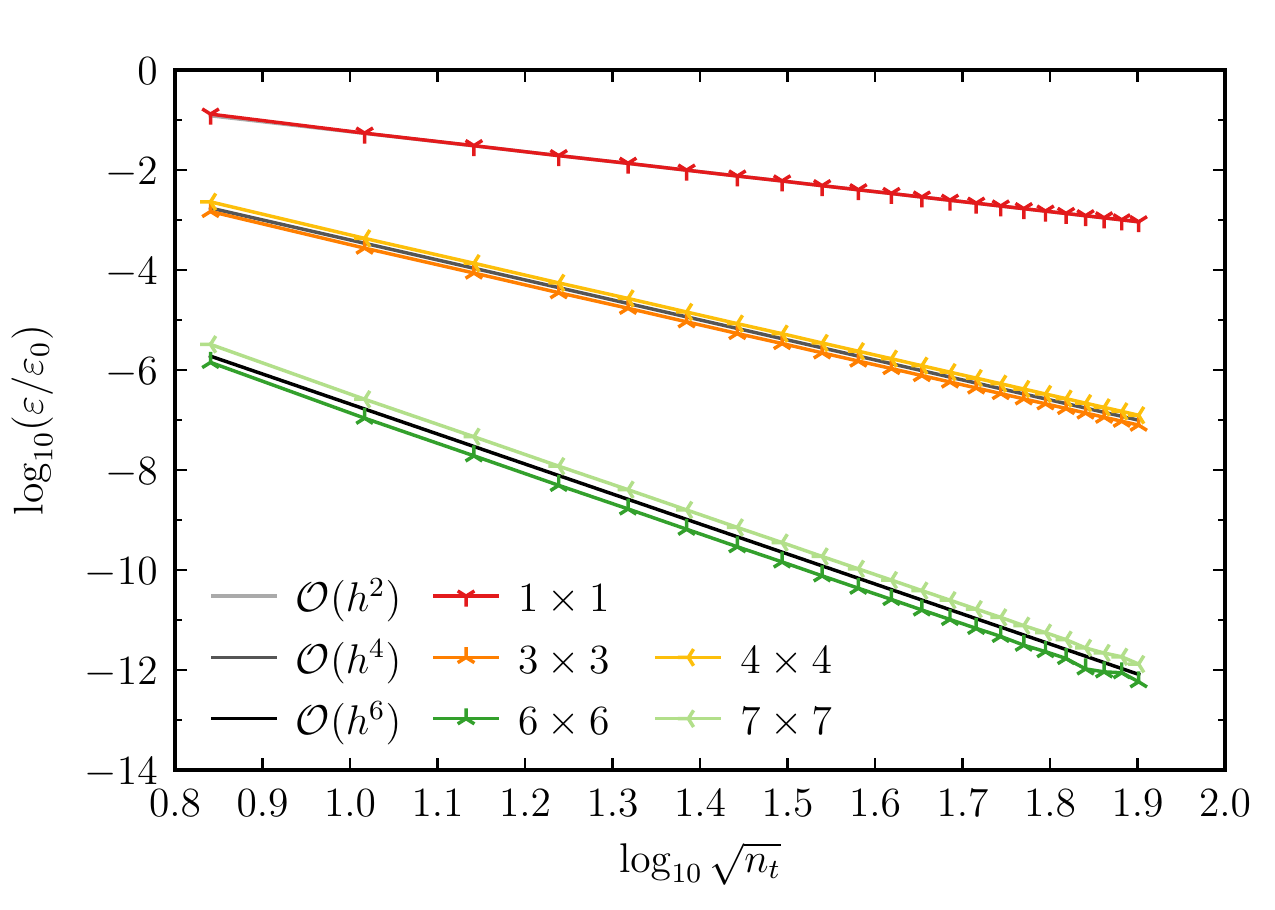}
\caption{$\varepsilon={\|\mathbf{e}_\mathbf{J}\|}_\infty$ from~\eqref{eq:opt_a}, $d=3$\vpad}
\label{fig:p7_d3}
\end{subfigure}
\caption{Domain-Discretization Error: $\varepsilon$ for different amounts of quadrature points.}
\vskip-\dp\strutbox
\label{fig:part57_sphere}
\end{figure}

Figure~\ref{fig:part24_sphere} shows $e_b(\mathbf{J}_\text{MS})$~\eqref{eq:b_error_eliminate} for $d=\{1,\,2,\,3\}$.  
For the left column (\ref{fig:p24_d1}, \ref{fig:p24_d2}, \ref{fig:p24_d3}), $b^q\big(\mathbf{H}^\mathcal{I}_\text{MS},\mathbf{J}_\text{MS}\big)$ in~\eqref{eq:b_error_eliminate} is computed on meshes composed of spherical triangles.  For the right column (\ref{fig:p14_d1}, \ref{fig:p14_d2}, \ref{fig:p14_d3}), $b^q\big(\mathbf{H}^\mathcal{I}_\text{MS},\mathbf{J}_\text{MS}\big)$ is approximated by $b_h^q\big(\mathbf{H}^\mathcal{I}_\text{MS},\mathbf{J}_\text{MS}\big)$ and computed on meshes composed of planar triangles.
By accounting for the curvature, $e_b(\mathbf{J}_\text{MS})$ converges at the expected rates listed in Table~\ref{tab:dunavant_properties} \reviewerTwo{(until the round-off error exceeds the domain-discretization error)} for $b^q\big(\mathbf{H}^\mathcal{I}_\text{MS},\mathbf{J}_\text{MS}\big)$, whereas its rate is limited to $\mathcal{O}(h^2)$ for $b_h^q\big(\mathbf{H}^\mathcal{I}_\text{MS},\mathbf{J}_\text{MS}\big)$.  

Finally, we can neglect curvature as described in Section~\ref{sec:nc}.  Figure~\ref{fig:part57_sphere} shows $e_a(\mathbf{J}_{h_\text{MS}})$~\eqref{eq:a_error_cancel} in the left column (\ref{fig:p5_d1}, \ref{fig:p5_d2}, \ref{fig:p5_d3}), and $\mathbf{e}_\mathbf{J}$~\eqref{eq:solution_error} arising from~\eqref{eq:opt_a} in the right column (\ref{fig:p7_d1}, \ref{fig:p7_d2}, \ref{fig:p7_d3}).  This assessment isolates the numerical-integration error, which converges at the expected rates listed in Table~\ref{tab:dunavant_properties}.

\section{Conclusions} 
\label{sec:conclusions}

In this paper, we presented code-verification approaches for the MoM implementation of the magnetic-field integral equation that account for the three sources of numerical error: domain discretization, solution discretization, and numerical integration.  

We isolated and measured the solution-discretization error by integrating exactly over the domain.  To integrate exactly, we manufactured the Green's function, and presented optimization approaches to select a unique solution when the manufactured Green's function makes the matrix practically singular.

We isolated and measured the numerical-integration error by canceling and eliminating the solution-discretization error.  Canceling the solution-discretization error uses the basis functions, whereas eliminating does not.   

To address the domain-discretization error, we accounted for curvature and we neglected the curvature.  We accounted for curvature by modifying the integrand to effectively integrate over curved triangular elements, such that we could measure the numerical-integration error.  We neglected the curvature by using planar triangular elements to isolate and measure the numerical-integration error.





We considered cases with and without coding errors to demonstrate the efficacy of these approaches.

\reviewerOne{To account for the actual Green's function, this work can be complemented with unit tests that assess the evaluation of the (nearly) singular integrals that arise from the actual Green's function.}
\section*{Acknowledgments} 
\label{sec:acknowledgments}

The authors thank Timothy Smith, Justin Owen, and Robert Pfeiffer for their insightful feedback. 
This article has been authored by employees of National Technology \& Engineering Solutions of Sandia, LLC under Contract No.~DE-NA0003525 with the U.S.~Department of Energy (DOE). The employees own all right, title, and interest in and to the article and are solely responsible for its contents. The United States Government retains and the publisher, by accepting the article for publication, acknowledges that the United States Government retains a non-exclusive, paid-up, irrevocable, world-wide license to publish or reproduce the published form of this article or allow others to do so, for United States Government purposes. The DOE will provide public access to these results of federally sponsored research in accordance with the DOE Public Access Plan \url{https://www.energy.gov/downloads/doe-public-access-plan}.

%
\clearpage
\phantomsection\addcontentsline{toc}{section}{\refname}
\bibliographystyle{elsarticle-num}
\bibliography{../quadrature_manuscript/quadrature.bib}

\end{document}